\documentclass[aps,prd,showpacs,amssymb,superscriptaddress, notitlepage,floats,floatfix,nofootinbib,10pt]{revtex4-2}
\usepackage[utf8]{inputenc}
\usepackage{graphicx}
\usepackage{dcolumn}
\usepackage{bm}
\usepackage{amsmath}
\usepackage{color}
\usepackage[dvipsnames]{xcolor}
\usepackage{hyperref}
\hypersetup{colorlinks=true, citecolor=MidnightBlue,linkcolor=CornflowerBlue, urlcolor=CornflowerBlue, linktocpage=true}
\usepackage{xfrac}

\begin{document}

\title{Tunneling method for Hawking quanta in analogue gravity}

\author{Francesco Del Porro}
\email{fdelporr@sissa.it}
\affiliation{SISSA, Via Bonomea 265, 34136 Trieste, Italy}
\affiliation{INFN Sezione di Trieste, Via Valerio 2, 34127 Trieste, Italy}
\affiliation{IFPU - Institute for Fundamental Physics of the Universe, Via Beirut 2, 34014 Trieste, Italy}
\author{Stefano Liberati}
\email{liberati@sissa.it}
\affiliation{SISSA, Via Bonomea 265, 34136 Trieste, Italy}
\affiliation{INFN Sezione di Trieste, Via Valerio 2, 34127 Trieste, Italy}
\affiliation{IFPU - Institute for Fundamental Physics of the Universe, Via Beirut 2, 34014 Trieste, Italy}
\author{Marc Schneider}
\email{mschneid@sissa.it}
\affiliation{SISSA, Via Bonomea 265, 34136 Trieste, Italy}
\affiliation{INFN Sezione di Trieste, Via Valerio 2, 34127 Trieste, Italy}
\affiliation{IFPU - Institute for Fundamental Physics of the Universe, Via Beirut 2, 34014 Trieste, Italy}

\date{\today}

\begin{abstract}
Analogue Hawking radiation from acoustic horizons is now a well-established phenomenon, both theoretically and experimentally. Its persistence, despite the modified dispersion relations characterising analogue models, has been crucial in advancing our understanding of the robustness of this phenomenon against ultraviolet modifications of our spacetime description. However, previous theoretical approaches, such as the Bogoliubov transformation relating asymptotic states, have somewhat lacked a straightforward physical intuition regarding the origin of this robustness and its limits of applicability. To address this, we revisit analogue Hawking radiation using the tunneling method. We present a unified treatment that allows us to consider flows with and without acoustic horizons and with superluminal or subluminal dispersion relations. This approach clarifies the fundamental mechanism behind the resilience of Hawking radiation in these settings and explains the puzzling occurrence of excitations even in subcritical (supercritical) flows with subluminal (superluminal) dispersion relations.
\end{abstract}

\maketitle

\section{Introduction}

The discovery of Hawking radiation from black holes set a pivotal milestone in our understanding of gravity and its intricate relationship with quantum physics. This phenomenon has not only deepened our insight into black hole thermodynamics but also raised profound questions such as the information loss issue and the trans-Planckian problem. The latter concerns the apparent dependence of Hawking radiation on the ultraviolet (UV) completion of quantum field theory and the underlying structure of spacetime.

To explore these questions, analogue gravity models were developed, beginning in 1981 with William Unruh's seminal paper~\cite{Unruh:1981cg}. These models simulate quantum field theory (QFT) phenomena in curved spacetime within laboratory settings. In addition, they offer a concrete example where the UV completion of the theory is explicitly known.

Indeed, about ten years after Unruh's paper, it was realised in~\cite{Jacobson:1991gr} that analogue gravity could provide a physical model for the ``trans-Planckian modes'' believed to be relevant for the Hawking effect. Shortly thereafter, the investigation of Hawking radiation in the presence of modified dispersion relations was further explored~\cite{Jacobson:1993hn, Unruh:1995je}.

It was soon recognised that analogue gravity systems (see~\cite{Barcelo_2005_review} for an extensive review) provide an ideal testing ground for the robustness of Hawking radiation against high-energy modifications. This is due to their theoretical simplicity and versatility, as well as their capability to offer explicit tabletop experimental settings to test such predictions.

Following the aforementioned pioneering works, the resilience of Hawking radiation in analogue systems was later investigated and confirmed through numerous theoretical studies (see e.g.~\cite{Unruh:1995je, Brout:1995wp, Corley:1996ar, Corley:1997pr, Himemoto:1999kd, Saida:1999ap, Unruh:2004zk}). Nonetheless, it was mainly due to the exhaustive investigations carried out by Parentani and collaborators in the second decade of this century that a more comprehensive understanding of this phenomenon was accomplished~\cite{Macher:2009tw, Macher:2009nz, Finazzi:2010yq, Finazzi:2011jd, Coutant:2011in, Finazzi:2012iu, PhysRevD.90.044033,Michel:2015aga}. However, these theoretical approaches relied predominantly on the Bogoliubov coefficient method, complemented by semi-analytical and numerical analyses.

Although Bogoliubov transformations provide a versatile method, they imply the comparison of asymptotic states and assumed certain boundary conditions, such as asymptotic flatness, to solve the integrals of the overlap matrix elements. While the mathematical standing of this method is unrivaled, its intrinsically non-local nature comes with the drawback of being oblivious to the local physics behind the effects it describes.

Here, to address these shortcomings, we employ a quasi-local technique: the tunneling picture for Hawking radiation~\cite{parikh2000hawking, srinivasan1999particle}. This method uses a geometric optics approximation to describe the particle creation process as a quantum tunneling event, involving complex paths across the causal boundary. In contrast to the Bogoliubov coefficients, no knowledge of the system's asymptotic boundary is necessary, allowing particle creation to be studied almost entirely locally, wherever the event may happen in spacetime. This property is especially useful for localising effects involving effective horizons, such as those associated with modified dispersion relations that characterise fields in analogue spacetimes. Still, the tunneling method has its own limitations such as the local nature of its insights, which remain oblivion to propagation effects (e.g.~gray body factors), or the intrinsic assumption of the validity of the WKB approximation (something shared with the Bogoliubov based calculations and well established at the horizon also in presence of modified dispersion relations).

Building upon the method of characteristics introduced in~\cite{brout1995primer}, we show that the application of the tunneling method allows us to reproduce most of the known results in the literature through simple analytic calculations. Furthermore, it provides a physically elegant interpretation of the Hawking effect and its robustness in acoustic analogue gravity. Our analysis will encompass dispersion relations with both superluminal and subluminal group velocities, with respect to the low-energy speed of the phononic excitations, propagating on fluid flows (analogue spacetimes) with or without acoustic horizons. In particular, we shall explain in a simple and intuitive way the puzzling persistence of particle creation (depending on the form of the dispersion relation) observed in flows that fail by a little to generate an acoustic horizon and are everywhere subsonic or supersonic.

Note, the scope is of this paper is to be a primer into the tunneling formalism for analogue systems and to illustrate its usefulness in capturing the salient features of the Hawking effect in acoustic flows. As such, we will limit ourselves to the process of local particle production, while a full agreement with the complexity of the analysis of the Bogoliubov coefficients would require additional steps that go beyond the aim of this work.

In what follows, we start our investigation with a brief recap of the general setup in section \ref{secGS}, covering the acoustic metric as well as the fields evolving with a generalised dispersion relation. In section \ref{secHE}, we introduce the tunneling picture and show how the Hawking effect is derived for modes subjected to subluminal as well as superluminal dispersion. Having established the general framework, we particularise our treatment to the three different cases in flow velocity: supercritical (cf. section \ref{secSUP}), critical (cf. section \ref{secCRIT}), and subcritical (cf. section \ref{secSUB}). In these sections, we perform a deep dive into the expected particle production, concluding with insights into some remaining subtle but intricate issues related to the formalism in section \ref{secSO}. Finally, in the conclusion, we discuss our results and contextualise them with respect to previous findings.

Throughout the article, we will use a mostly positive metric signature and take (unless otherwise stated) $\hbar=k_B=1$. Additional conventions adopted later will be introduced at the appropriate positions.

\section{General setting}\label{secGS}
Let us consider a $(1+1)$-dimensional stationary geometry, in which the background acoustic metric is given by the line-element
\begin{equation}\label{eq:acoustic_metric}
    \mathrm{d}s^2=-c_s^2(x) \mathrm{d}t^2 + (\mathrm{d}x-v(x) \mathrm{d}t)^2=-(c_s^2(x)-v^2(x))\mathrm{d}t^2-2v(x) \mathrm{d}x\mathrm{d}t + \mathrm{d}x^2 \,,
\end{equation}
where $c_s(x)$ describes the speed of sound $c^2_s\equiv \partial p/\partial \rho$, while $v(x)$ represents the velocity profile of the fluid \cite{Barcelo_2005_review}. 
Normally, in analogue gravity experiments, sonic horizons are realised by keeping $c_s$ constant and varying the velocity profile or vice versa (for example, by a Feshbach resonance in a Bose--Einstein condensate). In what follows, we shall adopt for simplicity (and easier transposition to Lorentz breaking spacetime settings) the first ansatz and assume $c_s={\rm constant}$.

In \eqref{eq:acoustic_metric}, the sign of $v(x)$ determines the nature of this spacetime. This line element has the same structure as the Schwarzschild one in Painlev'e-Gullstrand coordinates, which covers a black hole region or a white hole one depending on if $v(x) \le 0$ or $v(x) \ge 0$, respectively \cite{Barcelo_2004_Causal}. This is why, as in most of the extant literature, we shall adopt the convention of a negative velocity flow transitioning from subsonic to supersonic as $x$ goes from positive to negative, i.e.~with a black hole exterior on the r.h.s.~of the $x$-axis).  Note also that hereinafter we will omit to write the $x-$dependence explicitly, unless necessary. 

We now notice that this geometry enjoys a Killing vector, associated to time translation invariance in the \textit{lab frame}, given by $\chi^a\partial_a= \partial_t$. 
Together with such a lab frame, it will be useful to introduce the so-called \textit{preferred frame}, which is the system of reference where the fluid is at rest, described by the two normalised vectors
\begin{equation}
    u^a \partial_a\equiv \left( \partial_t + v \partial_x \right) \,, \qquad s^a \partial_a\equiv\partial_x \,.
    \label{eq:vect}
\end{equation}
Here, $u^a$ represents a notion of preferred time direction, while $s^a$ is the preferred notion of space. 
Note also, that the metric \eqref{eq:acoustic_metric} admits a Killing horizon located at
\begin{equation}\label{eq:KH}
    |\chi|^2= v^2-c_s^2=0 \,.
\end{equation}
It is due to the presence of such an horizon, where the fluid velocity overcomes the speed of sound, that we refer to this system as sonic black hole.

\subsection{Perturbations with modified dispersion relation}

In acoustic gravity, long wavelength perturbations, the so-called phonons, behave like a massless scalar field $\phi$ on the background \eqref{eq:acoustic_metric} (see e.g.~\cite{Barcelo:2005fc}). They obey, therefore, the Klein-Gordon equation $\Box\phi=0$ and enjoy a linear dispersion relation $\omega^2(k)=c^2_sk^2$. 
However, at short wavelengths the microstructure of the effective spacetime, characterised by some scale $\Lambda$, emerges from more general dispersion relations of the form $\omega^2(k)=c^2_{s}F(k^2,\Lambda)$ (where we are assuming invariance under parity and that for low $k$ one has $F(k^2,\Lambda)=k^2+{O} (k^4,\Lambda)$ to recover the relativistic limit). Hence, the equation of motion for the perturbations becomes  \cite{schutzhold2008origin}
\begin{equation}
-(\partial_t+\partial_x v)(\partial_t+v\partial_x )\phi+c^2_{{s}}F(\partial_x^2,\Lambda)\phi=0.
\end{equation}
This setup accommodates two free functions: the fluid's velocity profile $v(x)$ and the dispersion relation function $F(k^2,\Lambda)$.  
For what concerns our stationary flow's velocity profile $v(x)$, we shall assume that this quantity can in general be described by a monotonous function and we will treat setups in which the flow admits a subsonic and a supersonic region (super-critical flow) or a purely subsonic setup (sub-critical flow) separately.\footnote{Later on, we shall also discuss purely supersonic flows, but only as a dual case with respect to the purely subsonic ones.} We address the limiting case in which the flow is everywhere sub-critical except for a sonic point (critical flow) too.

For what concerns the dispersion function, $F(k^2,\Lambda)$ we limit ourselves to the first order corrections to the phononic dispersion relation, i.e.~to modified dispersion relations of the form $\omega^2=c^2_{s}\left(k^2+{ O} (k^4, \Lambda)\right)$. Such dispersion relations could stem from a modified Klein-Gordon equation of the form
\begin{equation}\label{eq:EOM}
    \left(\square + \xi \frac{\Delta^2}{\Lambda^2} \right) \phi=0 \,.
\end{equation}
Here $\Delta= (g^{ab} + u^a u^b)\nabla_a \nabla_b$ is a purely spatial operator in the preferred frame provided via the flow four-velocity $u^a$ which plays the role of an aether field in the analogue model. The parameter $\xi= \pm 1$ determines the sign of the higher derivative operators. However, different analogue systems predict different values for such parameters.

In what follows, we shall call the dispersion relations with $\xi=\pm 1$ superluminal (upper sign) and subluminal (lower sign) respectively, as they correspond to cases for which the group velocity of perturbations is always larger/smaller than the speed of sound $c_{s}$. This is somewhat an abuse of language as these scenarios would be more precisely labelled as supersonic/subsonic but we decided to avoid such labelling to not risk any confusion with the nature of the underlying hydrodynamic flow.

\subsection{Particles}
Now we delve further into the nature of the modified Klein-Gordon equation \eqref{eq:EOM}, and provide a particle interpretation to the field $\phi$. To this aim, we adopt a WKB approximation
\begin{equation}
    \phi=\phi_0 e^{i S} \quad \mbox{and} \quad k_a=-\partial_a S
\end{equation}
where $\phi_0$ is a slowly varying amplitude and $S$ is a phase that represents the point-particle action.

The trajectory of the associated ray is determined by minimising the classical action. Particles on this curve will move with the group velocity determined from the dispersion relation of the system. Introducing the four-momentum $k_a$ enables us to rewrite the field equation as a dispersion relation for a point particle, at the leading order in the WKB formalism\footnote{For the sake of simplicity, we limit our analysis to dispersion relations without a deep minimum in the ultraviolet. Indeed, such minima are known to cause roton instabilities that are in tension with the universality of the Hawking effect and deserve a more situational analysis of the particle production, cf. \cite{fischPhysRevD.105.124066,fischPhysRevD.107.L121502}.}
\begin{equation}\label{eq:DR}
    \omega^2=c_{s}^2 \left(k^2+\xi \frac{k^4}{\Lambda^2} \right)\,.
\end{equation}
The relation for $\omega(k)$ given in \eqref{eq:DR} has been written in the preferred frame, such that $\omega= k_a u^a$ becomes the preferred notion of energy and $k=k_a s^a$ the preferred (spatial) momentum.

Of course, due to the stationarity of our flow, the system features a timelike Killing vector $\partial_t$ and thus provides a notion of Killing energy $\Omega=k_a \chi^a$ for the particle. This can be linked to the preferred frame's energy $\omega$ in the very simple way
\begin{equation}
    \Omega=\omega-vk \,.
    \label{eq:Omom}
\end{equation}
Since $\Omega$ is associated to a translational symmetry, the idea is to find mode solutions $\phi_\Omega$ of \eqref{eq:DR} at fixed Killing energy, which can be proven to be a conserved quantity even for a modified dispersion~\cite{Cropp_2014_ray_tracing}. This, together with the dependence of our modified dispersion relation on the effective field theory scale $\Lambda$, strongly suggest the introduction of a dimensionless parameter  $\alpha\equiv\Omega/\Lambda$ that captures most of the mild deviations from the relativistic behaviour (recovered in the limit $\alpha\to 0$ if one excludes the special case of the zero-mode). 

Although one could explore \eqref{eq:DR} in principle also for $\alpha\geq 1$, its interpretation as the lowest order of a Taylor expansion renders this limit physically and logically unsound. For this reason we shall trust our results only for small $\alpha$ values, say $\alpha\lesssim 0.1$. Actually, for the subluminal branch of \eqref{eq:DR}, there is in any case a ``hard limit'', provided by  $\alpha=0.5$. This marks the upper value of $\alpha$ for which the subluminal dispersion relation remains meaningful, that is, $\omega\in\mathbb{R}$. For this reason, in what follows, we shall use such ``objective'' value as a reference upper bound $\alpha_{\rm max}$ for both the superluminal and subluminal dispersion relation branches, albeit, as said, in realistic analogue models lower values of $\alpha_{\rm max}$ would certainly have to be considered.

\subsection{Particle trajectories on the background}

Once we have found our WKB set of modes $\{ \phi_\Omega \}$, a particle notion is given by their superposition, peaked around some energy $\Omega$. The resulting particle then travels with the group velocity 
\begin{equation} \label{eq:cg}
    c_g= \frac{\partial \omega}{\partial k}= c_s^2 \frac{k}{\omega} \left( 1+2\xi \frac{k^2}{\Lambda^2} \right)\approx  c_s \left( 1+\frac{3}{2}\xi \frac{k^2}{\Lambda^2} \right) \,,
\end{equation}
where the last step holds in the physically relevant limit $k\ll \Lambda$. Let us point out that, as anticipated, from the above expression follows immediately that for any $k$ one has $|c_g|\ge c_s$ for $\xi >0$ and $|c_g|\le c_s$ for $\xi <0$. The corresponding trajectory is locally given by
\begin{equation}\label{eq:dtdx}
(c_g u_a+s_a)\mathrm{d}x^a=0 \iff \frac{\mathrm{d}t}{\mathrm{d}x}= \frac{1}{c_g + v} \,,
\end{equation}
where the dual vectors $u_a=(-1,0)$ and $s_a=(-v,1)$ are deduced from \eqref{eq:vect}.

The above expression also implies that the action for such point particle will take the form
\begin{equation}
    S=-\Omega \int {\frac{(c_g u_a+s_a)}{(c_g u_t+s_t)}}\mathrm{d}x^a=-\Omega\left( t+\int{\frac{\mathrm{d}x}{c_g+v}}\right)
    \label{eq:Action}
\end{equation}
The last expression can be formally integrated, to obtain the shape of the trajectory $t(x, \alpha)$ in the $(x,t)$-plane, which, as we shall see, is a function that depends on the Killing energy of the particle through $\alpha$. For a relativistic particle, $c_g= \pm c_s$ which exhibits that \eqref{eq:dtdx} describes an everywhere regular, ingoing mode as well as an outgoing mode with a simple pole at the Killing horizon, where $c_s+v=0$. 

However, for modified dispersion relations the solution space is larger \cite{Del_Porro_2023_Hawking}: 
analysing \eqref{eq:DR}, while keeping $\Omega$ fixed, amounts to solving a 4${}^{\rm th}$ order algebraic equation. Nonetheless, the number of solutions at any given point $x$ is not always the same, and in particular it depends on the norm of the Killing vector. This becomes clear when plugging equation \eqref{eq:Omom} into the dispersion relation \eqref{eq:DR} so to obtain
\begin{equation} \label{eq:DR2}
    \xi {c_s^2}\frac{k^4}{\Lambda^2} - \left(v^2-c_s^2\right) k^2 - 2 {v} \Omega k - {\Omega^2} =0 \,.
\end{equation}

If we consider a flow, modelling a black hole, i.e. $v(x)\le 0$, we find that all the coefficients in front of the various $k^n$-terms are of fixed-sign, with the exception of the coefficient in front of $k^2$. This particular one is proportional to $|\chi|^2$ and thus changes sign at the Killing horizon. Therefore, there will always be a region of spacetime with four real solutions for $k$ and another one where this number reduces to two. 

\begin{figure}[htb]
\includegraphics[width=0.48\linewidth]{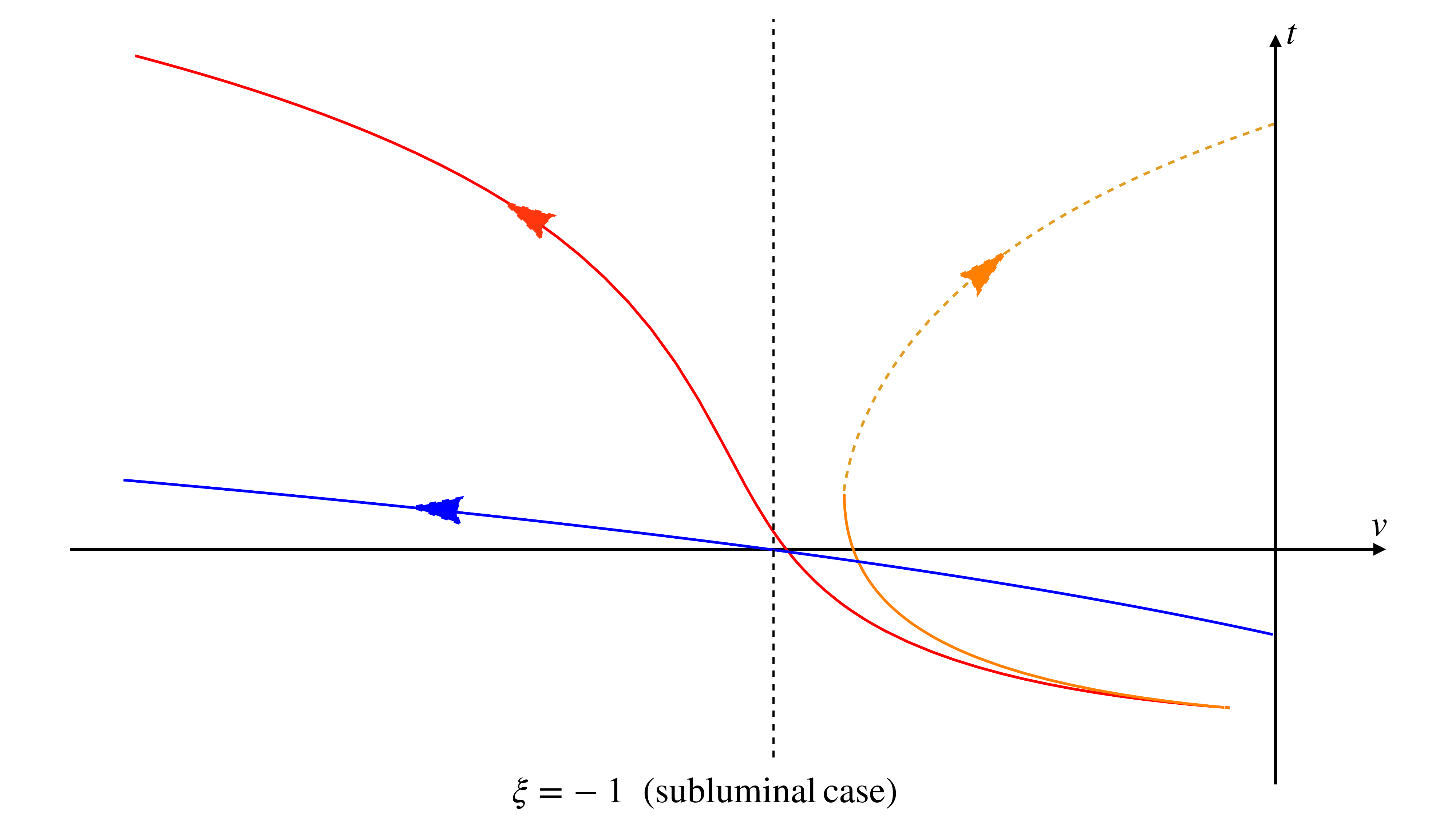}
\includegraphics[width=0.48\linewidth]{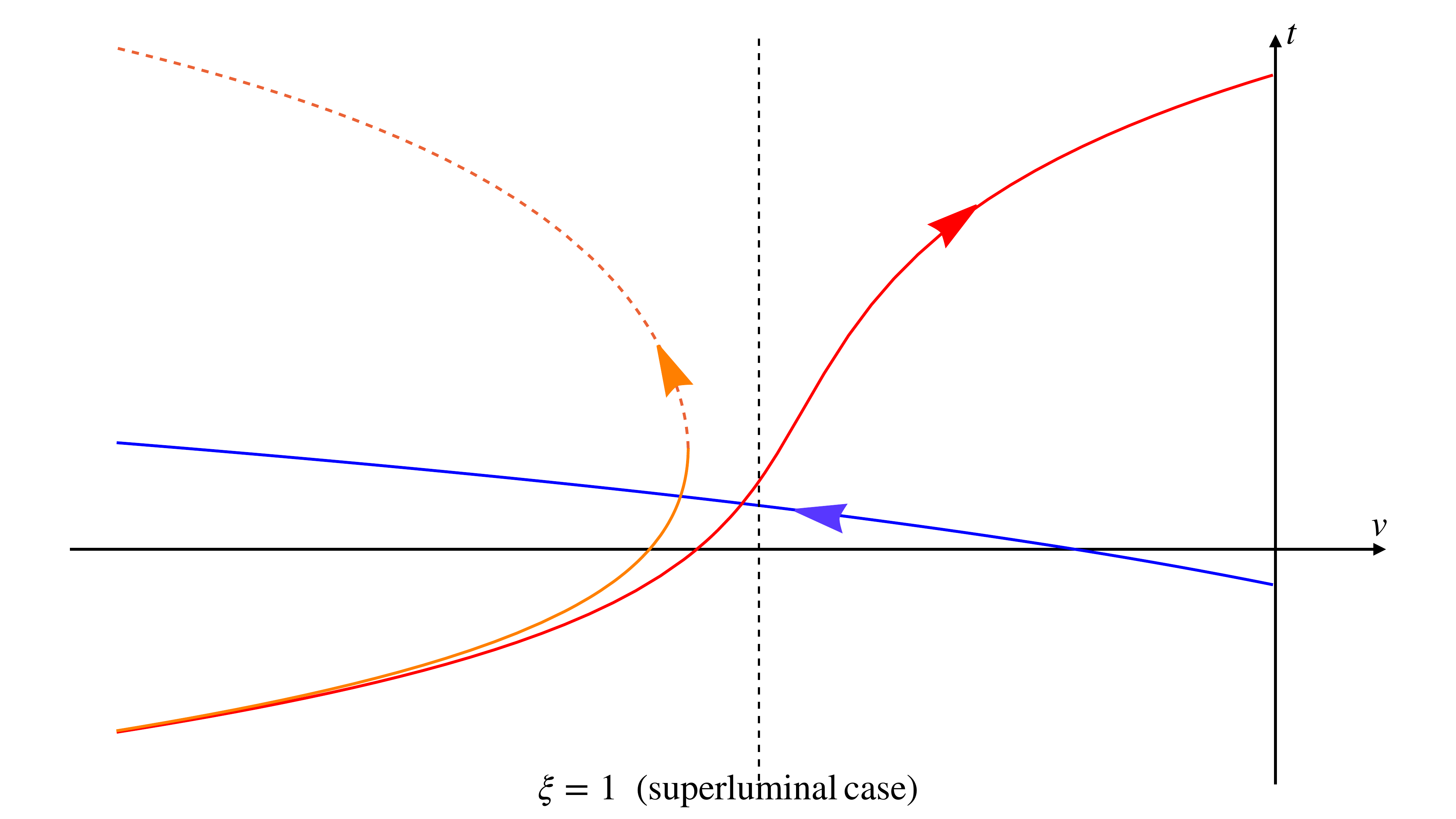}
 \caption{Numerical repesentation of the four different solution of \eqref{eq:DR2} at fixed $\Omega$. The horizontal axis shows $v$, without specifying any profile $v(x)$ yet, while $c_s=1$ everywhere. The dashed black vertical line is the Killing horizon $v=-1$ in both figures. In all plots, we have taken $\alpha=0.02$ for each ray. Left panel: subluminal case ($\xi=-1$); we see that the turning point, where the dashed and the solid orange lines meet, lies outside the Killing horizon. Right panel: superluminal case ($\xi=1$) for which we find turning point to be inside the Killing horizon. Both cases share a regular mode (in blue) which travels inwards and a mode (in red) which lingers at the horizon. The latter changes its direction depending on the sign of $\xi$, while the blue one remains qualitatively unchanged.}
 \label{f:traj}
 \end{figure}

The boundary between these two regions depends on the energy and is located at the point $x_{\textsc{tp}}(\alpha)$ (with TP standing for ``turning point'') where two of the four solutions become degenerate.
In terms of trajectories, this represents two smoothly merging trajectories at $x_{\textsc{tp}}(\alpha)$. The name turning point can be understood graphically in the $(t,x)$-plane. The curves at the meeting point can always be interpreted as two branches of a single trajectory that, in the $(t,x)$-plane, turns back at $x_{\textsc{tp}}(\alpha)$. Additionally, the position of $x_{\textsc{tp}}(\alpha)$ depends crucially on the sign of $\xi$. In the superluminal case, $x_{\textsc{tp}}(\alpha)$ can always be found inside the horizon, where $|\chi|^2>0$, while in the subluminal case $x_{\textsc{tp}}(\alpha)$ lies outside. This follows directly from the discriminant of \eqref{eq:DR2}. The shape of the trajectories is sketched for both cases in Fig.~\ref{f:traj}.

In the limit for which $\alpha \to 0$ ($\Lambda \to \infty$), the sub- and superluminal cases degenerate and we recover the relativistic behaviour, that is, two of the four solutions cease to exist, leaving us only with the upper branch of the turning mode (the dashed part of Fig.\ref{f:traj}), on one side of the Killing horizon, and with half of the lingering mode on the other side. These two represent the usual outgoing-ingoing relativistic particles (the would-be Hawking pairs) that peel infinitely at the horizon.

\section{Hawking radiation by tunnel method}\label{secHE}
In relativistic settings, i.e.~ when considering phononic dispersion relations, the presence of a sonic/Killing horizon is sufficient for sonic black holes to emit particles. Essentially, such a process reflects the analogue of the standard Hawking effect \cite{unruh1995sonic}. Within a tunneling approach \cite{parikh2000hawking,srinivasan1999particle}, this correspondence can easily be seen by analysing the outgoing particle peeling off outwards from the horizon and its counterpart peeling off inwards on the other side. This approach has been substantiated by several investigations, e.g. \cite{moretti2012state} where its connection with the quantum correlation function across the horizon is shown by means of algebraic quantum field theory. 

Let us stress, that while original derivation through null-geodesics \cite{parikh2000hawking} relied on specific coordinates and had to include the effect on the horizon of the Hawking quanta emission --- so mixing the cause-and-effect relation between Hawking radiation and the shrinking of the horizon --- here we shall adopt an Hamilton--Jacobi formalism providing a physically more sensible framework which does not require to include such backreaction.\footnote{It is interesting nevertheless to note, that in some analog systems like in a canonical acoustic black hole in Bose-Einstein condensates, the Hawking radiation backreaction can been studied --- simply it is ruled by geometrodynamical equations different from the Einstein ones --- and it was found to lead to a shrinking of the horizon as in general relativity~ \cite{Liberati:2020mdr}. This is not so surprising: for example it is what one would expect for an idealized radial flow in a sink endowed with an acoustic horizon. For a regular isolated system, the energy extracted via the related Hawking process can only be provided by the flow kinetic energy. The Hawking process would then induce a reduction of the flow speed (a change in the velocity profile) implying the realization of the sonic horizon deeper down in the sink geometry, i.e.~a shrinking of the acoustic black hole.} For a more detailed discussion, and a proof of the equivalence of the two methods, see~\cite{Vanzo_2011_Tunneling}.


If $x_{\textsc{kh}}$ be the horizon's location, the tunneling rate will be given by \cite{Vanzo_2011_Tunneling,Giavoni:2020gui}
\begin{equation} \label{eq:tunneling_rate}
    \Gamma=e^{-2 {\rm Im} (S)}, \quad \mbox{where} \quad {\rm Im}(S)=- \lim_{\varepsilon \to 0^+} \int_{x_{\textsc{kh}}- \varepsilon}^{x_{\textsc{kh}}+ \varepsilon} k_x \mathrm{d}x= - \lim_{\varepsilon \to 0^+} \Omega  \int_{x_{\textsc{kh}}- \varepsilon}^{x_{\textsc{kh}}+ \varepsilon} \frac{\mathrm{d}t(x)}{\mathrm{d}x} \mathrm{d}x \,,
\end{equation}
where we have used Eq.~\eqref{eq:Action} and, from here on, set $c_s=1$, for the sake of simplicity.

Formula \eqref{eq:tunneling_rate} can be heuristically understood as a complex path in $x$ followed by the two Hawking partners, one peeling off the horizon from inside and the other from outside (cf. \cite{Vanzo_2011_Tunneling,Giavoni:2020gui} for details). 
In the relativistic limit, i.e.~taking \eqref{eq:dtdx} with $c_g=1$, we get
\begin{equation} \label{eq:tunneling_rate_2}
   {\rm Im}(S)= - {\rm Im}\left[\lim_{\varepsilon \to 0^+} \frac{\Omega}{\partial_x (1+v)|_{x_{\textsc{kh}}}}  \ln(x-x_{\textsc{kh}}) \biggl|_{x_{\textsc{kh}}- \varepsilon}^{x_{\textsc{kh}}+ \varepsilon} \right] = \frac{\Omega \pi}{\kappa_{\textsc{kh}}}\,.
\end{equation}
Here, we have defined $\kappa_{\textsc{kh}}:=\partial_x (1+v)|_{x_{\textsc{kh}}}$, that is the surface gravity of the Killing horizon. Plugging \eqref{eq:tunneling_rate_2} into the rate $\Gamma$ we obtain
\begin{equation} \label{eq:tunneling_rate_3}
   \Gamma= \exp \left[ -\frac{2 \pi \Omega}{\kappa_{\textsc{kh}}} \right]\,.
\end{equation}
which can be compared with a Boltzmann factor $e^{-E/T}$ to extract the thermal behaviour driven by the usual Hawking temperature $T_{\textsc{h}}= \kappa_{\textsc{kh}}/2 \pi$. 

It is important to note that, at least locally, the tunnelling method and the Bogoliubov coefficient approach are equivalent (cf. Appendix \ref{app:tunnVSbogo}). However, when considering asymptotic fluxes, the Bogoliubov method is more comprehensive as it accounts for all propagation-related effects. In this sense, our present work should be viewed as a pedagogical model for capturing the basic mechanisms of particle production, rather than a method intended to match experimental data in realistic scenarios.

\subsection{Non-relativistic case}
\label{sec:nonrel} 

Mathematically speaking, in the above relativistic treatment, non-zero transition rates presume the presence of a simple pole in the expression of $\frac{\mathrm{d}t}{\mathrm{d}x}$. However, in a non relativistic scenario, nothing comparable will ever occur, given that the modified dispersion relation, never produces an exact peeling (see Fig.~\ref{f:traj}) and therefore the particle trajectories at the Killing horizon remain analytic. In turn, one would expect an absence of Hawking radiation in this settings (or, at least, that this effect cannot be described via the tunneling method).

However, even in the presence of modified dispersion relations, analogue systems have shown to be able to reproduce Hawking radiation  both theoretically and experimentally~\cite{Coutant_2016_Hawking_subcritical,Weinfurtner:2013zfa,Coutant_2018_Hawking_superluminal,Finazzi_2011_robustness}.
Also, a similar situation has been studied by the authors in the context of a gravitational black hole in Lorentz violating gravity \cite{Del_Porro_2023_Hawking}, supporting this conclusion. Finally, such an effect is expected just by a continuity argument: if we analyse energies well below the cutoff, say $\Omega \ll \Lambda$, the relativistic behaviour should be recovered approximately.

In the following section, we shall focus on how to describe this effect in a simple, analytical way via the tunneling method. To this aim, we assume the validity of the Hamilton-Jacobi approach in the case of dispersive modes\footnote{An extension of the Hamiltonian geometry  of the phase space to particles with modified dispersion relations can be found in~\cite{PhysRevD.92.084053}.}. As we shall see, the extension of the latter to setups with modified dispersion relations will prove itself useful and informative, thus, nicely complementary to previous, semi-analytical or numerical studies ~\cite{Macher:2009tw, Macher:2009nz, Finazzi:2010yq, Finazzi:2011jd, Coutant:2011in, Finazzi:2012iu, PhysRevD.90.044033,coutant2014hawking,Michel:2015aga,Coutant_2018_Hawking_superluminal}, based on the Bogoliubov coefficients analysis.

\subsection{Approximating the modes}

The idea behind the calculation from the previous section was to find a trajectory which approximates the two partners while enjoying a simple pole structure. In the case of $c_g=1+ \delta c_g$, we recover that this trajectory behaves almost like a relativistic null trajectory with the pole at the Killing horizon. 
One can go beyond such a crude approximation, however, to do so, we shall need first to analyse the asymptotic behaviour of the solutions of Eq.~\eqref{eq:DR}.

\subsubsection{Outside the horizon}
Let us consider a flow velocity profile $v(x)$ that has a subsonic region at large $x>0$ where it asymptotes $v=0$. For $c_s=1$ this is equivalent to impose that our geometry is asymptotically flat.
If we try to solve \eqref{eq:DR} in the regime where $|v| \ll 1$ at the leading order we get
\begin{equation} \label{eq:v_small}
    \omega= \Omega + O(v) \,, \qquad \Omega^2 =k^2+ \xi \frac{k^4}{\Lambda^2} + O(v)\,.
\end{equation}
Since for $v=0$ the dispersion relation becomes an equation for $k^2$ we obtain only two real solutions with opposite signs that we name $k_0^\pm(\Omega)$, with $k_0^+=-k_0^-$. Only these solutions will reach the $v=0$ line for both superluminal and subluminal dispersion relations. 

Specifically, with reference to Fig.\ref{f:traj}, such solutions describe either the regular ingoing mode and the upper branch of the turning mode for the subluminal dispersion relation, or the regular ingoing one and the outside branch of the lingering one for the superluminal case. If we name the three group velocities associated to the aforementioned modes respectively $c_g^{\rm reg}$, $c_g^{\rm turn}$ and $c_g^{\rm ling}$, we find at $v=0$ that
\begin{equation} \label{eq:v_small_cg}
   c_g^{\rm reg}=\begin{cases} -c_g^{\rm turn} & \mbox{if } \quad \xi=-1\\ 
   - c_g^{\rm ling} & \mbox{if } \quad \xi=1 \end{cases}\,.
\end{equation}
In both, the subluminal and the superluminal case, $-c_g^{\rm reg}$ describes the trajectory of our Hawking quanta in the $|v| \ll 1$ region. Looking back at \ref{eq:DR2}, we can easily see that this approximation, which is exact at $v=0$, can be extended beyond this region, and is valid whenever $\Omega \gg k v$, thus
\begin{equation} \label{eq:v_small_validity}
   v^2 \ll \frac{1+\sqrt{1+4 \xi \alpha^2}}{2} \,.
\end{equation}

\subsubsection{Inside the horizon}
Let us now assume that for some intermediate value of $x$ our flow admits an unique acoustic horizon. Inside the latter, the asymptotic region will be described by the regime $|v| \gg 1$. If we solve \eqref{eq:DR2} in this limit, we have again two solutions with opposite group velocities such that
\begin{equation} \label{eq:v_big}
    k_\infty = - \frac{\Omega}{v+1} \,, \qquad \omega_\infty^\pm= \pm \frac{\Omega}{v+1} \,.
\end{equation}
These two solution are as well associated to the regular mode and to the Hawking partner of Fig.\ref{f:traj}. For the choice $\xi=1$ the latter is represented by the upper branch of the turning mode, while for $\xi =-1$ this role is taken by the lingering mode. In summary, at the leading order we have
\begin{equation} \label{eq:v_big_cg}
   c_g^{\rm reg}=\begin{cases} -c_g^{\rm turn} & \mbox{if } \quad \xi=1\\ 
   - c_g^{\rm ling} & \mbox{if } \quad \xi=-1 \end{cases}\,
\end{equation}
Again, looking at Eq.~\eqref{eq:DR} and \eqref{eq:Omom} one can realize that the above approximation is still valid whenever $k^2 \ll \Lambda^2$ which translates into the following condition for the flow velocity.
\begin{equation} \label{eq:v_big_validity}
   (v+1)^2 \gg \alpha^2 \,.
\end{equation}
\subsubsection{The approximant trajectory}

Putting all of the previous analysis together, we realise that \eqref{eq:v_small_cg} and \eqref{eq:v_big_cg} tell us that the trajectory defined through $-c_g^{\rm reg}$ describes always the modes associated to the ``effective Hawking pair'' in both the regions $|v|\ll 1$ and $|v| \gg 1$, independently from the nature of the dispersion relation. In other words, this path effectively interpolates between these two different solutions. Hence, we call such an effective trajectory the ``approximant''.

\begin{figure}[htb]
\includegraphics[width=0.48\linewidth]{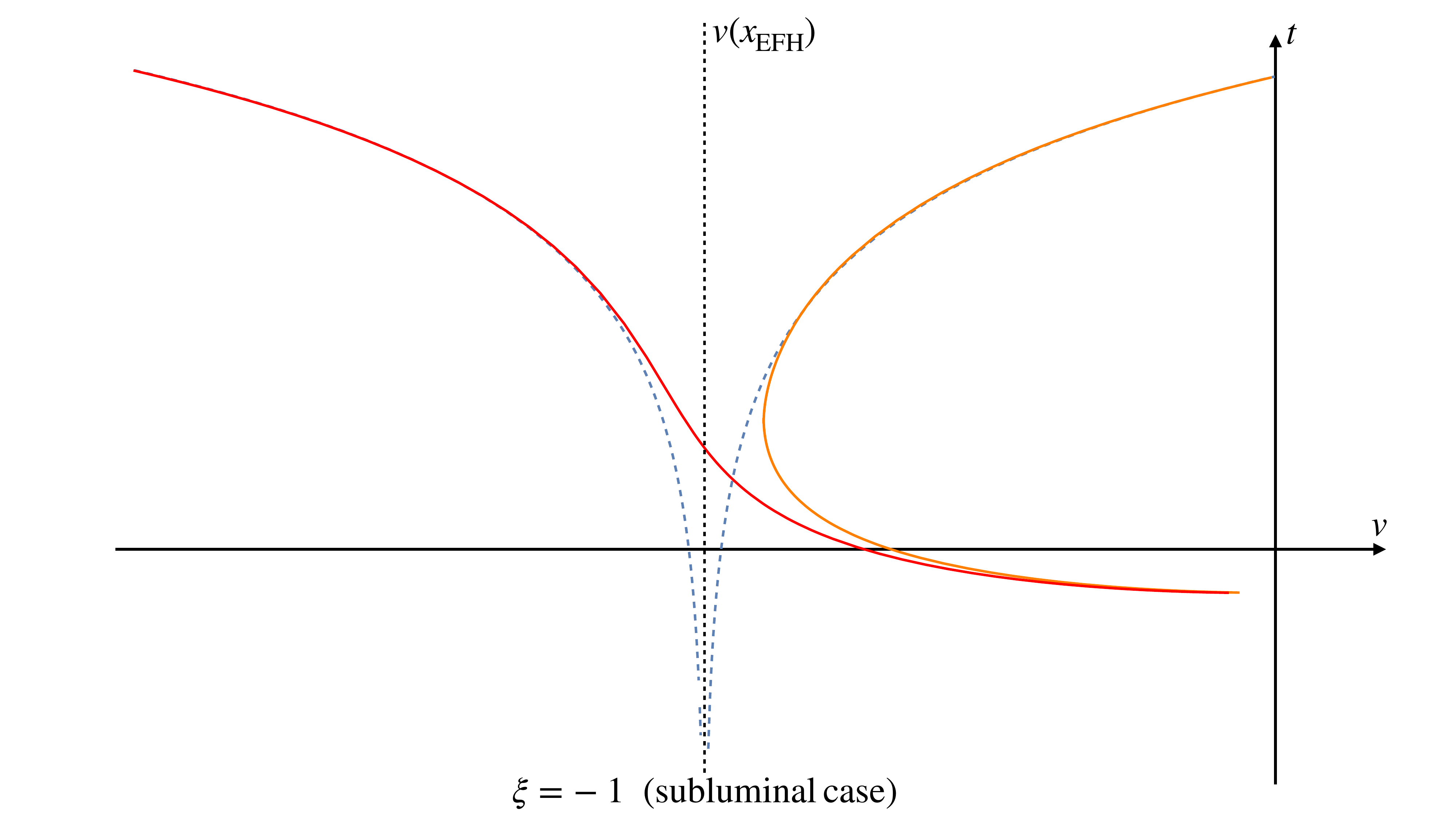} \includegraphics[width=0.48\linewidth]{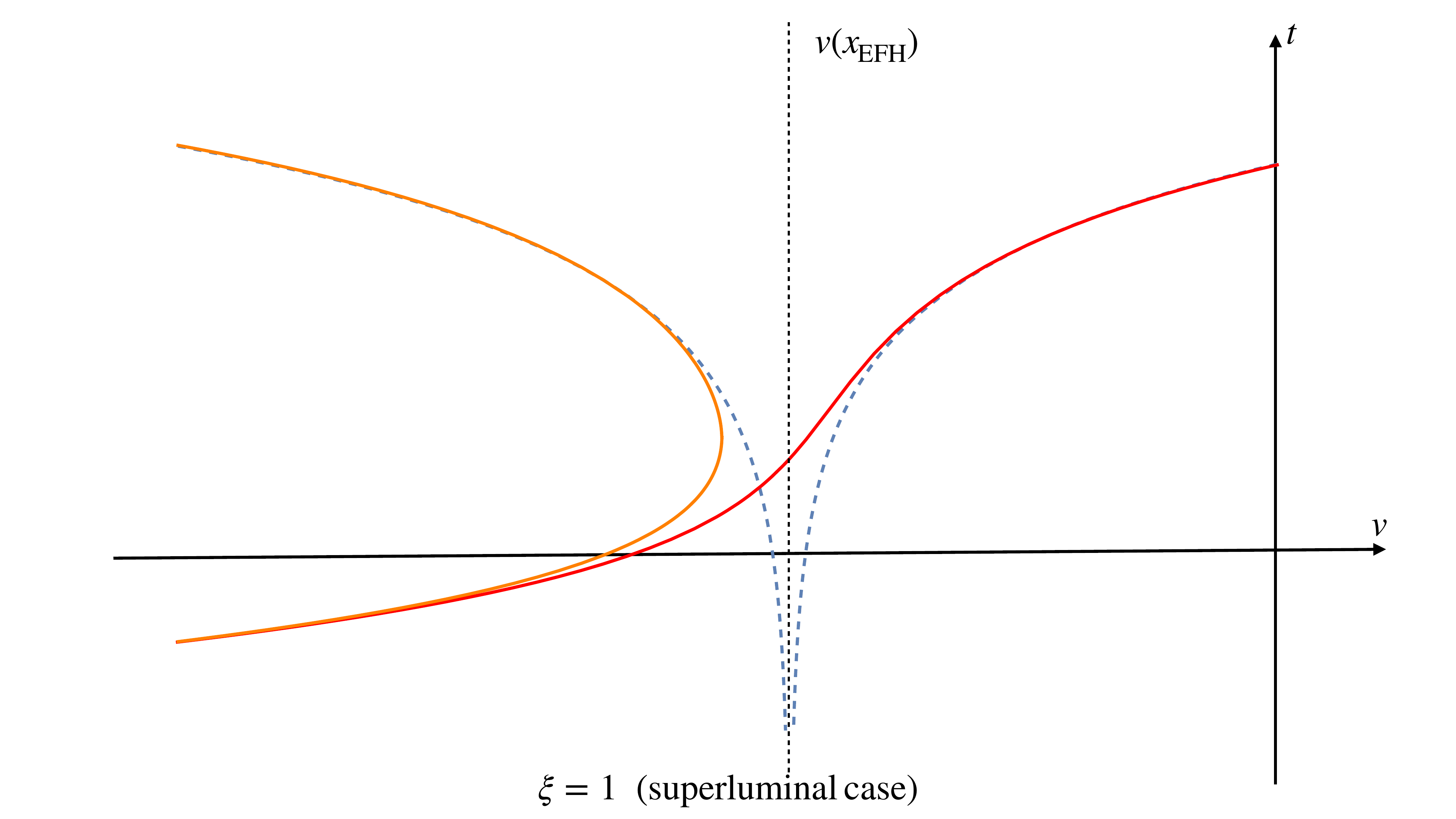}
 \caption{Approximant trajectory (dashed blue) versus actual solutions. The dashed trajectories mimic the branches responsible for the particle production in the subluminal (left) and superluminal (right) case asymptotically. Here again $\alpha=2 \times 10^{-2}$. The dashed black vertical line is the point where the approximant peels, i.e. the effective horizon $v(x_{\rm EFH})$.} 
 \label{f:approx}
\end{figure}

We shall return to this later when we assert the range of validity of such an approximation as well as the question why it is sufficient to reproduce the Hawking radiation derived via a full Bogoliubov approach.
For the moment let us see how the adoption of the approximant as a proxy for the trajectory of the Hawking pairs enables the emergence of a simple pole structure. Fig.~\ref{f:approx} provides a plot of this trajectory for both subcritical and supercritical flows and shows clearly the capacity of the approximant to uncover the presence of the effective horizon experienced by the modes associated to the Hawking process. Consequently, we can apply the tunneling method even though $\alpha$ is not perturbatively close to zero.

\subsection{Tunneling method via the approximant}
Let us now apply the ideas from the previous section to the calculation of the tunneling rate. To do so, we consider modes with energies $\alpha\Lambda$, such that the trajectories of the Hawking partners are effectively estimated by our approximant.
In particular, we take the outgoing ray to travel with speed $-c_g^{\rm reg}(x,\alpha)$ in the preferred frame. For this trajectory we find
\begin{equation} \label{eq:app_traj}
  \frac{\mathrm{d}t}{\mathrm{d}x}= \frac{1}{-c_g^{\rm reg} + v}\,.
\end{equation}
The position of the simple pole associated to this expression localises $x_{\textsc{efh}}(\alpha)$ of the ``effective horizon'' (EFH) for a particle of energy $\alpha$, so that 
\begin{equation} \label{eq:EH_cond}
  c_g^{\rm reg}(x_{\textsc{efh}},\alpha) = v(x_{\textsc{efh}}) \,.
\end{equation}
Please note, that the solution of this equation depends on $\alpha$, on the range of values of $v(x)$, and on $\xi$. If a solution to \eqref{eq:EH_cond} exists, it will allow us to define the trajectory outside of the EFH as
\begin{equation} \label{eq:app_traj_1}
  t(x)= \frac{1}{v'(x_{\textsc{efh}}) (1-\partial_v c_g^{\rm reg})|_{\textsc{efh}}} \ln[x-x_{\textsc{efh}}(\alpha)] \,.
\end{equation}
In analogy to our calculations in \eqref{eq:tunneling_rate}, we use \eqref{eq:app_traj_1} to calculate the tunneling rate as
\begin{equation} \label{eq:app_traj_2}
 \Gamma=\exp \left[- \frac{\Omega}{T(\alpha)} \right] \quad \mbox{where} \quad T(\alpha)=\frac{v'(x_{\textsc{efh}}) (1-\partial_v c_g^{\rm reg})|_{\textsc{efh}}}{2 \pi}\equiv\frac{\kappa(\alpha)}{2 \pi}\,,
\end{equation}
where we have defined $\kappa(\alpha)$ as the peeling factor of the EFH and $T(\alpha)$ as the associated ``effective temperature''. Let us stress that, despite the name, $T(\alpha)$ is energy dependent, and so the rate $\Gamma$ cannot be considered as a true Boltzmann factor; thus the emission is not purely thermal.

After this general treatment, we are going to discuss next the tunneling rate of our effective trajectory. In doing so, we address the subluminal and superluminal dispersion relations individually and specifically distinguish further between subcritical ($|v|<1$, without a horizon) and supercritical (with a horizon) flow (in doing so we also comment on the critical flow). As we shall see soon, the resulting cases have remarkable similarities but also striking differences.

\section{Particle production: supercritical flow}\label{secSUP}

\begin{figure}[htb]
 \includegraphics[width=0.5\linewidth]{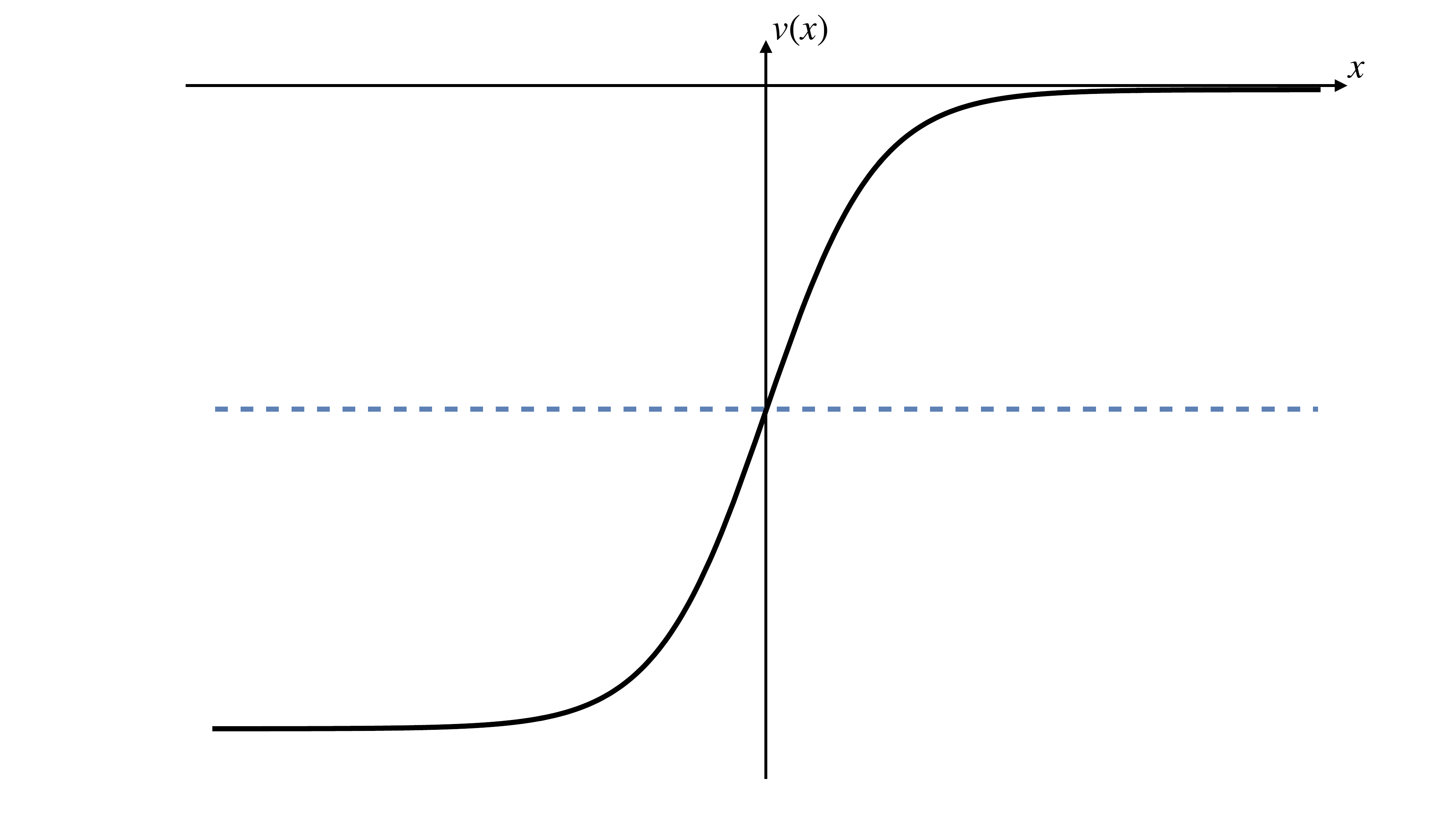}
 \caption{
 Supercritical velocity profile \eqref{eq:v_profile} for a left-going flow with one subsonic region $(x > 0)$ and one supersonic region $(x < 0)$. The dashed line marks the location of the sonic/Killing horizon, for which $x_{\textsc{kh}}=0$ and $v(x_{\textsc{kh}})=-1$.}
 \label{f:v_profile_supercritical}
\end{figure}

Our starting point will be the supercritical flow, that is to say, the permitted range for $v(x)$ supports the presence of a horizon, i.e. $|v(x)|>1$ for some $x$. For now, we keep our treatment as general as possible without particularising to a specific profile for $v(x)$. We demand: i) $v(x)$ to be monotonous to avoid inner Killing horizons (namely, $v(x)=-1$ has a single root), ii) $v \to 0$ for $x \to - \infty$, and iii) conditions~\eqref{eq:v_small_validity} and \eqref{eq:v_big_validity} hold almost everywhere apart from a small region around the effective horizon. A typical example of such flow is the one considered in~\cite{Schutzhold_2013_breakdown_WKB} and shown in figure \ref{f:v_profile_supercritical}
\begin{equation} \label{eq:v_profile}
   v(x)=  \tanh(\kappa_{\textsc{kh}} x) -1 \,.
\end{equation}
This profile interpolates between $v(-\infty):=v_{-\infty}=-2$ and $v(\infty):=v_\infty=0$ while the Killing horizon is located at $x_{\textsc{kh}}=0$, such that $\kappa_{\textsc{kh}}=v'(x_{\textsc{kh}})$ denotes the horizon's surface gravity. As long as the surface gravity (the profile steepness at the KH) is large, the region around the horizon, where the approximant will deviate from the real trajectory of the Hawking pair, will be small.

As a general feature, the supercritical flow connects our calculation with the relativistic limit in both, the subluminal and in the superluminal case, because the particle production for relativistic fields happens only in the presence of a horizon. As already mentioned, the relativistic behaviour appears when the higher order of the dispersion relation can be neglected, or in other words when $\alpha \to 0$. Now we explore such corrections for our specific dispersion relation and flow further; in doing so, we extend our investigation to a broader range of $\alpha$.

\subsection{Superluminal dispersion relation}

If we choose $\xi=1$ in \eqref{eq:DR}, we will get $|c_g^{\rm reg}| \ge 1$ everywhere. Hence, in this case, the solutions of~\eqref{eq:EH_cond} must always be located inside the Killing horizon (i.e.~for negative $x$), where $v\leq -1$ (with the equality valid for $\alpha=0$).

\begin{figure}[htb]
 \includegraphics[width=0.48\linewidth]{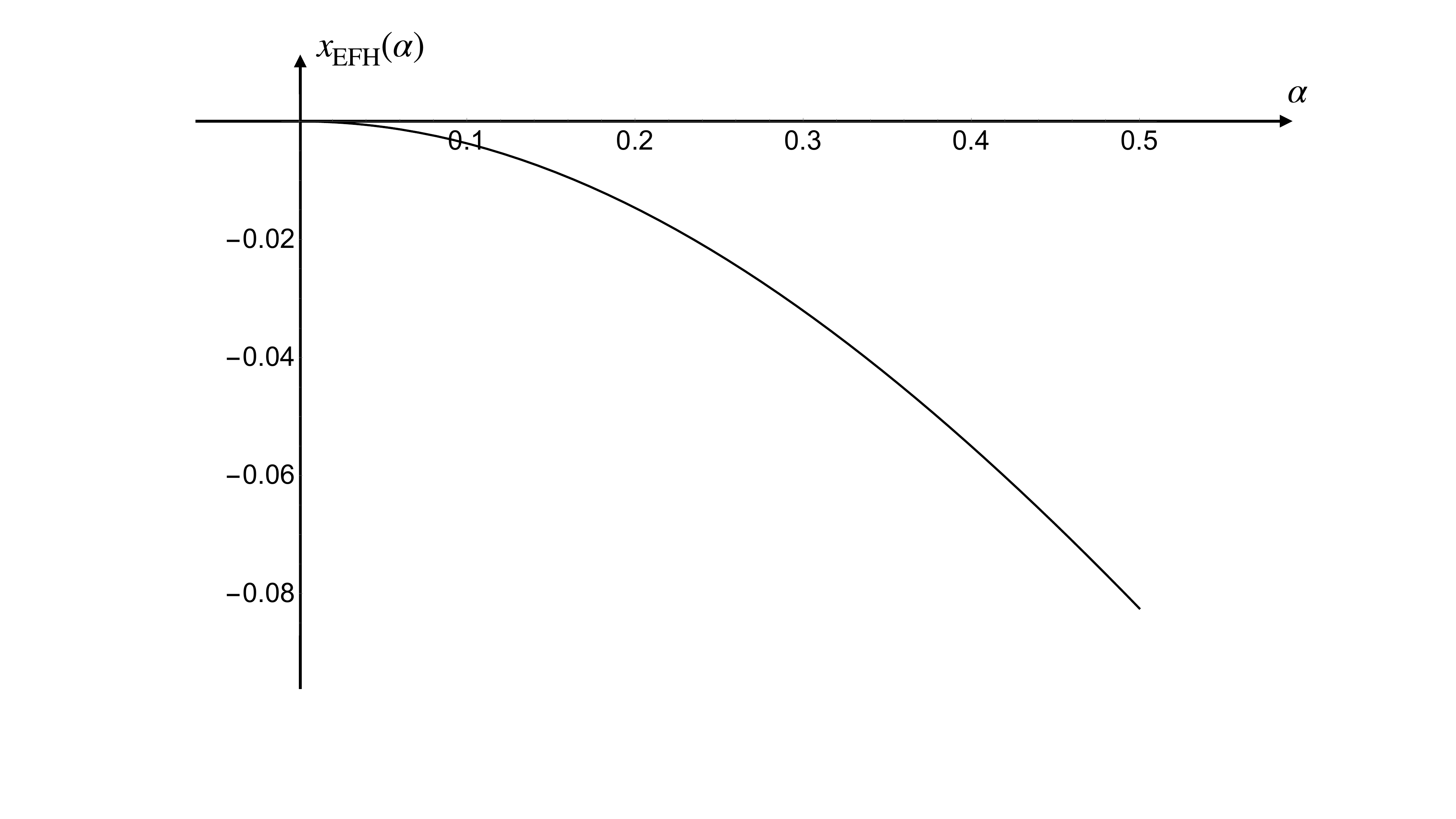}
 \includegraphics[width=0.48\linewidth]{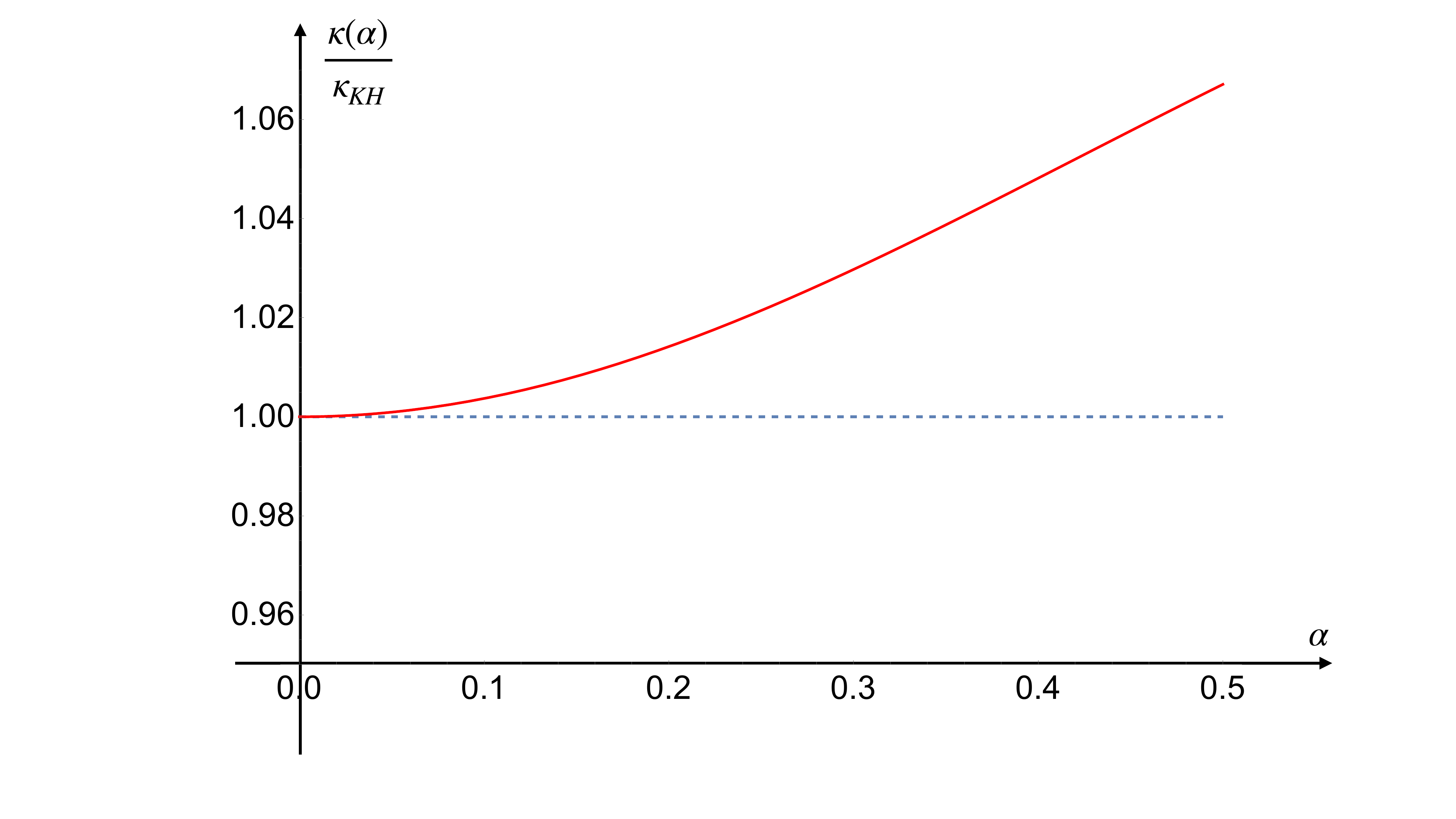}
  \caption{On the left: shape of $x_{\textsc{efh}}(\alpha)$ for the profile \eqref{eq:v_profile} up to $\alpha=1/2$. On the right: shape of the ratio $\kappa(\alpha)/\kappa_{\textsc{kh}}$. We see that, since $|x_{\textsc{efh}}(\alpha)| \ge |x_{\textsc{kh}}|$, so the EFH is inside the KH, the temperature of it appears to be slightly hotter than $T_{\rm H}$ at low $\alpha$ and then showing a $O(5\%)$ deviation from thermality for $\alpha\sim 0.5$. Note that we can directly compare the positions due to the monotony of $v(x)$}.
 \label{f:xEH_super_super}
\end{figure}

Nominally, the allowed range for $\alpha$ spans from $\alpha=0$, solving $c_g^{\rm reg}=-1$ up to $\alpha=\alpha_{\rm max}$, which is when the group velocity reaches the lower bound of $v(x)$, namely, $c_g^{\rm reg}=v_{ -\infty}$. In the case at hand, that is, $v_{-\infty}=-2$, we find $\alpha_{\rm max} \simeq 3$. This is clearly in conflict with the effective meaning given to \eqref{eq:DR}. Hence, as anticipated, we limit ourselves to values of $\alpha\leq 0.5$. In the left panel of Fig.~\ref{f:xEH_super_super}, we show the location of the effective horizon, determined numerically by Eq.~\eqref{eq:EH_cond}, for different values of $\alpha$ within the allowed range.

Similarly, once we know the shape of $x_{\textsc{efh}}(\alpha)$, one can evaluate $\kappa(\alpha)$ via Eq.~\eqref{eq:app_traj_2} and contrast it with $\kappa_{\textsc{kh}}$ of \eqref{eq:v_profile}. The ratio $\kappa(\alpha)/\kappa_{\textsc{kh}}$ is plotted in the left panel of Fig.~\ref{f:xEH_super_super}: its deviation from unity and constancy can be taken as a measure of the deviation from thermality induced by the dispersive behaviour. As we can see, in agreement with the previous studies, the Hawking effect displays a remarkable robustness given that values of $\alpha$ of order or larger than $0.1$ would have to be considered already in the far UV (as they corresponds to energies close to the effective field theory scale $\Lambda$).

\subsubsection{Low-energy regime}
Here, we focus on the low energy regime where an analytical treatment is possible. Let us start by solving \eqref{eq:EH_cond} perturbatively in $\alpha$, one obtains
\begin{equation} \label{eq:v_super_low}
    v(x_{\textsc{efh}}(\alpha))=-1-\frac{3}{8} \alpha^2 + O(\alpha^3) \,,
\end{equation}
or, if \eqref{eq:v_profile} is taken into account
\begin{equation} \label{eq:x_super_low}
    x_{\textsc{efh}}(\alpha)= - \frac{3 \alpha^2}{8 \kappa_{\textsc{kh}} }+ O(\alpha^3) \,.
\end{equation}
In both cases we find $x_{\textsc{efh}}$ to be perturbatively close to $x_{\textsc{kh}}$ with corrections starting at $O(\alpha^2)$. As already mentioned, for the superluminal case $|x_{\textsc{efh}}|>|x_{\textsc{kh}}|$. Given \eqref{eq:v_super_low}, we can compute the correction to $\kappa(\alpha)$. Since $1-\partial_v c_g^{\rm reg}= 1 + \delta_\alpha v = 1 + 3 \alpha^2/8 + O(\alpha^3)$ and $v'(x_{\textsc{efh}})= \kappa_{\textsc{kh}}+ O(\alpha^4)$ (with chosen profile \eqref{eq:v_profile}), we get
\begin{equation} \label{eq:k_super_low}
    \kappa(\alpha)= \kappa_{\textsc{kh}} \left( 1 + \frac{3}{8} \alpha^2 \right) + O(\alpha^3)\,.
\end{equation}
The result shows a similar qualitative behaviour was predicted at the Killing horizon of a Ho\v{r}ava black hole in~\cite{Del_Porro_2023_Hawking}. We find the first corrections to arise at $O(\alpha^2)$, highlighting how the emission still remains quasi-thermal for low-energy particles. The coefficient in front of the correction is different from the one found in the black hole case in \cite{Del_Porro_2023_Hawking}, however, this is not surprising because its value depends crucially on the specific geometry, in particular on the shape of $c_s(x)$ and $v(x)$.

\subsection{Subluminal dispersion relation}
\begin{figure}[htb]
 \includegraphics[width=0.48\linewidth]{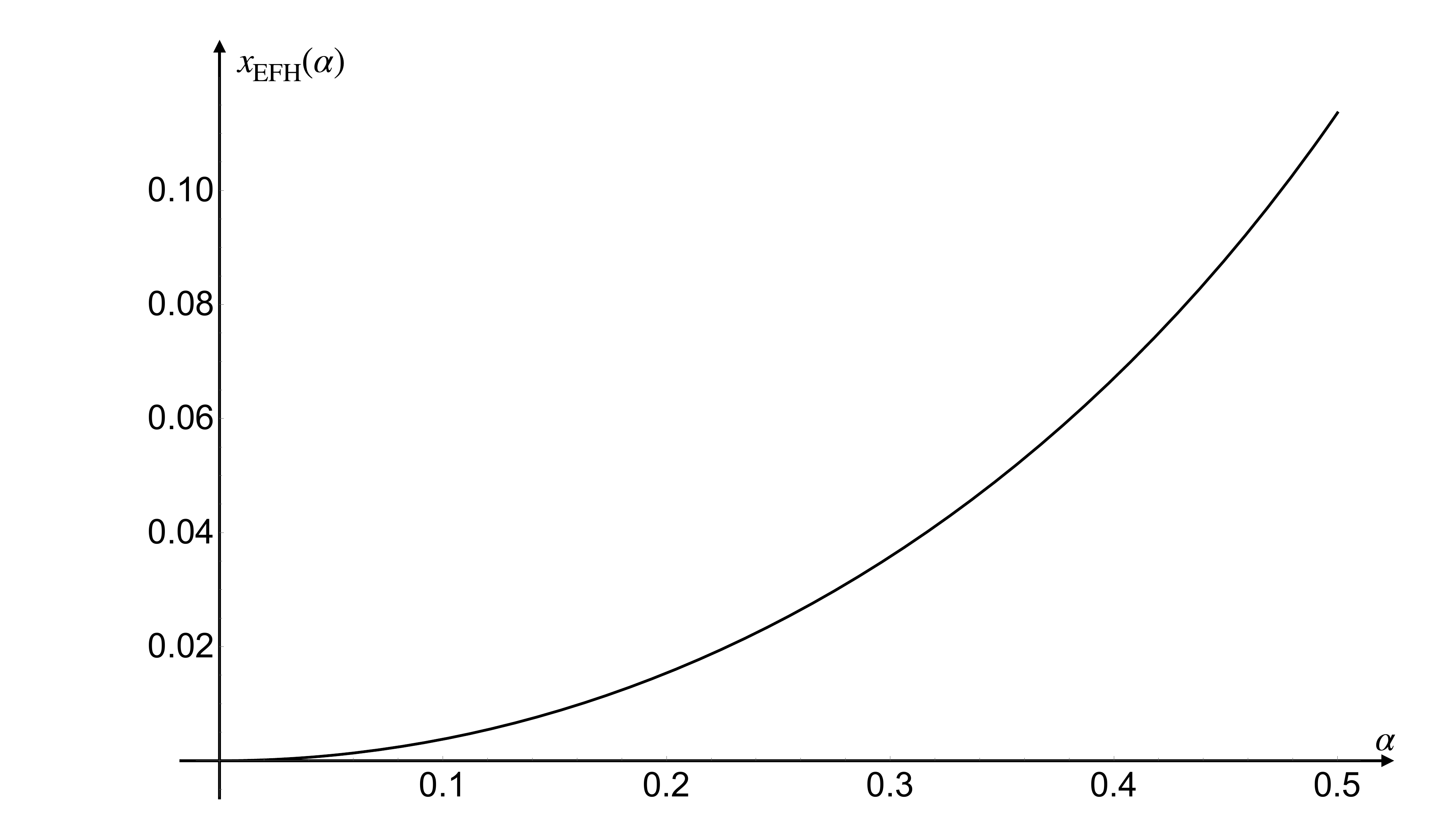}
 \includegraphics[width=0.48\linewidth]{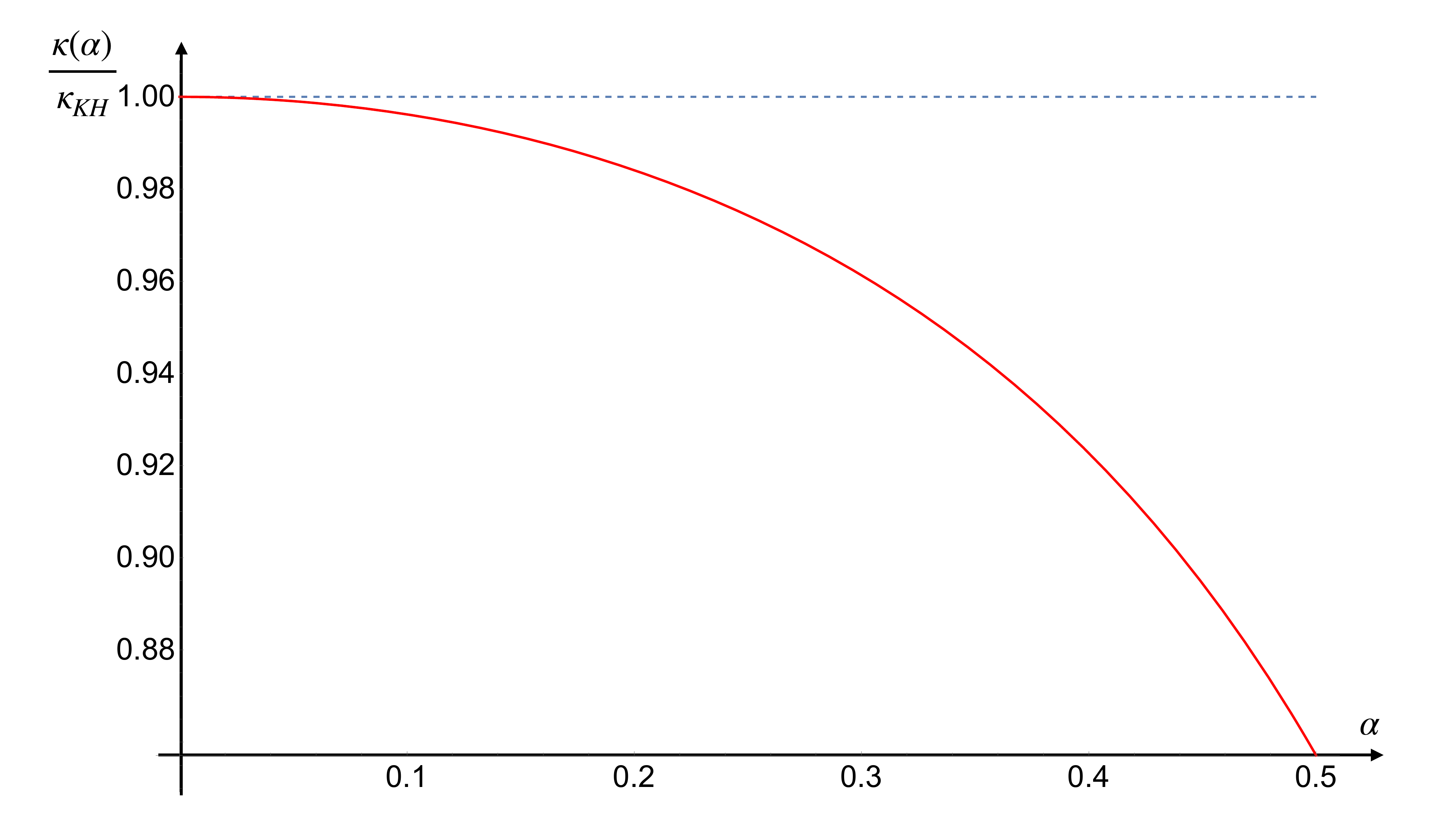}
 \caption{On the left: shape of $x_{\textsc{efh}}(\alpha)$ for the profile \eqref{eq:v_profile} up to $\alpha=1/2$. On the right: shape of the ratio $\kappa(\alpha)/\kappa_{\textsc{kh}}$. We see that, since $|x_{\textsc{efh}}(\alpha)| \le |x_{\textsc{kh}}|$, the effective horizon is inside the Killing one, the temperature of it being slightly colder than $T_{\rm H}$ at low $\alpha$ and then showing a $O(10\%)$ deviation from thermality for $\alpha\gtrsim 0.4$.}
 \label{f:xEH_super_sub}
\end{figure}

For what regards the subluminal case, $\xi=-1$, we have $|c_g^{\rm reg}|\le 1$ always, as such, \eqref{eq:EH_cond} admits possible solutions only outside the Killing horizon (i.e.~for positive $x$), where $|v| \le 1$. Again, the equality is valid for $\alpha=0$, but for values close to this minimum, perturbative analyses around the Killing horizon can be safely performed. For what concerns the upper bound in the $\alpha$-range we can scrutinise \eqref{eq:DR2} evaluated at $v=0$. The solutions are given by \eqref{eq:v_small} with $\xi=-1$. As anticipated, we find real solutions exclusively for $\alpha \le 1/2$. Once again, we compute the position of $x_{\textsc{efh}} (\alpha)$ as well as the value of the ratio $\kappa(\alpha)/\kappa_{\textsc{kh}}$ numerically so to test the robustness of Hawking radiation. The results are collected in Fig.~\ref{f:xEH_super_sub} for the allowed range. We can see again that for $\alpha\ll 0.1$ the spectrum is basically thermal with small deviations from the relativistic result.

\subsubsection{Low energy regime}
{Similarly to the treatment in previous section, we can now analyse the low-energy regime for the subluminal case. Effectively the calculations change only for the sign of $\xi$, hence we find}
\begin{equation} \label{eq:v_sub_low}
    v(x_{\textsc{efh}}(\alpha))=-1+\frac{3}{8} \alpha^2 + O(\alpha^3) \,,
\end{equation}
suggesting an effective horizon that lies outside the Killing horizon. Recalling the velocity profile \eqref{eq:v_profile} we get
\begin{equation} 
\label{eq:x_sub_low}
    x_{\textsc{efh}}(\alpha)= 
    \frac{3 \alpha^2}{8\kappa_{\textsc{kh}} }+ O(\alpha^3) \,.
\end{equation}
which consequently leads to 
\begin{equation} \label{eq:k_sub_low}
    \kappa(\alpha)= \kappa_{\textsc{kh}} \left( 1 - \frac{3}{8} \alpha^2 \right) + O(\alpha^3)\,.
\end{equation}
showing a similar correction but with the opposite sign with respect to the superluminal case. Once again, we discover that the low-energy regime admits a quasi-thermal behaviour, as expected.

\section{Particle creation: critical flow} 
\label{secCRIT}

\begin{figure}[h]
 \includegraphics[width=0.5\linewidth]{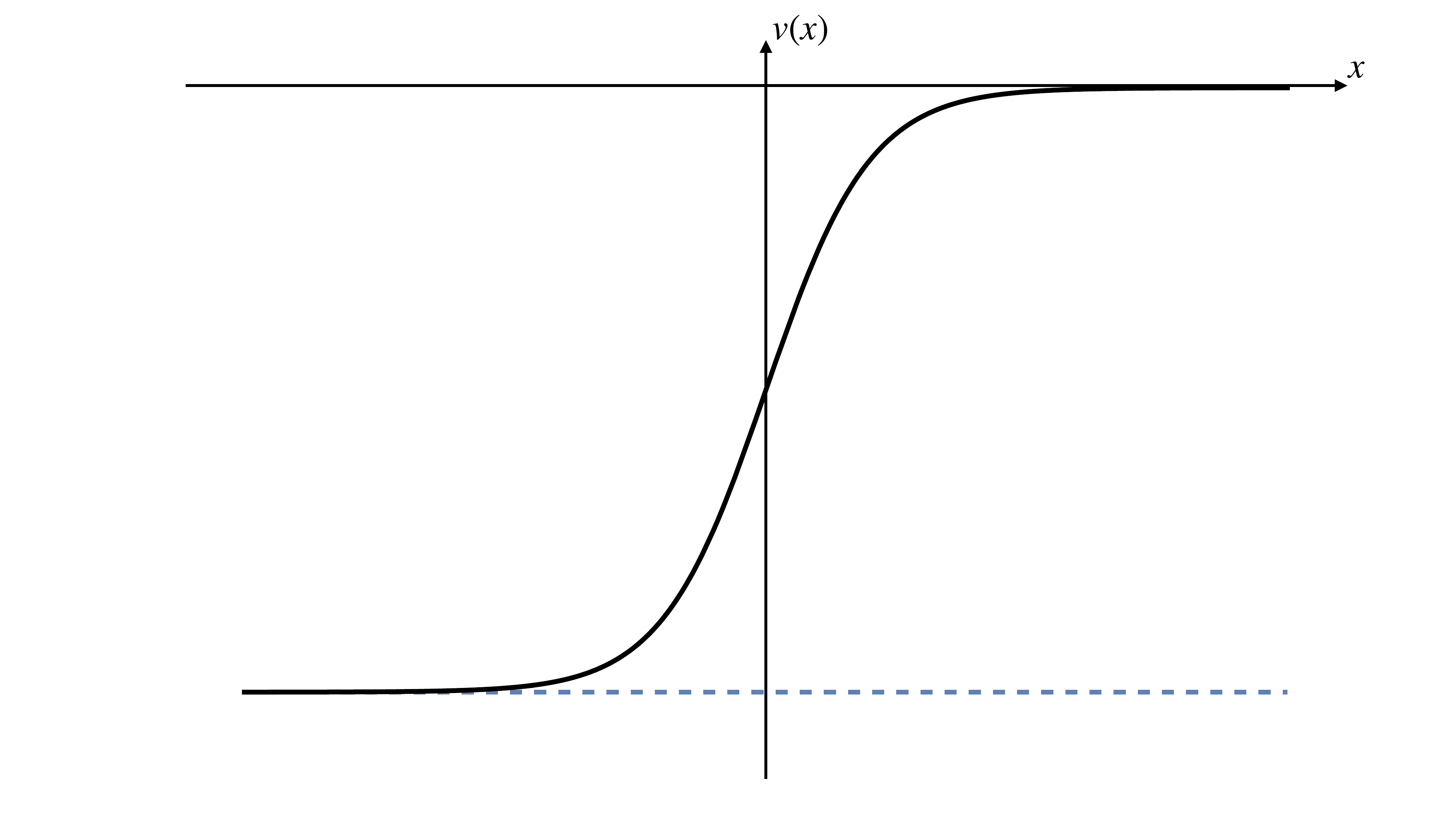}
 \caption{
 Critical velocity profile \eqref{eq:v_profile} for a left-going flow with one subsonic region $(x > -\infty)$ and one sonic point at $(x_{\textsc{kh}} =-\infty)$.
 The dashed line marks the location of the sonic/Killing extremal horizon, for which $x_{\textsc{kh}}=-\infty$ and $v(x_{\textsc{kh}})=0$.}
 \label{f:v_profile_critical}
\end{figure}

This section discusses a second possible regime for the fluid velocity: the critical flow for which $v_{-\infty}=-1$ and $|v|<1$ otherwise. An exemplary profile could be
\begin{equation} \label{eq:v_profile_critical}
   v(x)=  \frac{1}{2}[\tanh(\kappa_\circ x) -1] \,,
\end{equation}
which we plot in Fig.~\ref{f:v_profile_critical}. Note, $\kappa_\circ$ is a fiducial scale to compensate the dimension of $x$.

\begin{figure}[h]
 \includegraphics[width=0.45\linewidth]{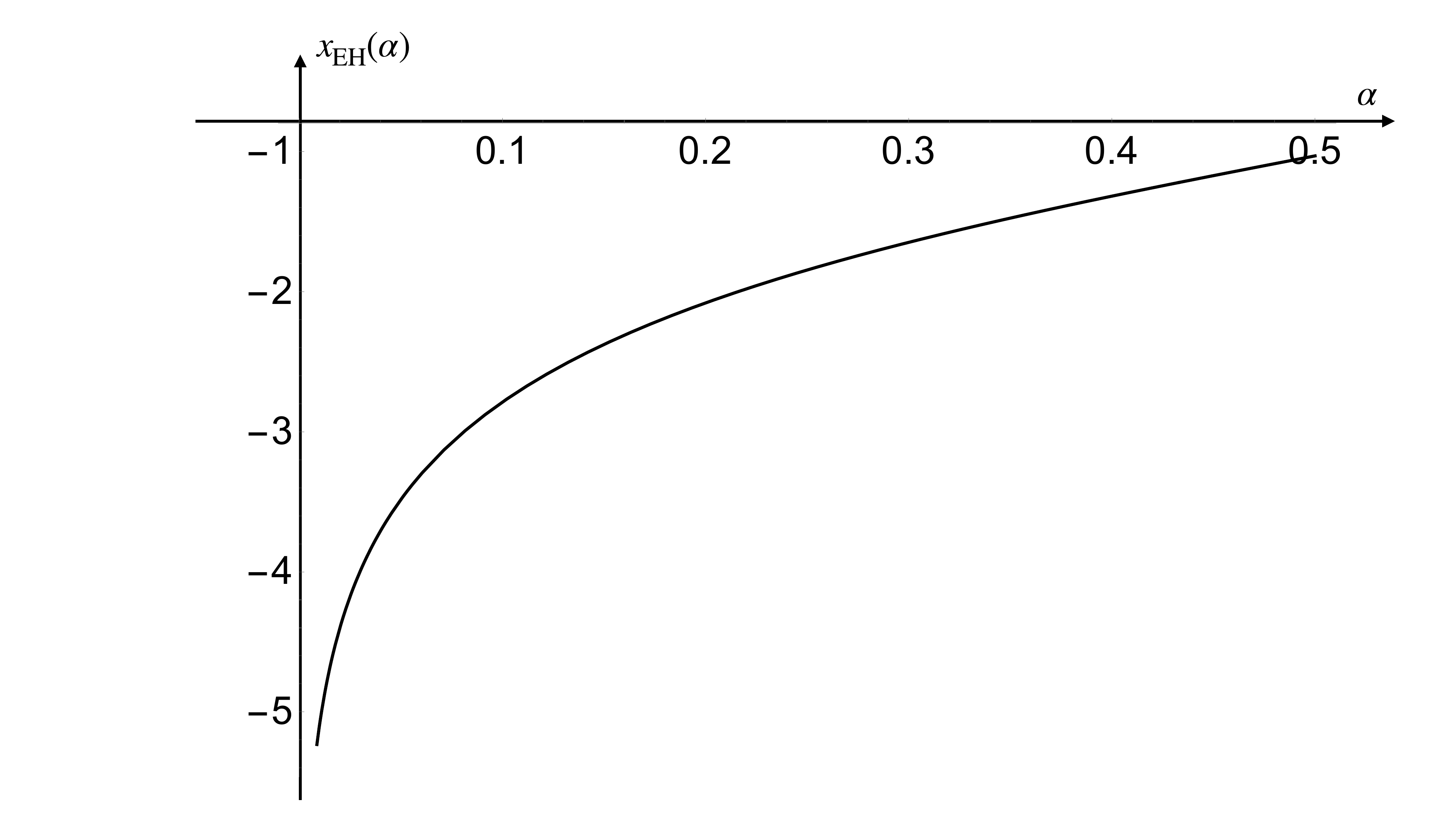}
 \includegraphics[width=0.45\linewidth]{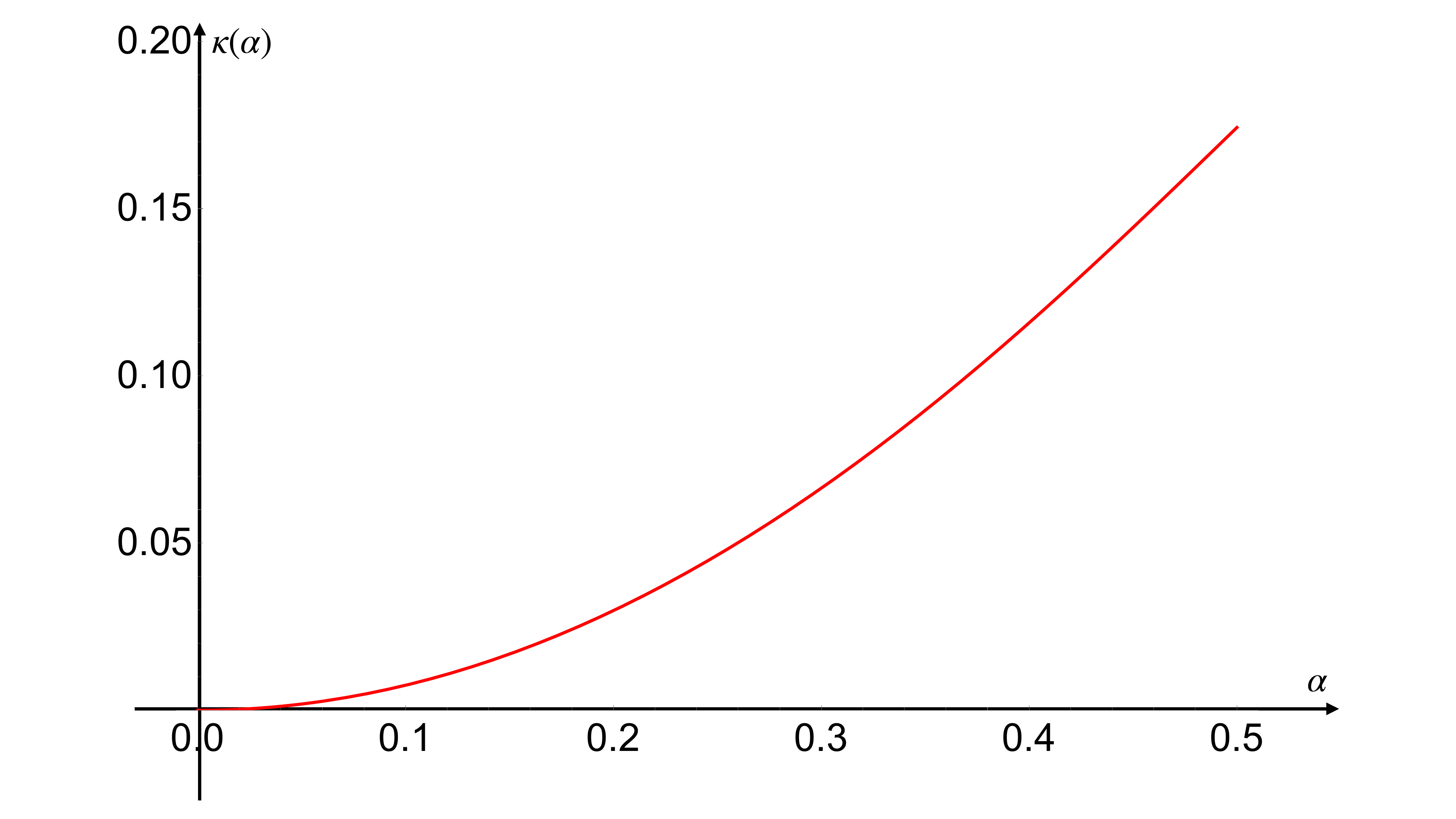}
 \caption{On the left: shape of $x_{\rm EH}(\alpha)$ for the profile \eqref{eq:v_profile_critical} up to $\alpha=1/2$. One can see that $x_{\textsc{efh}}\to - \infty$ when $\alpha \to 0$. On the right: shape of the ratio $\kappa(\alpha)$. This quantity goes to $0$ (which is the relativistic value) for very low energies but it still shows some nonzero value for $\alpha>0$.}
 \label{f:xEH_crit_sub}
\end{figure}

Here, the Killing horizon moved to $x_{\textsc{kh}}=-\infty$ where the surface gravity vanishes $\kappa_{\textsc{kh}}=v'_{-\infty}=0$. The analysis of \eqref{eq:EH_cond} can be split into various cases based on the type of dispersion relation as follows: 
\begin{itemize}
    \item superluminal case: the only solution to \eqref{eq:EH_cond} is found for $\alpha \to 0$. However, since $v'_\infty=0$, $\kappa(\alpha)=0$ and no particle production takes place, whatsoever. 
    \item subluminal case: we instead solve \eqref{eq:EH_cond} for $0\le \alpha \le 1/2$ and numerically compute $x_{\textsc{efh}}(\alpha)$ and $\kappa(\alpha)$ (see Fig. \ref{f:xEH_crit_sub}).
\end{itemize}
Interestingly, the subluminal case still supports particle production processes even when the relativistic mechanism has ceased. Let us point out that both setups, subluminal and superluminal, resonate with what we have found in the perturbative analysis of \eqref{eq:k_super_low} and \eqref{eq:k_sub_low} around $\alpha=0$ which returned $\kappa(\alpha)=O(\alpha^3)$.

\section{Particle creation: subcritical flow}
\label{secSUB}

Now, we complete our analysis by investigating a subcritical flow for which $|v|<1$ besides $|v_\mathrm{max}|\sim O(1)$. Then, no Killing/sonic horizon exists, hence such a flow mimics an horizonless ultra compact object (a quasi-black hole) rather than a black hole, cf. \cite{albuquerque2023inverse,albuquerque2024inverse} for further references. This is a quite interesting case, given that experimental problems in shallow water waves experiments concerning the stability of horizons led to the prevalent realisation of such ``near-critical'' flows ~\cite{Weinfurtner:2010nu,Weinfurtner:2013zfa,Euve:2014aga,Euve:2015vml}. Experiments that nonetheless observed particle production. This led to several numerical~\cite{PhysRevD.90.044033,Michel:2015aga} and analytical investigations~\cite{Coutant_2016_Hawking_subcritical} which verified the presence of a particle flux albeit generically characterised by substantial deviations from thermality.

Here we shall re-analyse this phenomenon using the tunneling method with the above introduced approximant trajectory.
As we shall see, our technique will be able to capture the spontaneous emission of an analogue ultra compact object in an ``Hawking-like'' picture, through the tunneling method in a non-relativistic setting like ours. The absence of an acoustic horizon, however, will reflect in a strong non-thermal character of such ``pre-Hawking" particle production phenomenon.

It is however important to stress that this effect will not be the only source of particle production for a subcritical flow in the presence of subluminal perturbations. What is described in the following reflects the fact that, for subluminal dispersion, and depending on the asymptotic values of the flow velocity $v(x)$, some high-energy (high $\alpha$) modes enjoy the presence of a turning point, as in figure~\ref{f:traj}. These modes, despite the absence of a sonic horizon, feel the presence of an EFH, and can be treated the same way as we have done in the supercritical case (because they admit two real solutions inside the EFH and four outside).

However, besides this, there is always a dominant set of low-energy (low $\alpha$) modes, which will not show any turning point, thus admitting four real solutions on the whole $(x,t)$ plane. For such modes, it is possible to show, by analyzing the Bogoliubov coefficients in a $4 \times 4$ scattering matrix, that a stimulated processes may occur as well~\cite{Robertson_2016}.
This latter effect has a completely different root to the one we focus here and could easily be dominant in a noisy experimental environment. 

Indeed, this would be more similar in nature to the so called ``Klein-paradox'' which describes a special case of superradiant amplification for scalar fields impinging on a potential barrier~\cite{brito2020superradiance}. Such a mechanism requires an incoming wave to start with, while the high-energy modes experiencing an EFH will also probe an effective ergoregion which allows for a spontaneous emission (vacuum instability).

Nonetheless, we feel it is quite interesting to see how a spontaneous channel can be present even in the absence of a full fledged acoustic horizon, as it predicts that even ultra compact horizonless object could be evaporating in non-relativistic matter frameworks (like, for example, the so called Standard Model Extension, see e.g~\cite{Liberati:2013xla}).

Let us then start our analysis with the EFH condition \eqref{eq:EH_cond}.
\begin{itemize}
    \item superluminal dispersion relation ($\xi=1$): particles fail to see any effective horizon, in fact, \eqref{eq:EH_cond} can never be fulfilled, since $|c_g^{\rm reg}|\ge 1$ always and $|v|<1$ everywhere,
    \item subluminal dispersion relation ($\xi=-1$): $|c_g^{\rm reg}|\le 1$ always, hence, there can be solutions of \eqref{eq:EH_cond}, even if $|v|<1$ everywhere.
\end{itemize}
Before particularising to a specific profile of $v(x)$, we quantify first how far the flow’s most negative value can be from the sonic point, $v=-1$,  while still admitting some solution to~\eqref{eq:EH_cond} for our subluminal dispersion relation. Within our setting this yields numerically the upper bound (namely, the solution of \eqref{eq:EH_cond} for $\alpha=1/2$)\footnote{The value for $\alpha$ is in principle arbitrarily chosen. In this setup, the threshold is determined by the value where the dispersion relation becomes multivalued. From a physical point of view, the minimal value represents that basically every object, no matter how far from forming a horizon, would radiate modes $k\sim\Lambda$.}
\begin{equation}
    v_{\rm min} \simeq -0.88 \,.
    \label{eq:vmin}
\end{equation}
For flows for which the maximally negative value remains smaller in norm than this $|v_{\rm min}|$ there cannot be any particle production, at least within the here considered formalism. Actually, particle creation will take place for a subluminal dispersion relation only if $|v_{-\infty}|>|v_{\rm min}|$. Indeed, the set of solutions to \eqref{eq:EH_cond} is consequently limited to the range of $x$ in which $|v_{-\infty}|\ge |v| \ge |v_{\rm min}|$ instead of $1>|v|>|v_{\rm min}|$.
\begin{figure}[h]
 \includegraphics[width=0.45\linewidth]{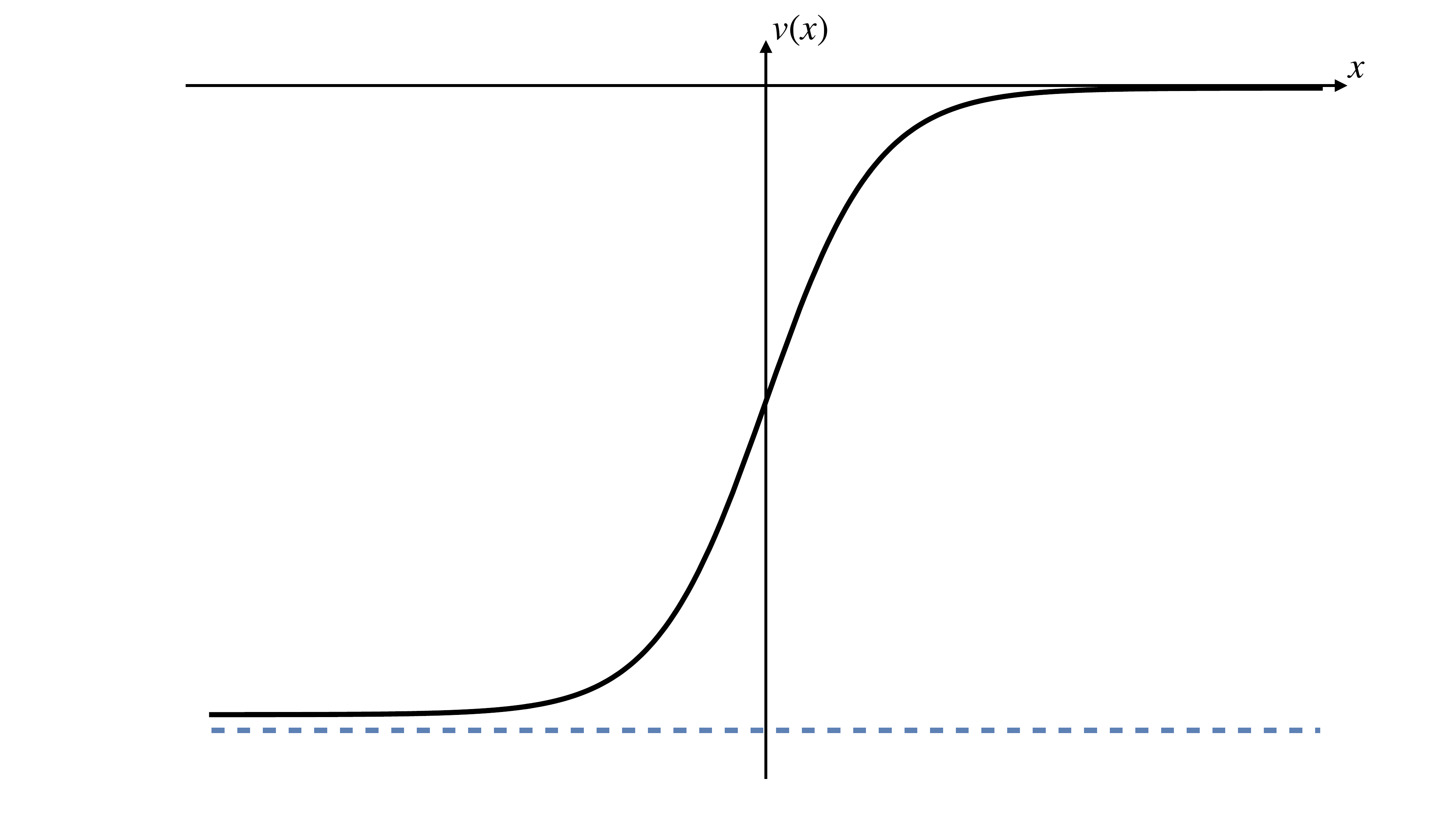}
 \caption{Profile for $v_\varepsilon(x)$ of eq.~\eqref{eq:v_profile_sub} with $\varepsilon=5 \times 10^{-2}$. The dashed line represents $v(x)=-1$.}\label{f:v_profile35}
\end{figure}

\begin{figure}[h]
 \includegraphics[width=0.50\linewidth]{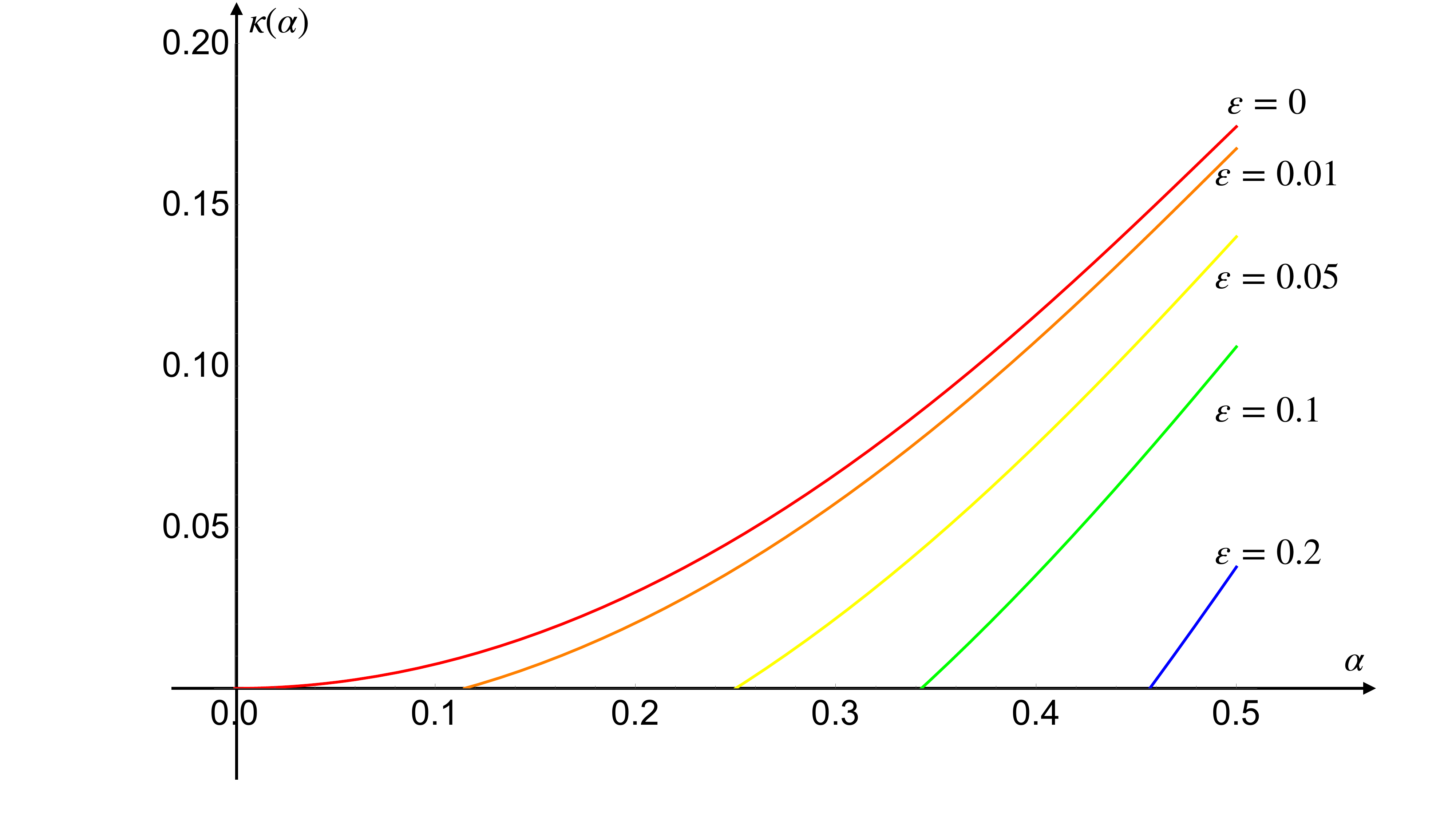}
 \caption{$\kappa(\alpha)$ for different values of $\varepsilon$: increasing $\varepsilon$ reduces the range of $\alpha$ for which \eqref{eq:EH_cond} has a solution. Moreover, $\kappa(\alpha)$ starts from $0$ with a non-vanishing derivative (apart from the critical case $\varepsilon=0$) which implies a deviation from thermality in the emission that will be the topic of the last section}
 \label{f:Talpha_sub_sub}
\end{figure}

In order to proceed further in our investigation, let us now assume a velocity profile like in figure \ref{f:v_profile35} given by the form
\begin{equation} \label{eq:v_profile_sub}
   v_\varepsilon(x)=  \frac{1}{2+\varepsilon}[\tanh(\kappa_\circ x) -1] \,,
\end{equation}
where $\varepsilon>0$ and $\kappa_\circ$ is again a fiducial scale to compensate the dimension of $x$. 
Note, \eqref{eq:v_profile_sub} shows a similar shape than the critical one shown in Figure~\ref{f:xEH_crit_sub}, however $|v_{-\infty}|<1$. Indeed, we have 
\begin{equation} \label{eq:v_profile_sub_limit}
  \lim_{x \to - \infty} v_\varepsilon(x)=  - \frac{2}{2+\varepsilon} > -1 \,.
\end{equation}
In order to achieve somewhere $|v|>|v_{\rm min}|$, we need $\varepsilon \leq 0.27$. In Fig.~\ref{f:Talpha_sub_sub} we have plotted $\kappa(\alpha)$ for different values of $\varepsilon$.

As a final remark, we stress that the presence of particle production in absence of a Killing horizon has been noticed also in the ``dual'' case for which the flow is everywhere supercritical and the dispersion relation is superluminal~\cite{Finazzi_2011_robustness}. Remarkably, it is easy to see that within our framework, this case is analogous to the just considered subluminal-subcritical case.

{Indeed, for $|v|>1$ everywhere, one still obtains roots for \eqref{eq:EH_cond} {only} when $\xi=1$ and, if the flow is not ``too much'' supercritical (a dual condition of that implied by Eq.~\eqref{eq:vmin}), this effect can be detected in the range $\alpha \le 0.5$, obtaining a behaviour which resembles closely the one found in Fig.~\ref{f:Talpha_sub_sub} for subluminal-subcritical case. This correspondence in behaviour is indeed just another manifestation of the duality between supercritical-superluminal and subcritical-subluminal settings already noticed in previous analogue gravity investigations.}

\section{Phenomenological Considerations}\label{secSO}
{In this section, we highlight some important aspects and subtleties related to our approach to provide a complete and comprehensive picture. For instance, we analyse the properties of the spectrum and the internal consistency of the methodology.}

\subsection{Deviation from thermality}
The first question addresses the problem whether the predicted particle production leads to a thermal (or approximately thermal) spectrum or not. To answer this, we define the quantity
\begin{equation} \label{eq:delta_alpha}
    \delta(\alpha)= \left|\frac{\partial_\alpha \kappa(\alpha)}{\kappa(\alpha)}\right| \,.
\end{equation}
Whenever $\delta(\alpha) \ll 1$, the change in $\kappa(\alpha)$ remains negligible with respect to $\kappa(\alpha)$ itself; therefore the emission will be considered as thermal. 

\subsubsection{Supercritical flow}
\begin{figure}[h]
 \includegraphics[width=0.48\linewidth]{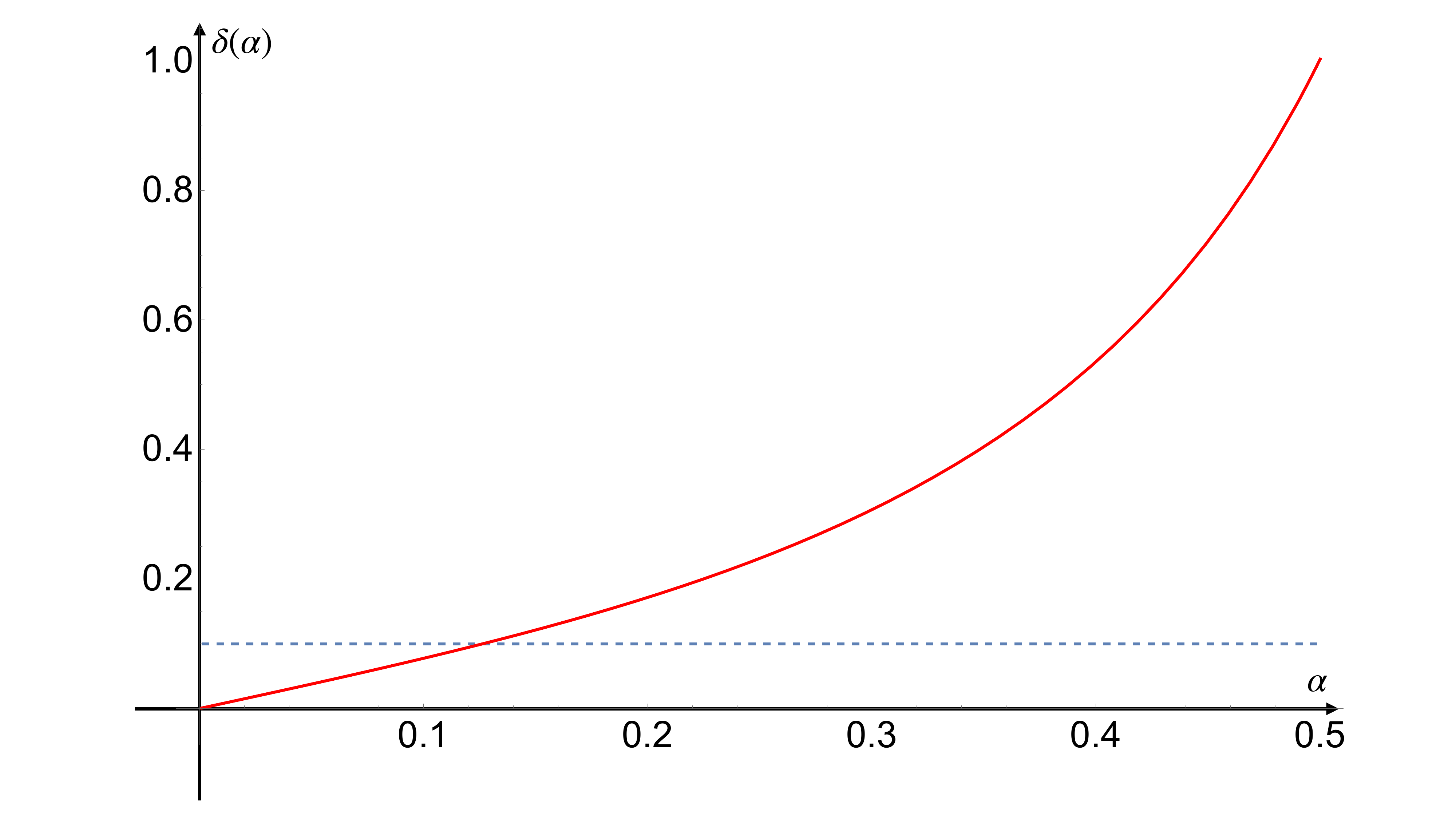}
 \includegraphics[width=0.48\linewidth]{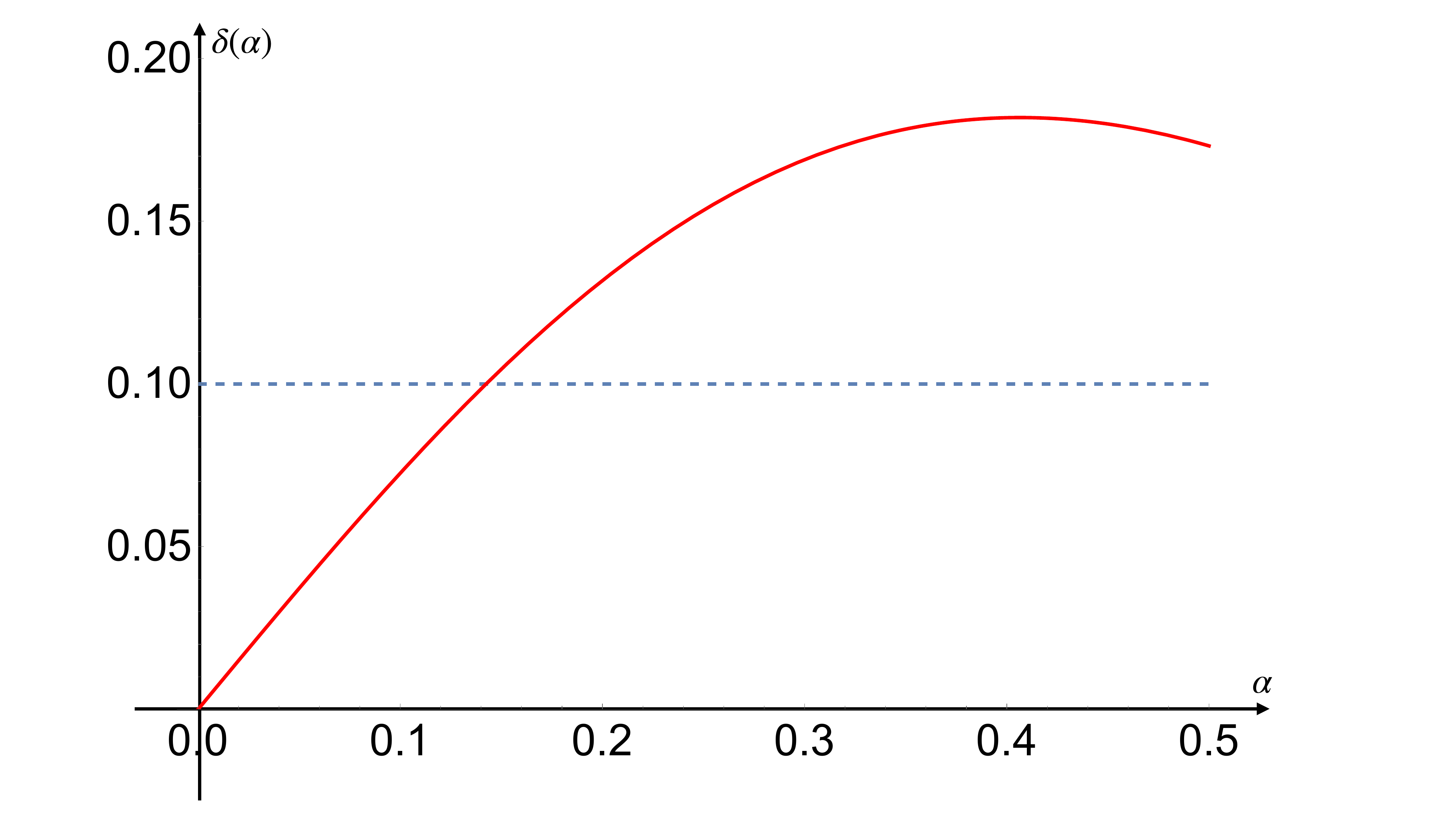}
 \caption{Deviation from thermatity in a supercritical flow.} On the left: $\delta(\alpha)$ for the subluminal case. On the right: $\delta(\alpha)$ for the superluminal case. The dashed line is $f(\alpha)=0.1$.
 \label{f:delta_super}
\end{figure}

For the supercritical regime, we can see, from \eqref{eq:k_sub_low} and \eqref{eq:k_super_low} that within the low energy region, the surface gravity behaves as
\begin{equation} \label{eq:delta_alpha_low}
    \kappa(\alpha) \simeq \kappa_{\textsc{kh}} + O(\alpha^2) \quad \mbox{and} \quad \partial_\alpha \kappa(\alpha) \simeq \frac{3}{4} \xi \alpha\kappa_{\textsc{kh}} + O(\alpha^3)\,.
\end{equation}
As a consequence, for $\alpha \simeq 0$, $\delta(\alpha)= O(\alpha)$ implying that the emitted spectrum shows perturbatively thermal features. This can be also extracted from Fig.\ref{f:delta_super}. 

\subsubsection{Subcritical flow}
While we perturbatively recover thermality in the supercritical case, we see that a subcritical (and subluminal) setup displays the complete opposite behaviour. Albeit having generically $\partial_\alpha \kappa(\alpha) \ne 0$ for $\alpha \ne 0$, we find that $\kappa(\alpha)=0$ at the minimum value of the subcritical profile $v=-2/(2+\varepsilon)$.\footnote{This particular feature depends on the fact that $v$ approaches the value $-2/(2+\varepsilon)$ with vanishing derivative. So it depends crucially on the detailed shape of $v$.}  This, in turn, means that $\delta(\alpha)$ diverges, thus maximising the deviation from thermality. This should come as no surprise given the essential role of the presence of a Killing/sonic horizon to which the EFH has to be close, in order to assure the approximate constancy of $T(\alpha)$.

\subsection{Spectrum}
\begin{figure}[htb]
 \includegraphics[width=0.50\linewidth]{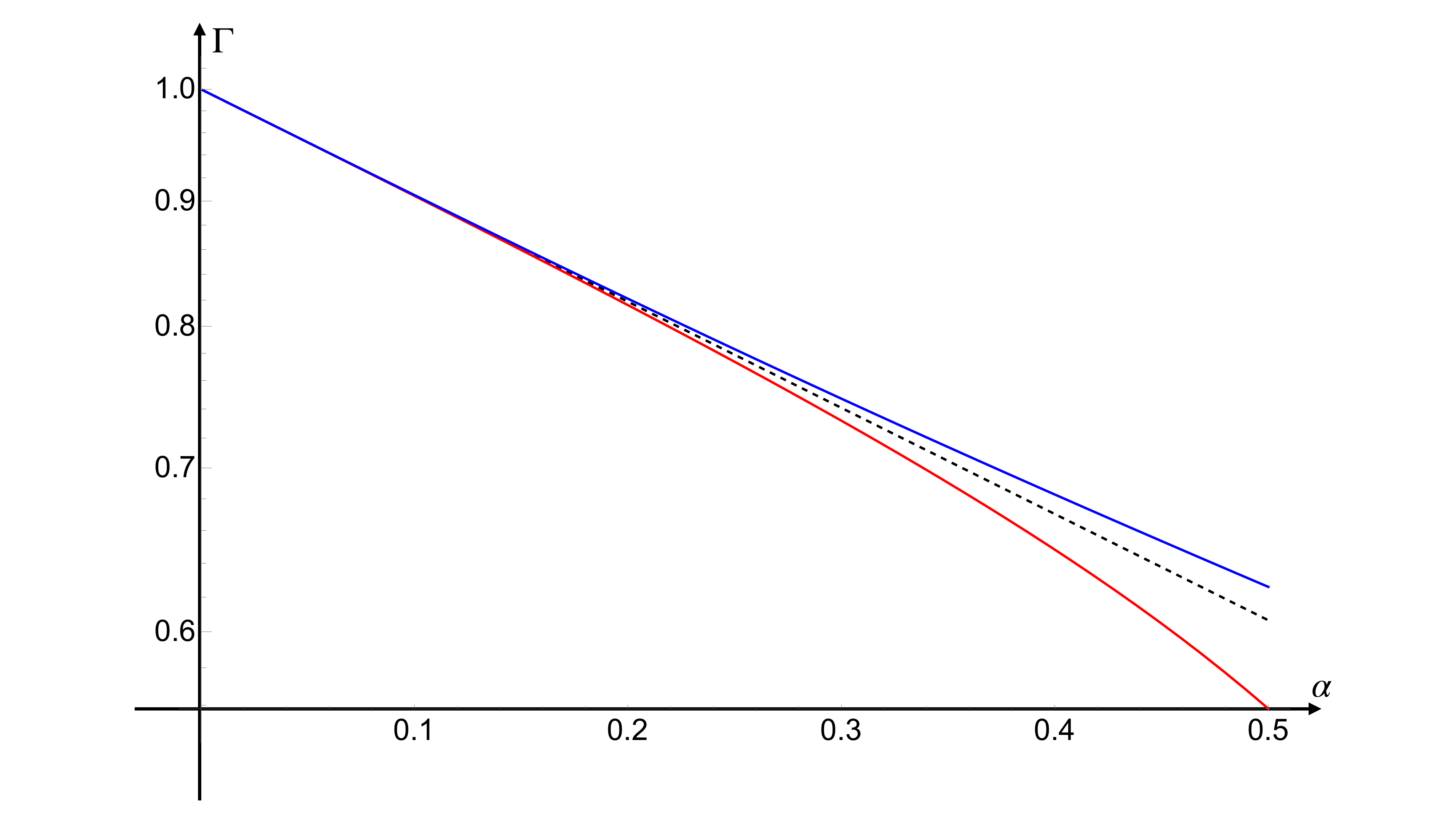}
 \caption{Plot (the vertical axis is displayed in a logarithmic scaling) of $\Gamma$ for a supercritical flow. The dashed black line depicts the relativistic rate, while the blue solid line represents the superluminal and the red the subluminal case. For convenience, here we have set $2 \pi \Lambda=1$, such that $\Gamma=e^{- \alpha/\kappa(\alpha)}$ and $\kappa_{\textsc{kh}}=1$.}
 \label{f:rate_super}
\end{figure}

\begin{figure}[htb]
 \includegraphics[width=0.50\linewidth]{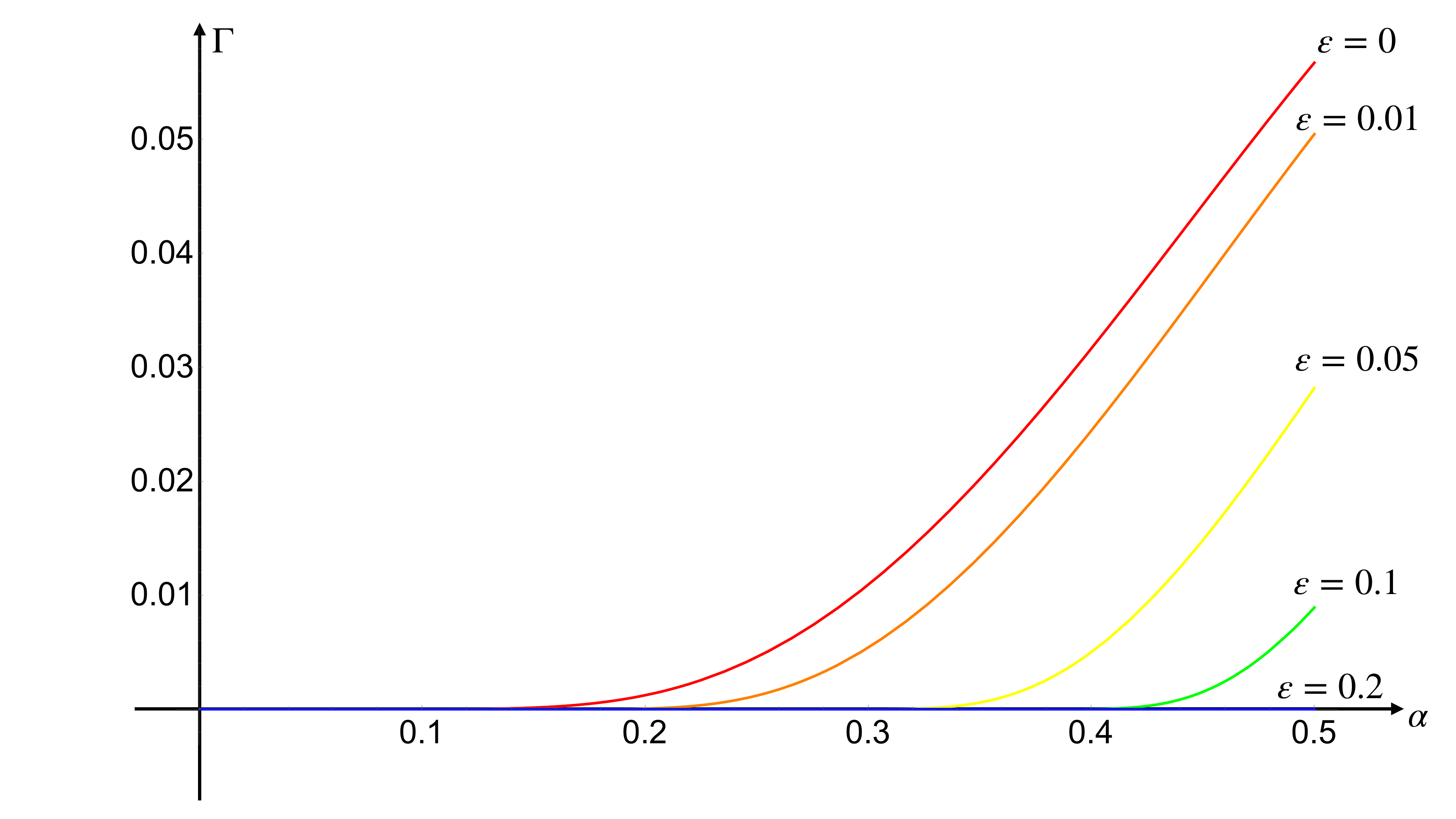}
 \caption{Plot of $\Gamma$ in the subluminal case for some values of $\varepsilon$. The case $\varepsilon=0$ represents the critical regime. Again we have set $2 \pi \Lambda=1$.}
 \label{f:rate_sub}
\end{figure}

{Whether or not thermal, the emission spectrum is determined by the tunneling rate \eqref{eq:tunneling_rate}. In Fig.\ref{f:rate_super}, we plotted $\Gamma$ for the supercritical regime, while the subcritical case (subluminal dispersion relation) is shown in Fig.\ref{f:rate_sub}.}

Comparing both spectra, we observe that $\Gamma$ stays close to the relativistic tunneling rate -- at low energies in the supercritical phase independently from the nature of the dispersion relation. However, when increasing $\alpha$, the subluminal as well as the superluminal scenarios depart further and further from the relativistic behaviour, describing respectively a colder and a hotter object. 

Finally, let us stress that even if subluminal dispersion relations predict particle production for super- as well as subcritical flows, it remains true that
\begin{equation}
    |\Gamma_{\rm subcrit}| \ll |\Gamma_{\rm supercrit}| \,.
\end{equation}
I.e.~even if subluminal dispersion allows for an in-principle particle production in the absence of a sonic/Killing horizon, this effect is strongly suppressed with respect to the particle production in the presence of an horizon.

\subsection{Energy conservation: subcritical flow-subluminal dispersion relation}

The particle creation found in Sect.~\ref{secSUB} confirms previous results based on Bogoliubov coefficients~\cite{PhysRevD.90.044033,Michel:2015aga,Coutant_2016_Hawking_subcritical}. However, given that in stationary geometries the spontaneous particle creation from the vacuum always requires the presence of an ergoregion to ensure energy conservation, this result might seem puzzling at first sight. Indeed, for this production to be consistent, the ingoing Hawking partner must carry a negative Killing energy to compensate for the positive one carried to infinity by the Hawking quantum. However, this can happen only within an ergoregion apparently absent in the considered subcritical flow. In what follows, we shall explain how this apparent paradox is resolved by the peculiar nature of subluminal dispersion relations.

Starting with Eq.~\eqref{eq:DR} (with $c_s(x)\equiv1$), the requirement for the existence of a quantum with $\Omega<0$ can be recast into the condition $\omega<k v$ (considering positive preferred energies amounts to $k<0$) which in turn means
\begin{equation}
   0> c_{\rm ph}= \frac{\omega}{k} > v \,,
\end{equation}
that is, the phase velocity's absolute value must be smaller than the fluid velocity.

For superluminal dispersion relations this requires $|v|>1$ which implies the necessity of the presence of a sonic/Killing horizon. However, for subluminal dispersion relations we can write
\begin{equation}
    c_{\rm ph}^2= \frac{\omega^2}{k^2} = 1- \frac{k^2}{\Lambda^2}<v^2 \,.
\end{equation}
This inequality is satisfied whenever
\begin{equation}
    \Lambda^2>k^2>\Lambda^2 (1-v^2) \,,
    \label{eq:condEC}
\end{equation}
where the upper bound $\Lambda^2>k^2$ was added to respect the perturbative interpretation of the dispersion relation as well as to ensure $\omega^2(k)\ge0$. Equation \eqref{eq:condEC} then reveals two important features
\begin{itemize}
    \item the energy balance can be satisfied in the subluminal case regardless of the presence of any sonic/Killing horizon. In particular, if $|v_{\rm max}|$ is close to the speed of sound, the window of opportunity described by \eqref{eq:condEC} may allow for the presence of an EFH for $k<\Lambda$.  While for deeply subcritical flows, such window rapidly closes and only for $k \simeq \Lambda$ some mode can be excited\footnote{However, one should study the non-perturbative structure of the dispersion relation, to verify the reasoning in such regimes.}.
    \item the lower branch of the negative energy mode for the subluminal case (which is the lingering mode in Fig.\ref{f:traj}) is dispersive around $v=0$ (the origin of the axes in Fig.\ref{f:traj}). In \eqref{eq:v_small} we have shown that the regular solutions around $v \simeq 0$ consist in two positive-$\Omega$ solutions, one ingoing and one outgoing. However, none of the lower branches of the lingering or the turning mode reach $v=0$, meaning that those branches are non-regular in that region of spacetime. This can be explained by the following fact: for these two branches $|k|$ increases while approaching $v=0$, so much so that for the subluminal dispersion relation $\omega(k)$ becomes imaginary and no real mode of these branches can be further propagated towards $v=0$. 
\end{itemize}
Let us notice, that the possibility of having negative-energy modes only when $|v|>|c_{\rm ph}|$ is an established, well-known fact, see cf. \cite{Unruh_2003_slow_light,Novello_2002_ArtificialBH}

\subsection{Validity of the approximation}
As a last topic we would like to quantify the validity of our approximation. All calculations are based on finding an approximant trajectory to our non-relativistic particle. Since this fiducial curve experiences an effective horizon, it allows us to perform a tunneling calculation as it is associated with a simple pole which is instead is absent in the true trajectory. Nonetheless our results confirm the expectations based on the Bogoliubov methods. How can this be?

The crucial issue here is that the actual particle creation process does not happen arbitrarily close to the horizon, but rather when the partners of the Hawking pair are sufficiently stripped apart from tidal forces for them to be distinguishable ``on-shell'' particles. Such critical distance is usually identified with the de Broglie wavelength (or Compton, if they are massive) of the particles~\cite{Giddings_2016_Hawking,Dey_2017_quantum_atmosphere,Dey:2019ugf}.
Thus, in turn, we assume that if our process happens within a de Broglie wavelength, the calculation can be considered as trustworthy. 

In an analogue setting, the de Broglie wavelength of an acoustic excitation has to be defined using the speed of sound $c_s$ explicitly~\cite{Barcelo_2005_review}. After restoring all relevant physical quantities, we can write
\begin{equation}
    \lambda_s=\frac{h c_s}{\Omega} \,.
\end{equation}
The idea is the following: since $\lambda_s$ denotes the characteristic distance between the Hawking partners at which they go on-shell, we must require that the approximant trajectory fails to mimic the physical trajectories only when these are separated by a distance smaller than $\lambda_s$, in order for our method to apply. This is tantamount to say that the physical trajectories and the approximant are indistinguishable from the point of the particle creation process. In other words, if we define $x_1(\alpha)$ as the point where we violate \eqref{eq:v_small_validity} (the approximant fails to trace the ray outside the effective horizon) and $x_2(\alpha)$ as the point where \eqref{eq:v_big_validity} is violated (the approximant digresses strongly from the inside ray), then their distance $\Delta x(\alpha)$ has to be smaller than $\lambda_s$:
\begin{equation} \label{eq:deltax}
    |\Delta x(\alpha)|=|x_2(\alpha)- x_1(\alpha)| \le \lambda_s= \frac{h c_s}{\alpha \Lambda}\,.
\end{equation}

Since \eqref{eq:v_big_validity} and \eqref{eq:v_small_validity} involve $v(x)$, we have to specify a profile for the flow velocity to determine actual values for $\Delta x$. Let us then take the following profile
\begin{equation}
   v(a,b;x)= \frac{a}{2} \left[ \tanh(bx)-1 \right]\,.
\end{equation}
controlled by the two parameters $a$ and $b$ which are associated to alternative features of the flow:
\begin{itemize}
    \item $a= -\lim_{x \to - \infty} v(a,b;x) $ controls the lower limit of $v$ and can be identified as $a=|v_{-\infty}|$.
    \item  $b$ controls the slope, i.e. the bigger $b$ is, the steeper $v$ becomes in passing from 0 to $-a$.
\end{itemize}
This profile for $v$ allows us to include all investigated cases in the discussion and, given that we can invert the function
\begin{equation}
    x(a,b;v)=\frac{1}{b} {\rm artanh }\left( 1+\frac{2v}{a} \right)\,,
    \label{eq:xlim}
\end{equation}
we can use it to study $\Delta x$.

\subsubsection{Superluminal case}

Taking $\xi=1$, the effective horizon will always be located inside the Killing horizon, as such we have to consider the case $v \le -1$ in condition \eqref{eq:v_big_validity}). So let us consider, for  a given $\alpha$, the values of the flow velocity for which the conditions \eqref{eq:v_small_validity} and \eqref{eq:v_big_validity} are saturated
\begin{equation}
     v_1(\alpha)=- \sqrt{\frac{1+\sqrt{1+4 \alpha^2}}{2}} \,, \quad \mbox{and} \quad  v_2(\alpha)=-1-\alpha \,.
\end{equation}
Then, after multiplying \eqref{eq:deltax} by $\alpha$ and collecting the $\alpha$-dependence, we can use \eqref{eq:xlim} to write
\begin{equation}\label{eq:alpha}
     \alpha \cdot \Delta x(\alpha)= \frac{\alpha}{b} \left[ {\rm artanh }\left( 1+\frac{2}{a} v_2(\alpha) \right) - {\rm artanh }\left( 1+\frac{2}{a} v_1(\alpha) \right)\right] \,.
\end{equation}
Since we are in the superluminal case, particle production occurs only in the supercritical regime, namely for $a \ge 1$. In general, the expression for $\alpha \cdot \Delta x(\alpha)$ is parametrically small, depending on the value of $b$ as follows: if $b$ increases, then $v$ will change rapidly in a very narrow region in $x$, such that the particles will be produced in a small neighborhood around the Killing horizon. Note that $b$ for $a=2$ represents exactly the Killing surface gravity, as one can immediately see from \eqref{eq:v_profile}.

At low energies, one can expand \eqref{eq:alpha} for $\alpha \simeq 0^+$ obtaining
\begin{equation} \label{eq:deltax_super_low}
    \alpha \cdot \Delta x(\alpha)= -\frac{a \alpha^2}{2b (a-1)} + O(\alpha^3),
\end{equation}
which reveals that \eqref{eq:deltax} will be always satisfied at low energies for $a>1$, independently from the value of $b$. For $a=1$, representing the critical behaviour for the flow, \eqref{eq:deltax_super_low} must be analysed separately due to the obvious pole. If we set $a=1$ and then expand for $\alpha \simeq 0$, 
\begin{equation} \label{eq:deltax_super_low_crit}
    \alpha \cdot \Delta x(\alpha)= -\frac{\alpha}{2b} \ln\left(\frac\alpha 2\right) + O(\alpha^2),
\end{equation}
which is again perturbatively small and satisfies the bound given by \eqref{eq:deltax}. 

{The only point where the approximation fails independently from $b$, occurs when $v$ is a slightly supercritical flow ($a\gtrsim 1$) such that we can have particle production up to $v_{- \infty}$ for $\alpha \le 1/2$. This happens only\footnote{Notice that $|v_2(\alpha)|>|v_1(\alpha)|$.} when
\begin{equation} \label{eq:deltax_super_break}
    v_2(\alpha)=-a=v_{-\infty} \iff \alpha=a-1 \,.
\end{equation}
When \eqref{eq:deltax_super_break} is fulfilled at a finite $\alpha \ne 0$, the product $\alpha \cdot \Delta x(\alpha)$ diverges and we cannot satisfy \eqref{eq:deltax}. This can be understood, simply because for $\alpha=a-1$, particle production should happen for velocities $v_{-\infty}$, viz. in the point $x=-\infty$.}\\

\subsubsection{Subluminal case}

For the subluminal case the discussion resembles the previous one in most parts. By setting $\xi=-1$ while considering that the effective horizon lies outside of the Killing horizon, the breakdown of the approximation, outside and inside the EFH, is destined to happen respectively at
\begin{equation}
     v_1(\alpha)=- \sqrt{\frac{1+\sqrt{1-4 \alpha^2}}{2}} \,, \quad \mbox{and} \quad  v_2(\alpha)=-1+\alpha \,,
\end{equation}
which yields an expression for $\alpha \cdot  \Delta x(\alpha)$ very close to \eqref{eq:deltax_super_low}. In general, this expression changes antiproportionally with $b$. At low energy, for $a>1$ (supercritical flow), we find again \eqref{eq:deltax_super_low} while for the critical case $a=1$ we recover \eqref{eq:deltax_super_low_crit}.
This confirms that the low energy behaviour is well described by our approximant. 

{Again, here, we violate \eqref{eq:deltax} when $\Delta x(\alpha)$ diverges at a finite $\alpha$, which is at
\begin{equation} \label{eq:deltax_super_break_subluminal}
    v_2(\alpha)=-a=v_{-\infty} \iff \alpha=1-a \,.
\end{equation}
This is possible only for the subcritical scenario, since $\alpha>0$. In this case, we discover again that the approximation breaks down at $x=-\infty$ when particles fulfil $\alpha=a-1$, then our treatment becomes invalid.}

Let us then summarise:
\begin{itemize}
    \item our approximation is always valid for low energy particles, in the supercritical and critical case, both for super- and subluminal perturbations and independently of the model,
    \item for a generic value of $\alpha$, the validity of the approximation depends strongly on the model, namely, the energy scale $\Lambda$, the sound speed $c_s$ and the steepness of the profile $v(x)$,
    \item for a $v(x)$-profile of the $\tanh(x)$-type, there exists an $\alpha$ for the supercritical and superluminal case $(\alpha=a-1)$ and one for the subcritical and subluminal case $(\alpha=1-a)$, for which our approximation fails.
\end{itemize}

In conclusion, we see that quite generically the particle creation process described via the tunelling method and the approximant, is basically insensitive of the near-horizon behaviour which is where the details of the dispersive behaviour of our modes matter most. This resonates with the more complete, but also more convoluted, results of the Bogoliubov approach, where the same robustness to the short-distance details of the process has been found, see cf. \cite{coutant2014hawking}.

\section{Discussion}\label{secD}

This article was devised as a pedagogical primer for Hawking radiation in acoustic gravity, providing an accessible and physically intuitive description of this phenomenon via the tunneling method. Due to its simplicity and versatility, this formalism allowed us to explore several different, albeit sometime idealized, scenarios that may be relevant in analogue gravity set-ups. We investigated three different flow types --- supercritical, critical, and subcritical --- over which we analysed the particle creation for a scalar field with generalised dispersion relations of subluminal as well as superluminal type. As a result, we found that particle creation can be associated with an effective horizon experienced by a trajectory that interpolates those of the Hawking partners outside and inside the (effective) horizon. Such approximant only fails to track these particles in a region of size below their de Broglie wavelength (where they cannot be treated as separated on-shell particles). The aforementioned effective horizon is always located inside the Killing horizon for superluminal particles and always outside for subluminal ones.

For supercritical flows, endowed with a proper acoustic/Killing horizon, we confirmed the robustness of Hawking radiation for both types of dispersion relations. Indeed, we observed that the standard Hawking temperature acquires only very small corrections (of order $O(\alpha^2)$) given that values of $\alpha \gtrsim 0.1$ should not be considered as physically relevant in realistic analogue systems.

Another relevant insight from our analysis provides the simple discovery of how the creation of subluminal particles, described theoretically \cite{PhysRevD.90.044033,Michel:2015aga, Coutant_2016_Hawking_subcritical} and experimentally \cite{Weinfurtner:2010nu,Weinfurtner:2013zfa,Euve:2014aga,Euve:2015vml} in stationary subcritical flows, can have a spontaneous component, in addition to a dominant stimulated one. Indeed, we unveiled that in such set-ups, some high-energy modes can experience an effective horizon, if the flow gets sufficiently close to forming a Killing horizon. For such modes, we showed that the Hawking partner, propagating inside the effective horizon, is endowed with a negative Killing frequency, and hence energy conservation can be made to hold as usual in spite of being in vacuum and in a time independent geometry.

A final comment must be made on the vacuum state. In our analysis, we implicitly assumed a ``no drama'' scenario for observers crossing the effective horizon or the acoustic/Killing horizon along the ``approximant'' trajectory. One could interpret this as implicitly assuming the existence of the usual global Unruh vacuum state, similar to what is done in the Bogoliubov method for stationary flows. However, rather than comparing particular states, the tunneling method performs a time-resolved analysis that selects an instantaneous local vacuum to count particles \cite{Agullo:2014ica}. In other words, as long as a near-horizon vacuum remains valid, the method is predictive. This assumption is justified by the physical intuition of how such a vacuum would be produced in a realistic experiment generating a supersonic flow and is substantiated by experimental data \cite{MunozdeNova:2018fxv}.

In conclusion, we hope that the simple analysis presented here can be extended to more realistic settings and specific experimental set-ups for analogue Killing horizons in the future, possibly allowing the investigation of dynamical trapping ones \cite{Giavoni:2020gui}. We leave these investigations to future work.

\section*{Acknowledgments}

SL would like to thank Renaud Parentani for several invaluable discussion on the subject. This research is supported by the Italian Ministry of Education and Scientific Research (MIUR) under the grant PRIN MIUR 2017-MB8AEZ.

\appendix
\section{Tunneling vs. Bogoliubov} \label{app:tunnVSbogo}
This appendix clarifies the relation between the tunneling and the Bogoliubov approach. The main point of our discussion will be to understand why the interpretation of a tunneling rate $\Gamma$ as defined in \eqref{eq:tunneling_rate} can be interpreted as a signature of thermality for the observer at infinity ($v(x)=0$) where the particle content is usually determined by evaluating the Bogoliubov coefficients between the near-horizon modes and the modes at infinity. This analysis, being a proof of concept, will be conducted fully in the  relativistic case, but we will argue that it can be applied to fields with modified dispersion relations, since the short-range details of the modes will not matter in the evaluation of the Hawking effect. For further details on the derivations, see \cite{DelPorro_2024_thesis} (chapter 2) and for similar consideration see \cite{moretti2012state}.

Let us then consider a relativistic field $\phi$, solving the massless Klein-Gordon equation in a static, $(1+1)$ dimensional\footnote{The very same discussion can be made with a spherically symmetric black hole example }, asymptotically flat geometry with a Killing horizon, such as \eqref{eq:acoustic_metric},
\begin{equation}
    \square \phi =0 \,.
\end{equation}
In order to compute the Hawking effect we focus on the outgoing branch of the solutions. This can be described labelling the modes at fixed Killing energy $\Omega$. Both inside and outside the horizon, these modes assume the usual logarithmic non-analyticity in the near-KH limit 
\begin{equation} \label{eq:nearKHmodes}
   \phi_\Omega^{\textsc{kh}, \rm out}= \exp \left[ - i\Omega t + \Omega \frac{i}{\kappa_{\textsc{kh}}} \ln( x- x_{\textsc{kh}} )  \right] \,, \quad \phi_\Omega^{\textsc{kh}, \rm in}= \exp \left[ - i\Omega t + \Omega \frac{i}{\kappa_{\textsc{kh}}} \ln( x_{\textsc{kh}}-x )  \right] \,,
\end{equation}
while at infinity, where the geometry is flat, they assume the shape of a plane wave:
\begin{equation}
   \phi_\Omega^{\infty}= \exp \left[ - i\Omega (t - x ) \right] \,.
\end{equation}
The Bogoliubov relation between the two (inequivalent) bases $\{\phi_\Omega^{\textsc{kh}, \rm out}\}$ and $\{\phi_\Omega^{\infty}\}$, together with the assumption of the state to be vacuum for the freely-falling observer at horizon crossing, essentially captures the spontaneous particle production of a black hole, that is the Hawking effect.

An intuitive way to compute the Bogoliubov coefficients is to perform a trick \textit{\`a la Unruh} \cite{unruh1976notes}, where we get rid of the non-analyticity given in \eqref{eq:nearKHmodes}, building a new set of solutions which is thus analytical across the KH \cite{Jacobson:2003vx}:
\begin{align}
\Phi_\Omega^\pm= C^\pm \left[\phi_\Omega^{\textsc{kh}, \rm out} \pm e^{\mp \frac{\Omega}{2 \kappa_{\textsc{kh}}}} (\phi_\Omega^{\textsc{kh}, \rm in})^* \right] \,.
\end{align}
The exponential $e^{\mp \frac{\Omega}{2 \kappa_{\textsc{kh}}}}$ that appears in the expression above, represents nothing else than the analytical continuation of the logarithm in \eqref{eq:nearKHmodes} across the KH on the upper and lower half complex $x$-plane, depending on the sign of the exponent. The coefficients $C^\pm$, instead, are normalization coefficients. The analyticity of $\Phi^+_\Omega$ ($\Phi^-_\Omega$) allows us to express this as a sum of only positive (negative) frequency modes at infinity $\phi_\Omega^{\infty}$. This means that the vacuum state for the observer at infinity is the same vacuum defined through the $\Phi^+_\Omega$. So, the evaluation of the Bogoliubov coefficients can be equivalently made, instead of between $\{\phi_\Omega^{\textsc{kh}, \rm out}\}$ and $\{\phi_\Omega^{\infty}\}$, just between $\{\phi_\Omega^{\textsc{kh}, \rm out}\}$ and $\{\Phi_\Omega^{\pm}\}$.

As a matter of fact, the mean value of the number operator turns out to be given by \cite{DelPorro_2024_thesis}:
\begin{equation}
    -C^-=-\langle \Phi^-_\Omega,\phi_\Omega^{{\textsc{kh}},\rm out} \rangle = \int  \frac{{\rm d}  \bar{\Omega}}{2 \pi} |\beta_{\Omega \Bar{\Omega}}|^2= \langle \hat{N}_\Omega \rangle \,.
\end{equation}
Taking into account the normalization condition of $\Phi^\pm_\Omega$ we get:
\begin{equation}
   \langle \hat{N}_\Omega \rangle= \int  \frac{{\rm d}  \bar{\Omega}}{2 \pi} |\beta_{\Omega \Bar{\Omega}}|^2= \frac{1}{e^{2 \pi \Omega/\kappa_{{\textsc{kh}}}}-1}  \,,
\end{equation}
which is a Bose-Einstein distribution, underlying a thermal character of the particle number $\hat{N}_\Omega$ (the expetation value is taken on the freely falling vacuum) with temperature $T=\kappa_{{\textsc{kh}}}/2 \pi$, that is the Hawking temperature. However, this is not the distribution which an observer far from the horizon would detect since, due to the curved geometry, the propagation affects the particle content in an energy-dependent way through the so-called \textit{grey-body factor} $\Gamma_{\rm grey}(\Omega)$ that quantifying the probability of a single particle of energy $\Omega$ to overcome the gravitational potential and eventually reach infinity. The resulting spectrum is given by \cite{Jacobson:2003vx}:
\begin{equation}
    \frac{\Gamma_{\rm grey}(\Omega)}{e^{2 \pi \Omega/\kappa_{{\textsc{kh}}}}-1}  \,.
\end{equation}
Let us underline that the grey-body factor always affects the propagation of the outgoing modes, apart from the particular case of a massless relativistic particle in $(1+1)$-dimensional geometry, because any such metric is conformally flat \cite{Finazzi:2010yq}. 

The equivalence between the tunneling and the Bogoliubov approach relies on the fact that the source of thermality is mathematically given by the non-analyticity of the outgoing modes near the horizon. The tunneling rate \eqref{eq:tunneling_rate} represents exactly the computation of this factor, since it involves the imaginary part of the phase of $\phi$. This equivalence is possible due to the \textit{exact} WKB character of the solution near the horizon. In other words $\phi_\Omega^{{\textsc{kh}},\rm out}$, takes the form given in \eqref{eq:nearKHmodes} without relying on any approximation, due to the exponential peeling.

Usually, the tunneling rate $\Gamma$ has a statistical interpretation \cite{Senovilla_2014}. In this case, one can interpret the existence of an outgoing particle as the sum of two processes
\begin{equation}
    \phi_\Omega^{{\textsc{kh}},\rm out}= C^- \Phi^-_\Omega + C^+ \Phi^+_\Omega \,,
\end{equation}
which tells us that a particle $\phi_\Omega^{{\textsc{kh}},\rm out}$ can be originated by a the outgoing flux with probability $-C^-=P_{\rm em}$ and by the reverse process with probability $C^+=P_{\rm abs}$. The ratio between those two processes gives us exactly
\begin{equation}
    \Gamma=\frac{P_{\rm em}}{P_{\rm abs}}= \frac{\langle \phi_\Omega^{{\textsc{kh}},\rm out}, \Phi^-_\Omega \rangle}{\langle \phi_\Omega^{{\textsc{kh}},\rm out}, \Phi^+_\Omega \rangle}= e^{-2 \pi \Omega/\kappa_{\rm KH}} \,.
\end{equation}
From the formula above we recognize immediately that the numerator and the denominator of the ratio exactly correspond to the ratio of the $\alpha$ and $\beta$ Bogoliubov coefficients:
\begin{equation}
    \Gamma=\frac{\int \frac{{\rm d} \bar{\Omega}}{2 \pi} |\beta_{\Omega \bar{\Omega}}|^2}{\int \frac{{\rm d} \bar{\Omega}}{2 \pi} |\alpha_{\Omega \bar{\Omega}}|^2}= e^{-2 \pi \Omega/\kappa_{\rm KH}} \,.
\end{equation}
This, together with the completeness relation:
\begin{equation}
    1=\int \frac{{\rm d}  \bar{\Omega}}{2 \pi} \, (|\alpha_{\Omega \bar{\Omega}}|^2-|\beta_{\Omega \bar{\Omega}}|^2)
\end{equation}
allows us to derive
\begin{equation}
    \int  \frac{{\rm d}  \bar{\Omega}}{2 \pi} |\beta_{\Omega \Bar{\Omega}}|^2= \frac{1}{e^{2 \pi \Omega/\kappa_{{\textsc{kh}}}}-1}
\end{equation}
as obtained above. 

This confirms that, through the tunneling rate, one can directly read the relation between the non-analyticity of the modes and the thermal character of the state off. If $\Gamma$ turns out to be a Boltzmann factor, one finds the temperature automatically. Morevoer, this analysis helps to understand where the choice of vacuum plays a role in a context like the tunneling formalism. In fact, such a Boltzmann factor gives rise to a thermal state in the analytical basis only when the non-analytical one is set in vacuum and the number of particle can be identified as the integral $|\beta|^2$ only. If the state is not taken to be vacuum for the freely falling observer, this is not true anymore, and, despite $\Gamma$ assumes the same shape, the shape of the state at infinity cannot be directly inferred from it in the same way.

Additionally, let us emphasize that this calculation can be performed locally near the horizon. However, as we have already pointed out, this does not correspond exactly to the detection at infinity, due to propagation effects. This is of particular importance when we consider modified dispersion relations where, even for the massless case, one expects a nontrivial effect from the propagation. Therefore, the results of this article cannot be quantitatively compared, beyond the zeroth order in $\alpha$, with the Bogoliubov coefficient approach, without a previous estimation of the effect of the modes' propagation up to infinity.



\bibliography{SGbibliography}

\begin{thebibliography}{60}%
\makeatletter
\providecommand \@ifxundefined [1]{%
 \@ifx{#1\undefined}
}%
\providecommand \@ifnum [1]{%
 \ifnum #1\expandafter \@firstoftwo
 \else \expandafter \@secondoftwo
 \fi
}%
\providecommand \@ifx [1]{%
 \ifx #1\expandafter \@firstoftwo
 \else \expandafter \@secondoftwo
 \fi
}%
\providecommand \natexlab [1]{#1}%
\providecommand \enquote  [1]{``#1''}%
\providecommand \bibnamefont  [1]{#1}%
\providecommand \bibfnamefont [1]{#1}%
\providecommand \citenamefont [1]{#1}%
\providecommand \href@noop [0]{\@secondoftwo}%
\providecommand \href [0]{\begingroup \@sanitize@url \@href}%
\providecommand \@href[1]{\@@startlink{#1}\@@href}%
\providecommand \@@href[1]{\endgroup#1\@@endlink}%
\providecommand \@sanitize@url [0]{\catcode `\\12\catcode `\$12\catcode
  `\&12\catcode `\#12\catcode `\^12\catcode `\_12\catcode `\%12\relax}%
\providecommand \@@startlink[1]{}%
\providecommand \@@endlink[0]{}%
\providecommand \url  [0]{\begingroup\@sanitize@url \@url }%
\providecommand \@url [1]{\endgroup\@href {#1}{\urlprefix }}%
\providecommand \urlprefix  [0]{URL }%
\providecommand \Eprint [0]{\href }%
\providecommand \doibase [0]{https://doi.org/}%
\providecommand \selectlanguage [0]{\@gobble}%
\providecommand \bibinfo  [0]{\@secondoftwo}%
\providecommand \bibfield  [0]{\@secondoftwo}%
\providecommand \translation [1]{[#1]}%
\providecommand \BibitemOpen [0]{}%
\providecommand \bibitemStop [0]{}%
\providecommand \bibitemNoStop [0]{.\EOS\space}%
\providecommand \EOS [0]{\spacefactor3000\relax}%
\providecommand \BibitemShut  [1]{\csname bibitem#1\endcsname}%
\let\auto@bib@innerbib\@empty
\bibitem [{\citenamefont {Unruh}(1981)}]{Unruh:1981cg}%
  \BibitemOpen
  \bibfield  {author} {\bibinfo {author} {\bibfnamefont {W.~G.}\ \bibnamefont
  {Unruh}},\ }\bibfield  {title} {\bibinfo {title} {{Experimental black hole
  evaporation}},\ }\href {https://doi.org/10.1103/PhysRevLett.46.1351}
  {\bibfield  {journal} {\bibinfo  {journal} {Phys. Rev. Lett.}\ }\textbf
  {\bibinfo {volume} {46}},\ \bibinfo {pages} {1351} (\bibinfo {year}
  {1981})}\BibitemShut {NoStop}%
\bibitem [{\citenamefont {Jacobson}(1991)}]{Jacobson:1991gr}%
  \BibitemOpen
  \bibfield  {author} {\bibinfo {author} {\bibfnamefont {T.~A.}\ \bibnamefont
  {Jacobson}},\ }\bibfield  {title} {\bibinfo {title} {{Black-hole evaporation
  and ultrashort distances}},\ }\href
  {https://doi.org/10.1103/PhysRevD.44.1731} {\bibfield  {journal} {\bibinfo
  {journal} {Phys. Rev. D}\ }\textbf {\bibinfo {volume} {44}},\ \bibinfo
  {pages} {1731} (\bibinfo {year} {1991})}\BibitemShut {NoStop}%
\bibitem [{\citenamefont {Jacobson}(1993)}]{Jacobson:1993hn}%
  \BibitemOpen
  \bibfield  {author} {\bibinfo {author} {\bibfnamefont {T.~A.}\ \bibnamefont
  {Jacobson}},\ }\bibfield  {title} {\bibinfo {title} {{Black hole radiation in
  the presence of a short distance cutoff}},\ }\href
  {https://doi.org/10.1103/PhysRevD.48.728} {\bibfield  {journal} {\bibinfo
  {journal} {Phys. Rev. D}\ }\textbf {\bibinfo {volume} {48}},\ \bibinfo
  {pages} {728} (\bibinfo {year} {1993})},\ \Eprint
  {https://arxiv.org/abs/hep-th/9303103} {hep-th/9303103} \BibitemShut
  {NoStop}%
\bibitem [{\citenamefont {Unruh}(1995{\natexlab{a}})}]{Unruh:1995je}%
  \BibitemOpen
  \bibfield  {author} {\bibinfo {author} {\bibfnamefont {W.~G.}\ \bibnamefont
  {Unruh}},\ }\bibfield  {title} {\bibinfo {title} {{Sonic analog of black
  holes and the effects of high frequencies on black hole evaporation}},\
  }\href {https://doi.org/10.1103/PhysRevD.51.2827} {\bibfield  {journal}
  {\bibinfo  {journal} {Phys. Rev. D}\ }\textbf {\bibinfo {volume} {51}},\
  \bibinfo {pages} {2827} (\bibinfo {year} {1995}{\natexlab{a}})},\ \Eprint
  {https://arxiv.org/abs/gr-qc/9409008} {arXiv:gr-qc/9409008} \BibitemShut
  {NoStop}%
\bibitem [{\citenamefont {Barceló}\ \emph {et~al.}(2005)\citenamefont
  {Barceló}, \citenamefont {Liberati},\ and\ \citenamefont
  {Visser}}]{Barcelo_2005_review}%
  \BibitemOpen
  \bibfield  {author} {\bibinfo {author} {\bibfnamefont {C.}~\bibnamefont
  {Barceló}}, \bibinfo {author} {\bibfnamefont {S.}~\bibnamefont {Liberati}},\
  and\ \bibinfo {author} {\bibfnamefont {M.}~\bibnamefont {Visser}},\
  }\bibfield  {title} {\bibinfo {title} {Analogue gravity},\ }\bibfield
  {journal} {\bibinfo  {journal} {Living Reviews in Relativity}\ }\textbf
  {\bibinfo {volume} {8}},\ \href {https://doi.org/10.12942/lrr-2005-12}
  {10.12942/lrr-2005-12} (\bibinfo {year} {2005})\BibitemShut {NoStop}%
\bibitem [{\citenamefont {Brout}\ \emph
  {et~al.}(1995{\natexlab{a}})\citenamefont {Brout}, \citenamefont {Massar},
  \citenamefont {Parentani},\ and\ \citenamefont {Spindel}}]{Brout:1995wp}%
  \BibitemOpen
  \bibfield  {author} {\bibinfo {author} {\bibfnamefont {R.}~\bibnamefont
  {Brout}}, \bibinfo {author} {\bibfnamefont {S.}~\bibnamefont {Massar}},
  \bibinfo {author} {\bibfnamefont {R.}~\bibnamefont {Parentani}},\ and\
  \bibinfo {author} {\bibfnamefont {P.}~\bibnamefont {Spindel}},\ }\bibfield
  {title} {\bibinfo {title} {{Hawking radiation without transPlanckian
  frequencies}},\ }\href {https://doi.org/10.1103/PhysRevD.52.4559} {\bibfield
  {journal} {\bibinfo  {journal} {Phys. Rev. D}\ }\textbf {\bibinfo {volume}
  {52}},\ \bibinfo {pages} {4559} (\bibinfo {year} {1995}{\natexlab{a}})},\
  \Eprint {https://arxiv.org/abs/hep-th/9506121} {arXiv:hep-th/9506121}
  \BibitemShut {NoStop}%
\bibitem [{\citenamefont {Corley}\ and\ \citenamefont
  {Jacobson}(1996)}]{Corley:1996ar}%
  \BibitemOpen
  \bibfield  {author} {\bibinfo {author} {\bibfnamefont {S.}~\bibnamefont
  {Corley}}\ and\ \bibinfo {author} {\bibfnamefont {T.}~\bibnamefont
  {Jacobson}},\ }\bibfield  {title} {\bibinfo {title} {{Hawking spectrum and
  high frequency dispersion}},\ }\href
  {https://doi.org/10.1103/PhysRevD.54.1568} {\bibfield  {journal} {\bibinfo
  {journal} {Phys. Rev. D}\ }\textbf {\bibinfo {volume} {54}},\ \bibinfo
  {pages} {1568} (\bibinfo {year} {1996})},\ \Eprint
  {https://arxiv.org/abs/hep-th/9601073} {arXiv:hep-th/9601073} \BibitemShut
  {NoStop}%
\bibitem [{\citenamefont {Corley}(1998)}]{Corley:1997pr}%
  \BibitemOpen
  \bibfield  {author} {\bibinfo {author} {\bibfnamefont {S.}~\bibnamefont
  {Corley}},\ }\bibfield  {title} {\bibinfo {title} {{Computing the spectrum of
  black hole radiation in the presence of high frequency dispersion: An
  Analytical approach}},\ }\href {https://doi.org/10.1103/PhysRevD.57.6280}
  {\bibfield  {journal} {\bibinfo  {journal} {Phys. Rev. D}\ }\textbf {\bibinfo
  {volume} {57}},\ \bibinfo {pages} {6280} (\bibinfo {year} {1998})},\ \Eprint
  {https://arxiv.org/abs/hep-th/9710075} {arXiv:hep-th/9710075} \BibitemShut
  {NoStop}%
\bibitem [{\citenamefont {Himemoto}\ and\ \citenamefont
  {Tanaka}(2000)}]{Himemoto:1999kd}%
  \BibitemOpen
  \bibfield  {author} {\bibinfo {author} {\bibfnamefont {Y.}~\bibnamefont
  {Himemoto}}\ and\ \bibinfo {author} {\bibfnamefont {T.}~\bibnamefont
  {Tanaka}},\ }\bibfield  {title} {\bibinfo {title} {{A Generalization of the
  model of Hawking radiation with modified high frequency dispersion
  relation}},\ }\href {https://doi.org/10.1103/PhysRevD.61.064004} {\bibfield
  {journal} {\bibinfo  {journal} {Phys. Rev. D}\ }\textbf {\bibinfo {volume}
  {61}},\ \bibinfo {pages} {064004} (\bibinfo {year} {2000})},\ \Eprint
  {https://arxiv.org/abs/gr-qc/9904076} {arXiv:gr-qc/9904076} \BibitemShut
  {NoStop}%
\bibitem [{\citenamefont {Saida}\ and\ \citenamefont
  {Sakagami}(2000)}]{Saida:1999ap}%
  \BibitemOpen
  \bibfield  {author} {\bibinfo {author} {\bibfnamefont {H.}~\bibnamefont
  {Saida}}\ and\ \bibinfo {author} {\bibfnamefont {M.-a.}\ \bibnamefont
  {Sakagami}},\ }\bibfield  {title} {\bibinfo {title} {{Black hole radiation
  with high frequency dispersion}},\ }\href
  {https://doi.org/10.1103/PhysRevD.61.084023} {\bibfield  {journal} {\bibinfo
  {journal} {Phys. Rev. D}\ }\textbf {\bibinfo {volume} {61}},\ \bibinfo
  {pages} {084023} (\bibinfo {year} {2000})},\ \Eprint
  {https://arxiv.org/abs/gr-qc/9905034} {arXiv:gr-qc/9905034} \BibitemShut
  {NoStop}%
\bibitem [{\citenamefont {Unruh}\ and\ \citenamefont
  {Schutzhold}(2005)}]{Unruh:2004zk}%
  \BibitemOpen
  \bibfield  {author} {\bibinfo {author} {\bibfnamefont {W.~G.}\ \bibnamefont
  {Unruh}}\ and\ \bibinfo {author} {\bibfnamefont {R.}~\bibnamefont
  {Schutzhold}},\ }\bibfield  {title} {\bibinfo {title} {{On the universality
  of the Hawking effect}},\ }\href {https://doi.org/10.1103/PhysRevD.71.024028}
  {\bibfield  {journal} {\bibinfo  {journal} {Phys. Rev. D}\ }\textbf {\bibinfo
  {volume} {71}},\ \bibinfo {pages} {024028} (\bibinfo {year} {2005})},\
  \Eprint {https://arxiv.org/abs/gr-qc/0408009} {arXiv:gr-qc/0408009}
  \BibitemShut {NoStop}%
\bibitem [{\citenamefont {Macher}\ and\ \citenamefont
  {Parentani}(2009{\natexlab{a}})}]{Macher:2009tw}%
  \BibitemOpen
  \bibfield  {author} {\bibinfo {author} {\bibfnamefont {J.}~\bibnamefont
  {Macher}}\ and\ \bibinfo {author} {\bibfnamefont {R.}~\bibnamefont
  {Parentani}},\ }\bibfield  {title} {\bibinfo {title} {{Black/White hole
  radiation from dispersive theories}},\ }\href
  {https://doi.org/10.1103/PhysRevD.79.124008} {\bibfield  {journal} {\bibinfo
  {journal} {Phys. Rev. D}\ }\textbf {\bibinfo {volume} {79}},\ \bibinfo
  {pages} {124008} (\bibinfo {year} {2009}{\natexlab{a}})},\ \Eprint
  {https://arxiv.org/abs/0903.2224} {arXiv:0903.2224 [hep-th]} \BibitemShut
  {NoStop}%
\bibitem [{\citenamefont {Macher}\ and\ \citenamefont
  {Parentani}(2009{\natexlab{b}})}]{Macher:2009nz}%
  \BibitemOpen
  \bibfield  {author} {\bibinfo {author} {\bibfnamefont {J.}~\bibnamefont
  {Macher}}\ and\ \bibinfo {author} {\bibfnamefont {R.}~\bibnamefont
  {Parentani}},\ }\bibfield  {title} {\bibinfo {title} {{Black hole radiation
  in Bose-Einstein condensates}},\ }\href
  {https://doi.org/10.1103/PhysRevA.80.043601} {\bibfield  {journal} {\bibinfo
  {journal} {Phys. Rev. A}\ }\textbf {\bibinfo {volume} {80}},\ \bibinfo
  {pages} {043601} (\bibinfo {year} {2009}{\natexlab{b}})},\ \Eprint
  {https://arxiv.org/abs/0905.3634} {arXiv:0905.3634 [cond-mat.quant-gas]}
  \BibitemShut {NoStop}%
\bibitem [{\citenamefont {Finazzi}\ and\ \citenamefont
  {Parentani}(2011{\natexlab{a}})}]{Finazzi:2010yq}%
  \BibitemOpen
  \bibfield  {author} {\bibinfo {author} {\bibfnamefont {S.}~\bibnamefont
  {Finazzi}}\ and\ \bibinfo {author} {\bibfnamefont {R.}~\bibnamefont
  {Parentani}},\ }\bibfield  {title} {\bibinfo {title} {{Spectral properties of
  acoustic black hole radiation: broadening the horizon}},\ }\href
  {https://doi.org/10.1103/PhysRevD.83.084010} {\bibfield  {journal} {\bibinfo
  {journal} {Phys. Rev. D}\ }\textbf {\bibinfo {volume} {83}},\ \bibinfo
  {pages} {084010} (\bibinfo {year} {2011}{\natexlab{a}})},\ \Eprint
  {https://arxiv.org/abs/1012.1556} {arXiv:1012.1556 [gr-qc]} \BibitemShut
  {NoStop}%
\bibitem [{\citenamefont {Finazzi}\ and\ \citenamefont
  {Parentani}(2011{\natexlab{b}})}]{Finazzi:2011jd}%
  \BibitemOpen
  \bibfield  {author} {\bibinfo {author} {\bibfnamefont {S.}~\bibnamefont
  {Finazzi}}\ and\ \bibinfo {author} {\bibfnamefont {R.}~\bibnamefont
  {Parentani}},\ }\bibfield  {title} {\bibinfo {title} {{On the robustness of
  acoustic black hole spectra}},\ }\href
  {https://doi.org/10.1088/1742-6596/314/1/012030} {\bibfield  {journal}
  {\bibinfo  {journal} {J. Phys. Conf. Ser.}\ }\textbf {\bibinfo {volume}
  {314}},\ \bibinfo {pages} {012030} (\bibinfo {year} {2011}{\natexlab{b}})},\
  \Eprint {https://arxiv.org/abs/1102.1452} {arXiv:1102.1452 [gr-qc]}
  \BibitemShut {NoStop}%
\bibitem [{\citenamefont {Coutant}\ \emph {et~al.}(2012)\citenamefont
  {Coutant}, \citenamefont {Parentani},\ and\ \citenamefont
  {Finazzi}}]{Coutant:2011in}%
  \BibitemOpen
  \bibfield  {author} {\bibinfo {author} {\bibfnamefont {A.}~\bibnamefont
  {Coutant}}, \bibinfo {author} {\bibfnamefont {R.}~\bibnamefont {Parentani}},\
  and\ \bibinfo {author} {\bibfnamefont {S.}~\bibnamefont {Finazzi}},\
  }\bibfield  {title} {\bibinfo {title} {{Black hole radiation with short
  distance dispersion, an analytical S-matrix approach}},\ }\href
  {https://doi.org/10.1103/PhysRevD.85.024021} {\bibfield  {journal} {\bibinfo
  {journal} {Phys. Rev. D}\ }\textbf {\bibinfo {volume} {85}},\ \bibinfo
  {pages} {024021} (\bibinfo {year} {2012})},\ \Eprint
  {https://arxiv.org/abs/1108.1821} {arXiv:1108.1821 [hep-th]} \BibitemShut
  {NoStop}%
\bibitem [{\citenamefont {Finazzi}\ and\ \citenamefont
  {Parentani}(2012)}]{Finazzi:2012iu}%
  \BibitemOpen
  \bibfield  {author} {\bibinfo {author} {\bibfnamefont {S.}~\bibnamefont
  {Finazzi}}\ and\ \bibinfo {author} {\bibfnamefont {R.}~\bibnamefont
  {Parentani}},\ }\bibfield  {title} {\bibinfo {title} {{Hawking radiation in
  dispersive theories, the two regimes}},\ }\href
  {https://doi.org/10.1103/PhysRevD.85.124027} {\bibfield  {journal} {\bibinfo
  {journal} {Phys. Rev. D}\ }\textbf {\bibinfo {volume} {85}},\ \bibinfo
  {pages} {124027} (\bibinfo {year} {2012})},\ \Eprint
  {https://arxiv.org/abs/1202.6015} {arXiv:1202.6015 [gr-qc]} \BibitemShut
  {NoStop}%
\bibitem [{\citenamefont {Michel}\ and\ \citenamefont
  {Parentani}(2014)}]{PhysRevD.90.044033}%
  \BibitemOpen
  \bibfield  {author} {\bibinfo {author} {\bibfnamefont {F.}~\bibnamefont
  {Michel}}\ and\ \bibinfo {author} {\bibfnamefont {R.}~\bibnamefont
  {Parentani}},\ }\bibfield  {title} {\bibinfo {title} {Probing the thermal
  character of analogue hawking radiation for shallow water waves?},\ }\href
  {https://doi.org/10.1103/PhysRevD.90.044033} {\bibfield  {journal} {\bibinfo
  {journal} {Phys. Rev. D}\ }\textbf {\bibinfo {volume} {90}},\ \bibinfo
  {pages} {044033} (\bibinfo {year} {2014})}\BibitemShut {NoStop}%
\bibitem [{\citenamefont {Michel}\ and\ \citenamefont
  {Parentani}(2017)}]{Michel:2015aga}%
  \BibitemOpen
  \bibfield  {author} {\bibinfo {author} {\bibfnamefont {F.}~\bibnamefont
  {Michel}}\ and\ \bibinfo {author} {\bibfnamefont {R.}~\bibnamefont
  {Parentani}},\ }\bibfield  {title} {\bibinfo {title} {{Mode mixing in sub-
  and trans-critical flows over an obstacle: When should
  Hawking\textquoteright{}s predictions be recovered?}},\ }in\ \href
  {https://doi.org/10.1142/9789813226609_0173} {\emph {\bibinfo {booktitle}
  {{14th Marcel Grossmann Meeting on Recent Developments in Theoretical and
  Experimental General Relativity, Astrophysics, and Relativistic Field
  Theories}}}},\ Vol.~\bibinfo {volume} {2}\ (\bibinfo {year} {2017})\ pp.\
  \bibinfo {pages} {1709--1717},\ \Eprint {https://arxiv.org/abs/1508.02044}
  {arXiv:1508.02044 [gr-qc]} \BibitemShut {NoStop}%
\bibitem [{\citenamefont {Parikh}\ and\ \citenamefont
  {Wilczek}(2000)}]{parikh2000hawking}%
  \BibitemOpen
  \bibfield  {author} {\bibinfo {author} {\bibfnamefont {M.~K.}\ \bibnamefont
  {Parikh}}\ and\ \bibinfo {author} {\bibfnamefont {F.}~\bibnamefont
  {Wilczek}},\ }\bibfield  {title} {\bibinfo {title} {Hawking radiation as
  tunneling},\ }\href@noop {} {\bibfield  {journal} {\bibinfo  {journal}
  {Physical review letters}\ }\textbf {\bibinfo {volume} {85}},\ \bibinfo
  {pages} {5042} (\bibinfo {year} {2000})}\BibitemShut {NoStop}%
\bibitem [{\citenamefont {Srinivasan}\ and\ \citenamefont
  {Padmanabhan}(1999)}]{srinivasan1999particle}%
  \BibitemOpen
  \bibfield  {author} {\bibinfo {author} {\bibfnamefont {K.}~\bibnamefont
  {Srinivasan}}\ and\ \bibinfo {author} {\bibfnamefont {T.}~\bibnamefont
  {Padmanabhan}},\ }\bibfield  {title} {\bibinfo {title} {Particle production
  and complex path analysis},\ }\href@noop {} {\bibfield  {journal} {\bibinfo
  {journal} {Physical Review D}\ }\textbf {\bibinfo {volume} {60}},\ \bibinfo
  {pages} {024007} (\bibinfo {year} {1999})}\BibitemShut {NoStop}%
\bibitem [{\citenamefont {Brout}\ \emph
  {et~al.}(1995{\natexlab{b}})\citenamefont {Brout}, \citenamefont {Massar},
  \citenamefont {Parentani},\ and\ \citenamefont {Spindel}}]{brout1995primer}%
  \BibitemOpen
  \bibfield  {author} {\bibinfo {author} {\bibfnamefont {R.}~\bibnamefont
  {Brout}}, \bibinfo {author} {\bibfnamefont {S.}~\bibnamefont {Massar}},
  \bibinfo {author} {\bibfnamefont {R.}~\bibnamefont {Parentani}},\ and\
  \bibinfo {author} {\bibfnamefont {P.}~\bibnamefont {Spindel}},\ }\bibfield
  {title} {\bibinfo {title} {Hawking radiation without trans-planckian
  frequencies},\ }\href@noop {} {\bibfield  {journal} {\bibinfo  {journal}
  {Physical Review D}\ }\textbf {\bibinfo {volume} {52}},\ \bibinfo {pages}
  {4559} (\bibinfo {year} {1995}{\natexlab{b}})}\BibitemShut {NoStop}%
\bibitem [{\citenamefont {Barceló}\ \emph {et~al.}(2004)\citenamefont
  {Barceló}, \citenamefont {Liberati}, \citenamefont {Sonego},\ and\
  \citenamefont {Visser}}]{Barcelo_2004_Causal}%
  \BibitemOpen
  \bibfield  {author} {\bibinfo {author} {\bibfnamefont {C.}~\bibnamefont
  {Barceló}}, \bibinfo {author} {\bibfnamefont {S.}~\bibnamefont {Liberati}},
  \bibinfo {author} {\bibfnamefont {S.}~\bibnamefont {Sonego}},\ and\ \bibinfo
  {author} {\bibfnamefont {M.}~\bibnamefont {Visser}},\ }\bibfield  {title}
  {\bibinfo {title} {Causal structure of analogue spacetimes},\ }\href
  {https://doi.org/10.1088/1367-2630/6/1/186} {\bibfield  {journal} {\bibinfo
  {journal} {New Journal of Physics}\ }\textbf {\bibinfo {volume} {6}},\
  \bibinfo {pages} {186–186} (\bibinfo {year} {2004})}\BibitemShut {NoStop}%
\bibitem [{\citenamefont {Barcelo}\ \emph {et~al.}(2005)\citenamefont
  {Barcelo}, \citenamefont {Liberati},\ and\ \citenamefont
  {Visser}}]{Barcelo:2005fc}%
  \BibitemOpen
  \bibfield  {author} {\bibinfo {author} {\bibfnamefont {C.}~\bibnamefont
  {Barcelo}}, \bibinfo {author} {\bibfnamefont {S.}~\bibnamefont {Liberati}},\
  and\ \bibinfo {author} {\bibfnamefont {M.}~\bibnamefont {Visser}},\
  }\bibfield  {title} {\bibinfo {title} {{Analogue gravity}},\ }\href
  {https://doi.org/10.12942/lrr-2005-12} {\bibfield  {journal} {\bibinfo
  {journal} {Living Rev. Rel.}\ }\textbf {\bibinfo {volume} {8}},\ \bibinfo
  {pages} {12} (\bibinfo {year} {2005})},\ \Eprint
  {https://arxiv.org/abs/gr-qc/0505065} {arXiv:gr-qc/0505065} \BibitemShut
  {NoStop}%
\bibitem [{\citenamefont {Sch{\"u}tzhold}\ and\ \citenamefont
  {Unruh}(2008)}]{schutzhold2008origin}%
  \BibitemOpen
  \bibfield  {author} {\bibinfo {author} {\bibfnamefont {R.}~\bibnamefont
  {Sch{\"u}tzhold}}\ and\ \bibinfo {author} {\bibfnamefont {W.~G.}\
  \bibnamefont {Unruh}},\ }\bibfield  {title} {\bibinfo {title} {Origin of the
  particles in black hole evaporation},\ }\href@noop {} {\bibfield  {journal}
  {\bibinfo  {journal} {Physical Review D}\ }\textbf {\bibinfo {volume} {78}},\
  \bibinfo {pages} {041504} (\bibinfo {year} {2008})}\BibitemShut {NoStop}%
\bibitem [{\citenamefont {Ribeiro}\ \emph {et~al.}(2022)\citenamefont
  {Ribeiro}, \citenamefont {Baak},\ and\ \citenamefont
  {Fischer}}]{fischPhysRevD.105.124066}%
  \BibitemOpen
  \bibfield  {author} {\bibinfo {author} {\bibfnamefont {C.~C.~H.}\
  \bibnamefont {Ribeiro}}, \bibinfo {author} {\bibfnamefont {S.-S.}\
  \bibnamefont {Baak}},\ and\ \bibinfo {author} {\bibfnamefont {U.~R.}\
  \bibnamefont {Fischer}},\ }\bibfield  {title} {\bibinfo {title} {Existence of
  steady-state black hole analogs in finite quasi-one-dimensional bose-einstein
  condensates},\ }\href {https://doi.org/10.1103/PhysRevD.105.124066}
  {\bibfield  {journal} {\bibinfo  {journal} {Phys. Rev. D}\ }\textbf {\bibinfo
  {volume} {105}},\ \bibinfo {pages} {124066} (\bibinfo {year}
  {2022})}\BibitemShut {NoStop}%
\bibitem [{\citenamefont {Holanda~Ribeiro}\ and\ \citenamefont
  {Fischer}(2023)}]{fischPhysRevD.107.L121502}%
  \BibitemOpen
  \bibfield  {author} {\bibinfo {author} {\bibfnamefont {C.~C.}\ \bibnamefont
  {Holanda~Ribeiro}}\ and\ \bibinfo {author} {\bibfnamefont {U.~R.}\
  \bibnamefont {Fischer}},\ }\bibfield  {title} {\bibinfo {title} {Impact of
  trans-planckian excitations on black-hole radiation in dipolar condensates},\
  }\href {https://doi.org/10.1103/PhysRevD.107.L121502} {\bibfield  {journal}
  {\bibinfo  {journal} {Phys. Rev. D}\ }\textbf {\bibinfo {volume} {107}},\
  \bibinfo {pages} {L121502} (\bibinfo {year} {2023})}\BibitemShut {NoStop}%
\bibitem [{\citenamefont {Cropp}\ \emph {et~al.}(2014)\citenamefont {Cropp},
  \citenamefont {Liberati}, \citenamefont {Mohd},\ and\ \citenamefont
  {Visser}}]{Cropp_2014_ray_tracing}%
  \BibitemOpen
  \bibfield  {author} {\bibinfo {author} {\bibfnamefont {B.}~\bibnamefont
  {Cropp}}, \bibinfo {author} {\bibfnamefont {S.}~\bibnamefont {Liberati}},
  \bibinfo {author} {\bibfnamefont {A.}~\bibnamefont {Mohd}},\ and\ \bibinfo
  {author} {\bibfnamefont {M.}~\bibnamefont {Visser}},\ }\bibfield  {title}
  {\bibinfo {title} {Ray tracing einstein-Æther black holes: Universal versus
  killing horizons},\ }\bibfield  {journal} {\bibinfo  {journal} {Physical
  Review D}\ }\textbf {\bibinfo {volume} {89}},\ \href
  {https://doi.org/10.1103/physrevd.89.064061} {10.1103/physrevd.89.064061}
  (\bibinfo {year} {2014})\BibitemShut {NoStop}%
\bibitem [{\citenamefont {Del~Porro}\ \emph {et~al.}(2023)\citenamefont
  {Del~Porro}, \citenamefont {Herrero-Valea}, \citenamefont {Liberati},\ and\
  \citenamefont {Schneider}}]{Del_Porro_2023_Hawking}%
  \BibitemOpen
  \bibfield  {author} {\bibinfo {author} {\bibfnamefont {F.}~\bibnamefont
  {Del~Porro}}, \bibinfo {author} {\bibfnamefont {M.}~\bibnamefont
  {Herrero-Valea}}, \bibinfo {author} {\bibfnamefont {S.}~\bibnamefont
  {Liberati}},\ and\ \bibinfo {author} {\bibfnamefont {M.}~\bibnamefont
  {Schneider}},\ }\bibfield  {title} {\bibinfo {title} {Hawking radiation in
  lorentz violating gravity: a tale of two horizons},\ }\bibfield  {journal}
  {\bibinfo  {journal} {Journal of High Energy Physics}\ }\textbf {\bibinfo
  {volume} {2023}},\ \href {https://doi.org/10.1007/jhep12(2023)094}
  {10.1007/jhep12(2023)094} (\bibinfo {year} {2023})\BibitemShut {NoStop}%
\bibitem [{\citenamefont {Unruh}(1995{\natexlab{b}})}]{unruh1995sonic}%
  \BibitemOpen
  \bibfield  {author} {\bibinfo {author} {\bibfnamefont {W.~G.}\ \bibnamefont
  {Unruh}},\ }\bibfield  {title} {\bibinfo {title} {Sonic analogue of black
  holes and the effects of high frequencies on black hole evaporation},\
  }\href@noop {} {\bibfield  {journal} {\bibinfo  {journal} {Physical Review
  D}\ }\textbf {\bibinfo {volume} {51}},\ \bibinfo {pages} {2827} (\bibinfo
  {year} {1995}{\natexlab{b}})}\BibitemShut {NoStop}%
\bibitem [{\citenamefont {Moretti}\ and\ \citenamefont
  {Pinamonti}(2012)}]{moretti2012state}%
  \BibitemOpen
  \bibfield  {author} {\bibinfo {author} {\bibfnamefont {V.}~\bibnamefont
  {Moretti}}\ and\ \bibinfo {author} {\bibfnamefont {N.}~\bibnamefont
  {Pinamonti}},\ }\bibfield  {title} {\bibinfo {title} {State independence for
  tunnelling processes through black hole horizons and hawking radiation},\
  }\href@noop {} {\bibfield  {journal} {\bibinfo  {journal} {Communications in
  Mathematical Physics}\ }\textbf {\bibinfo {volume} {309}},\ \bibinfo {pages}
  {295} (\bibinfo {year} {2012})}\BibitemShut {NoStop}%
\bibitem [{\citenamefont {Liberati}\ \emph {et~al.}(2020)\citenamefont
  {Liberati}, \citenamefont {Tricella},\ and\ \citenamefont
  {Trombettoni}}]{Liberati:2020mdr}%
  \BibitemOpen
  \bibfield  {author} {\bibinfo {author} {\bibfnamefont {S.}~\bibnamefont
  {Liberati}}, \bibinfo {author} {\bibfnamefont {G.}~\bibnamefont {Tricella}},\
  and\ \bibinfo {author} {\bibfnamefont {A.}~\bibnamefont {Trombettoni}},\
  }\bibfield  {title} {\bibinfo {title} {{Back-reaction in canonical analogue
  black holes}},\ }\href {https://doi.org/10.3390/app10248868} {\bibfield
  {journal} {\bibinfo  {journal} {Appl. Sciences}\ }\textbf {\bibinfo {volume}
  {10}},\ \bibinfo {pages} {8868} (\bibinfo {year} {2020})},\ \Eprint
  {https://arxiv.org/abs/2010.09966} {arXiv:2010.09966 [gr-qc]} \BibitemShut
  {NoStop}%
\bibitem [{\citenamefont {Vanzo}\ \emph {et~al.}(2011)\citenamefont {Vanzo},
  \citenamefont {Acquaviva},\ and\ \citenamefont
  {Criscienzo}}]{Vanzo_2011_Tunneling}%
  \BibitemOpen
  \bibfield  {author} {\bibinfo {author} {\bibfnamefont {L.}~\bibnamefont
  {Vanzo}}, \bibinfo {author} {\bibfnamefont {G.}~\bibnamefont {Acquaviva}},\
  and\ \bibinfo {author} {\bibfnamefont {R.~D.}\ \bibnamefont {Criscienzo}},\
  }\bibfield  {title} {\bibinfo {title} {Tunnelling methods and hawking’s
  radiation: achievements and prospects},\ }\bibfield  {journal} {\bibinfo
  {journal} {Classical and Quantum Gravity}\ }\textbf {\bibinfo {volume}
  {28}},\ \href {https://doi.org/10.1088/0264-9381/28/18/183001}
  {10.1088/0264-9381/28/18/183001} (\bibinfo {year} {2011})\BibitemShut
  {NoStop}%
\bibitem [{\citenamefont {Giavoni}\ and\ \citenamefont
  {Schneider}(2020)}]{Giavoni:2020gui}%
  \BibitemOpen
  \bibfield  {author} {\bibinfo {author} {\bibfnamefont {C.}~\bibnamefont
  {Giavoni}}\ and\ \bibinfo {author} {\bibfnamefont {M.}~\bibnamefont
  {Schneider}},\ }\bibfield  {title} {\bibinfo {title} {{Quantum effects across
  dynamical horizons}},\ }\href {https://doi.org/10.1088/1361-6382/abb576}
  {\bibfield  {journal} {\bibinfo  {journal} {Class. Quant. Grav.}\ }\textbf
  {\bibinfo {volume} {37}},\ \bibinfo {pages} {215020} (\bibinfo {year}
  {2020})},\ \Eprint {https://arxiv.org/abs/2003.11095} {arXiv:2003.11095
  [gr-qc]} \BibitemShut {NoStop}%
\bibitem [{\citenamefont {Coutant}\ and\ \citenamefont
  {Weinfurtner}(2016)}]{Coutant_2016_Hawking_subcritical}%
  \BibitemOpen
  \bibfield  {author} {\bibinfo {author} {\bibfnamefont {A.}~\bibnamefont
  {Coutant}}\ and\ \bibinfo {author} {\bibfnamefont {S.}~\bibnamefont
  {Weinfurtner}},\ }\bibfield  {title} {\bibinfo {title} {The imprint of the
  analogue hawking effect in subcritical flows},\ }\bibfield  {journal}
  {\bibinfo  {journal} {Physical Review D}\ }\textbf {\bibinfo {volume} {94}},\
  \href {https://doi.org/10.1103/physrevd.94.064026}
  {10.1103/physrevd.94.064026} (\bibinfo {year} {2016})\BibitemShut {NoStop}%
\bibitem [{\citenamefont {Weinfurtner}\ \emph {et~al.}(2013)\citenamefont
  {Weinfurtner}, \citenamefont {Tedford}, \citenamefont {Penrice},
  \citenamefont {Unruh},\ and\ \citenamefont {Lawrence}}]{Weinfurtner:2013zfa}%
  \BibitemOpen
  \bibfield  {author} {\bibinfo {author} {\bibfnamefont {S.}~\bibnamefont
  {Weinfurtner}}, \bibinfo {author} {\bibfnamefont {E.~W.}\ \bibnamefont
  {Tedford}}, \bibinfo {author} {\bibfnamefont {M.~C.~J.}\ \bibnamefont
  {Penrice}}, \bibinfo {author} {\bibfnamefont {W.~G.}\ \bibnamefont {Unruh}},\
  and\ \bibinfo {author} {\bibfnamefont {G.~A.}\ \bibnamefont {Lawrence}},\
  }\bibfield  {title} {\bibinfo {title} {{Classical aspects of Hawking
  radiation verified in analogue gravity experiment}},\ }\href
  {https://doi.org/10.1007/978-3-319-00266-8_8} {\bibfield  {journal} {\bibinfo
   {journal} {Lect. Notes Phys.}\ }\textbf {\bibinfo {volume} {870}},\ \bibinfo
  {pages} {167} (\bibinfo {year} {2013})},\ \Eprint
  {https://arxiv.org/abs/1302.0375} {arXiv:1302.0375 [gr-qc]} \BibitemShut
  {NoStop}%
\bibitem [{\citenamefont {Coutant}\ and\ \citenamefont
  {Weinfurtner}(2018)}]{Coutant_2018_Hawking_superluminal}%
  \BibitemOpen
  \bibfield  {author} {\bibinfo {author} {\bibfnamefont {A.}~\bibnamefont
  {Coutant}}\ and\ \bibinfo {author} {\bibfnamefont {S.}~\bibnamefont
  {Weinfurtner}},\ }\bibfield  {title} {\bibinfo {title} {Low-frequency
  analogue hawking radiation: The bogoliubov-de gennes model},\ }\bibfield
  {journal} {\bibinfo  {journal} {Physical Review D}\ }\textbf {\bibinfo
  {volume} {97}},\ \href {https://doi.org/10.1103/physrevd.97.025006}
  {10.1103/physrevd.97.025006} (\bibinfo {year} {2018})\BibitemShut {NoStop}%
\bibitem [{\citenamefont {Finazzi}\ and\ \citenamefont
  {Parentani}(2011{\natexlab{c}})}]{Finazzi_2011_robustness}%
  \BibitemOpen
  \bibfield  {author} {\bibinfo {author} {\bibfnamefont {S.}~\bibnamefont
  {Finazzi}}\ and\ \bibinfo {author} {\bibfnamefont {R.}~\bibnamefont
  {Parentani}},\ }\bibfield  {title} {\bibinfo {title} {On the robustness of
  acoustic black hole spectra},\ }\href
  {https://doi.org/10.1088/1742-6596/314/1/012030} {\bibfield  {journal}
  {\bibinfo  {journal} {Journal of Physics: Conference Series}\ }\textbf
  {\bibinfo {volume} {314}},\ \bibinfo {pages} {012030} (\bibinfo {year}
  {2011}{\natexlab{c}})}\BibitemShut {NoStop}%
\bibitem [{\citenamefont {Barcaroli}\ \emph {et~al.}(2015)\citenamefont
  {Barcaroli}, \citenamefont {Brunkhorst}, \citenamefont {Gubitosi},
  \citenamefont {Loret},\ and\ \citenamefont {Pfeifer}}]{PhysRevD.92.084053}%
  \BibitemOpen
  \bibfield  {author} {\bibinfo {author} {\bibfnamefont {L.}~\bibnamefont
  {Barcaroli}}, \bibinfo {author} {\bibfnamefont {L.~K.}\ \bibnamefont
  {Brunkhorst}}, \bibinfo {author} {\bibfnamefont {G.}~\bibnamefont
  {Gubitosi}}, \bibinfo {author} {\bibfnamefont {N.}~\bibnamefont {Loret}},\
  and\ \bibinfo {author} {\bibfnamefont {C.}~\bibnamefont {Pfeifer}},\
  }\bibfield  {title} {\bibinfo {title} {Hamilton geometry: Phase space
  geometry from modified dispersion relations},\ }\href
  {https://doi.org/10.1103/PhysRevD.92.084053} {\bibfield  {journal} {\bibinfo
  {journal} {Phys. Rev. D}\ }\textbf {\bibinfo {volume} {92}},\ \bibinfo
  {pages} {084053} (\bibinfo {year} {2015})}\BibitemShut {NoStop}%
\bibitem [{\citenamefont {Coutant}\ and\ \citenamefont
  {Parentani}(2014)}]{coutant2014hawking}%
  \BibitemOpen
  \bibfield  {author} {\bibinfo {author} {\bibfnamefont {A.}~\bibnamefont
  {Coutant}}\ and\ \bibinfo {author} {\bibfnamefont {R.}~\bibnamefont
  {Parentani}},\ }\bibfield  {title} {\bibinfo {title} {Hawking radiation with
  dispersion: The broadened horizon paradigm},\ }\href@noop {} {\bibfield
  {journal} {\bibinfo  {journal} {Physical Review D}\ }\textbf {\bibinfo
  {volume} {90}},\ \bibinfo {pages} {121501} (\bibinfo {year}
  {2014})}\BibitemShut {NoStop}%
\bibitem [{\citenamefont {Schützhold}\ and\ \citenamefont
  {Unruh}(2013)}]{Schutzhold_2013_breakdown_WKB}%
  \BibitemOpen
  \bibfield  {author} {\bibinfo {author} {\bibfnamefont {R.}~\bibnamefont
  {Schützhold}}\ and\ \bibinfo {author} {\bibfnamefont {W.~G.}\ \bibnamefont
  {Unruh}},\ }\bibfield  {title} {\bibinfo {title} {Hawking radiation with
  dispersion versus breakdown of the wkb approximation},\ }\bibfield  {journal}
  {\bibinfo  {journal} {Physical Review D}\ }\textbf {\bibinfo {volume} {88}},\
  \href {https://doi.org/10.1103/physrevd.88.124009}
  {10.1103/physrevd.88.124009} (\bibinfo {year} {2013})\BibitemShut {NoStop}%
\bibitem [{\citenamefont {Albuquerque}\ \emph {et~al.}(2023)\citenamefont
  {Albuquerque}, \citenamefont {V{\"o}lkel}, \citenamefont {Kokkotas},\ and\
  \citenamefont {Bezerra}}]{albuquerque2023inverse}%
  \BibitemOpen
  \bibfield  {author} {\bibinfo {author} {\bibfnamefont {S.}~\bibnamefont
  {Albuquerque}}, \bibinfo {author} {\bibfnamefont {S.~H.}\ \bibnamefont
  {V{\"o}lkel}}, \bibinfo {author} {\bibfnamefont {K.~D.}\ \bibnamefont
  {Kokkotas}},\ and\ \bibinfo {author} {\bibfnamefont {V.~B.}\ \bibnamefont
  {Bezerra}},\ }\bibfield  {title} {\bibinfo {title} {Inverse problem of analog
  gravity systems},\ }\href@noop {} {\bibfield  {journal} {\bibinfo  {journal}
  {Physical Review D}\ }\textbf {\bibinfo {volume} {108}},\ \bibinfo {pages}
  {124053} (\bibinfo {year} {2023})}\BibitemShut {NoStop}%
\bibitem [{\citenamefont {Albuquerque}\ \emph {et~al.}(2024)\citenamefont
  {Albuquerque}, \citenamefont {V{\"o}lkel}, \citenamefont {Kokkotas},\ and\
  \citenamefont {Bezerra}}]{albuquerque2024inverse}%
  \BibitemOpen
  \bibfield  {author} {\bibinfo {author} {\bibfnamefont {S.}~\bibnamefont
  {Albuquerque}}, \bibinfo {author} {\bibfnamefont {S.~H.}\ \bibnamefont
  {V{\"o}lkel}}, \bibinfo {author} {\bibfnamefont {K.~D.}\ \bibnamefont
  {Kokkotas}},\ and\ \bibinfo {author} {\bibfnamefont {V.~B.}\ \bibnamefont
  {Bezerra}},\ }\bibfield  {title} {\bibinfo {title} {Inverse problem of analog
  gravity systems ii: rotation and energy-dependent boundary conditions},\
  }\href@noop {} {\bibfield  {journal} {\bibinfo  {journal} {arXiv preprint
  arXiv:2406.16670}\ } (\bibinfo {year} {2024})}\BibitemShut {NoStop}%
\bibitem [{\citenamefont {Weinfurtner}\ \emph {et~al.}(2011)\citenamefont
  {Weinfurtner}, \citenamefont {Tedford}, \citenamefont {Penrice},
  \citenamefont {Unruh},\ and\ \citenamefont {Lawrence}}]{Weinfurtner:2010nu}%
  \BibitemOpen
  \bibfield  {author} {\bibinfo {author} {\bibfnamefont {S.}~\bibnamefont
  {Weinfurtner}}, \bibinfo {author} {\bibfnamefont {E.~W.}\ \bibnamefont
  {Tedford}}, \bibinfo {author} {\bibfnamefont {M.~C.~J.}\ \bibnamefont
  {Penrice}}, \bibinfo {author} {\bibfnamefont {W.~G.}\ \bibnamefont {Unruh}},\
  and\ \bibinfo {author} {\bibfnamefont {G.~A.}\ \bibnamefont {Lawrence}},\
  }\bibfield  {title} {\bibinfo {title} {{Measurement of stimulated Hawking
  emission in an analogue system}},\ }\href
  {https://doi.org/10.1103/PhysRevLett.106.021302} {\bibfield  {journal}
  {\bibinfo  {journal} {Phys. Rev. Lett.}\ }\textbf {\bibinfo {volume} {106}},\
  \bibinfo {pages} {021302} (\bibinfo {year} {2011})},\ \Eprint
  {https://arxiv.org/abs/1008.1911} {arXiv:1008.1911 [gr-qc]} \BibitemShut
  {NoStop}%
\bibitem [{\citenamefont {Euv\'e}\ \emph {et~al.}(2015)\citenamefont {Euv\'e},
  \citenamefont {Michel}, \citenamefont {Parentani},\ and\ \citenamefont
  {Rousseaux}}]{Euve:2014aga}%
  \BibitemOpen
  \bibfield  {author} {\bibinfo {author} {\bibfnamefont {L.-P.}\ \bibnamefont
  {Euv\'e}}, \bibinfo {author} {\bibfnamefont {F.}~\bibnamefont {Michel}},
  \bibinfo {author} {\bibfnamefont {R.}~\bibnamefont {Parentani}},\ and\
  \bibinfo {author} {\bibfnamefont {G.}~\bibnamefont {Rousseaux}},\ }\bibfield
  {title} {\bibinfo {title} {{Wave blocking and partial transmission in
  subcritical flows over an obstacle}},\ }\href
  {https://doi.org/10.1103/PhysRevD.91.024020} {\bibfield  {journal} {\bibinfo
  {journal} {Phys. Rev. D}\ }\textbf {\bibinfo {volume} {91}},\ \bibinfo
  {pages} {024020} (\bibinfo {year} {2015})},\ \Eprint
  {https://arxiv.org/abs/1409.3830} {arXiv:1409.3830 [gr-qc]} \BibitemShut
  {NoStop}%
\bibitem [{\citenamefont {Euv\'e}\ \emph {et~al.}(2016)\citenamefont {Euv\'e},
  \citenamefont {Michel}, \citenamefont {Parentani}, \citenamefont {Philbin},\
  and\ \citenamefont {Rousseaux}}]{Euve:2015vml}%
  \BibitemOpen
  \bibfield  {author} {\bibinfo {author} {\bibfnamefont {L.~P.}\ \bibnamefont
  {Euv\'e}}, \bibinfo {author} {\bibfnamefont {F.}~\bibnamefont {Michel}},
  \bibinfo {author} {\bibfnamefont {R.}~\bibnamefont {Parentani}}, \bibinfo
  {author} {\bibfnamefont {T.~G.}\ \bibnamefont {Philbin}},\ and\ \bibinfo
  {author} {\bibfnamefont {G.}~\bibnamefont {Rousseaux}},\ }\bibfield  {title}
  {\bibinfo {title} {{Observation of noise correlated by the Hawking effect in
  a water tank}},\ }\href {https://doi.org/10.1103/PhysRevLett.117.121301}
  {\bibfield  {journal} {\bibinfo  {journal} {Phys. Rev. Lett.}\ }\textbf
  {\bibinfo {volume} {117}},\ \bibinfo {pages} {121301} (\bibinfo {year}
  {2016})},\ \Eprint {https://arxiv.org/abs/1511.08145} {arXiv:1511.08145
  [physics.flu-dyn]} \BibitemShut {NoStop}%
\bibitem [{\citenamefont {Robertson}\ \emph {et~al.}(2016)\citenamefont
  {Robertson}, \citenamefont {Michel},\ and\ \citenamefont
  {Parentani}}]{Robertson_2016}%
  \BibitemOpen
  \bibfield  {author} {\bibinfo {author} {\bibfnamefont {S.}~\bibnamefont
  {Robertson}}, \bibinfo {author} {\bibfnamefont {F.}~\bibnamefont {Michel}},\
  and\ \bibinfo {author} {\bibfnamefont {R.}~\bibnamefont {Parentani}},\
  }\bibfield  {title} {\bibinfo {title} {{Scattering of gravity waves in
  subcritical flows over an obstacle}},\ }\href
  {https://doi.org/10.1103/PhysRevD.93.124060} {\bibfield  {journal} {\bibinfo
  {journal} {Phys. Rev. D}\ }\textbf {\bibinfo {volume} {93}},\ \bibinfo
  {pages} {124060} (\bibinfo {year} {2016})},\ \Eprint
  {https://arxiv.org/abs/1604.07253} {arXiv:1604.07253 [gr-qc]} \BibitemShut
  {NoStop}%
\bibitem [{\citenamefont {Brito}\ \emph {et~al.}(2020)\citenamefont {Brito},
  \citenamefont {Cardoso}, \citenamefont {Pani} \emph
  {et~al.}}]{brito2020superradiance}%
  \BibitemOpen
  \bibfield  {author} {\bibinfo {author} {\bibfnamefont {R.}~\bibnamefont
  {Brito}}, \bibinfo {author} {\bibfnamefont {V.}~\bibnamefont {Cardoso}},
  \bibinfo {author} {\bibfnamefont {P.}~\bibnamefont {Pani}}, \emph {et~al.},\
  }\href@noop {} {\emph {\bibinfo {title} {Superradiance New Frontiers in Black
  Hole Physics}}}\ (\bibinfo  {publisher} {Springer},\ \bibinfo {year}
  {2020})\BibitemShut {NoStop}%
\bibitem [{\citenamefont {Liberati}(2013)}]{Liberati:2013xla}%
  \BibitemOpen
  \bibfield  {author} {\bibinfo {author} {\bibfnamefont {S.}~\bibnamefont
  {Liberati}},\ }\bibfield  {title} {\bibinfo {title} {{Tests of Lorentz
  invariance: a 2013 update}},\ }\href
  {https://doi.org/10.1088/0264-9381/30/13/133001} {\bibfield  {journal}
  {\bibinfo  {journal} {Class. Quant. Grav.}\ }\textbf {\bibinfo {volume}
  {30}},\ \bibinfo {pages} {133001} (\bibinfo {year} {2013})},\ \Eprint
  {https://arxiv.org/abs/1304.5795} {arXiv:1304.5795 [gr-qc]} \BibitemShut
  {NoStop}%
\bibitem [{\citenamefont {Unruh}\ and\ \citenamefont
  {Schützhold}(2003)}]{Unruh_2003_slow_light}%
  \BibitemOpen
  \bibfield  {author} {\bibinfo {author} {\bibfnamefont {W.~G.}\ \bibnamefont
  {Unruh}}\ and\ \bibinfo {author} {\bibfnamefont {R.}~\bibnamefont
  {Schützhold}},\ }\bibfield  {title} {\bibinfo {title} {On slow light as a
  black hole analogue},\ }\bibfield  {journal} {\bibinfo  {journal} {Physical
  Review D}\ }\textbf {\bibinfo {volume} {68}},\ \href
  {https://doi.org/10.1103/physrevd.68.024008} {10.1103/physrevd.68.024008}
  (\bibinfo {year} {2003})\BibitemShut {NoStop}%
\bibitem [{\citenamefont {Novello}\ \emph {et~al.}(2002)\citenamefont
  {Novello}, \citenamefont {Visser},\ and\ \citenamefont
  {Volovik}}]{Novello_2002_ArtificialBH}%
  \BibitemOpen
  \bibfield  {author} {\bibinfo {author} {\bibfnamefont {M.}~\bibnamefont
  {Novello}}, \bibinfo {author} {\bibfnamefont {M.}~\bibnamefont {Visser}},\
  and\ \bibinfo {author} {\bibfnamefont {G.}~\bibnamefont {Volovik}},\ }\href
  {https://doi.org/10.1142/4861} {\emph {\bibinfo {title} {Artificial Black
  Holes}}}\ (\bibinfo  {publisher} {WORLD SCIENTIFIC},\ \bibinfo {year}
  {2002})\ \Eprint
  {https://arxiv.org/abs/https://www.worldscientific.com/doi/pdf/10.1142/4861}
  {https://www.worldscientific.com/doi/pdf/10.1142/4861} \BibitemShut {NoStop}%
\bibitem [{\citenamefont {Giddings}(2016)}]{Giddings_2016_Hawking}%
  \BibitemOpen
  \bibfield  {author} {\bibinfo {author} {\bibfnamefont {S.~B.}\ \bibnamefont
  {Giddings}},\ }\bibfield  {title} {\bibinfo {title} {Hawking radiation, the
  stefan–boltzmann law, and unitarization},\ }\href
  {https://doi.org/10.1016/j.physletb.2015.12.076} {\bibfield  {journal}
  {\bibinfo  {journal} {Physics Letters B}\ }\textbf {\bibinfo {volume}
  {754}},\ \bibinfo {pages} {39–42} (\bibinfo {year} {2016})}\BibitemShut
  {NoStop}%
\bibitem [{\citenamefont {Dey}\ \emph {et~al.}(2017)\citenamefont {Dey},
  \citenamefont {Liberati},\ and\ \citenamefont
  {Pranzetti}}]{Dey_2017_quantum_atmosphere}%
  \BibitemOpen
  \bibfield  {author} {\bibinfo {author} {\bibfnamefont {R.}~\bibnamefont
  {Dey}}, \bibinfo {author} {\bibfnamefont {S.}~\bibnamefont {Liberati}},\ and\
  \bibinfo {author} {\bibfnamefont {D.}~\bibnamefont {Pranzetti}},\ }\bibfield
  {title} {\bibinfo {title} {The black hole quantum atmosphere},\ }\href
  {https://doi.org/10.1016/j.physletb.2017.09.076} {\bibfield  {journal}
  {\bibinfo  {journal} {Physics Letters B}\ }\textbf {\bibinfo {volume}
  {774}},\ \bibinfo {pages} {308–316} (\bibinfo {year} {2017})}\BibitemShut
  {NoStop}%
\bibitem [{\citenamefont {Dey}\ \emph {et~al.}(2019)\citenamefont {Dey},
  \citenamefont {Liberati}, \citenamefont {Mirzaiyan},\ and\ \citenamefont
  {Pranzetti}}]{Dey:2019ugf}%
  \BibitemOpen
  \bibfield  {author} {\bibinfo {author} {\bibfnamefont {R.}~\bibnamefont
  {Dey}}, \bibinfo {author} {\bibfnamefont {S.}~\bibnamefont {Liberati}},
  \bibinfo {author} {\bibfnamefont {Z.}~\bibnamefont {Mirzaiyan}},\ and\
  \bibinfo {author} {\bibfnamefont {D.}~\bibnamefont {Pranzetti}},\ }\bibfield
  {title} {\bibinfo {title} {{Black hole quantum atmosphere for freely falling
  observers}},\ }\href {https://doi.org/10.1016/j.physletb.2019.134828}
  {\bibfield  {journal} {\bibinfo  {journal} {Phys. Lett. B}\ }\textbf
  {\bibinfo {volume} {797}},\ \bibinfo {pages} {134828} (\bibinfo {year}
  {2019})},\ \Eprint {https://arxiv.org/abs/1906.02958} {arXiv:1906.02958
  [gr-qc]} \BibitemShut {NoStop}%
\bibitem [{\citenamefont {Agullo}\ \emph {et~al.}(2015)\citenamefont {Agullo},
  \citenamefont {Nelson},\ and\ \citenamefont {Ashtekar}}]{Agullo:2014ica}%
  \BibitemOpen
  \bibfield  {author} {\bibinfo {author} {\bibfnamefont {I.}~\bibnamefont
  {Agullo}}, \bibinfo {author} {\bibfnamefont {W.}~\bibnamefont {Nelson}},\
  and\ \bibinfo {author} {\bibfnamefont {A.}~\bibnamefont {Ashtekar}},\
  }\bibfield  {title} {\bibinfo {title} {{Preferred instantaneous vacuum for
  linear scalar fields in cosmological space-times}},\ }\href
  {https://doi.org/10.1103/PhysRevD.91.064051} {\bibfield  {journal} {\bibinfo
  {journal} {Phys. Rev. D}\ }\textbf {\bibinfo {volume} {91}},\ \bibinfo
  {pages} {064051} (\bibinfo {year} {2015})},\ \Eprint
  {https://arxiv.org/abs/1412.3524} {arXiv:1412.3524 [gr-qc]} \BibitemShut
  {NoStop}%
\bibitem [{\citenamefont {Mu\~noz~de Nova}\ \emph {et~al.}(2019)\citenamefont
  {Mu\~noz~de Nova}, \citenamefont {Golubkov}, \citenamefont {Kolobov},\ and\
  \citenamefont {Steinhauer}}]{MunozdeNova:2018fxv}%
  \BibitemOpen
  \bibfield  {author} {\bibinfo {author} {\bibfnamefont {J.~R.}\ \bibnamefont
  {Mu\~noz~de Nova}}, \bibinfo {author} {\bibfnamefont {K.}~\bibnamefont
  {Golubkov}}, \bibinfo {author} {\bibfnamefont {V.~I.}\ \bibnamefont
  {Kolobov}},\ and\ \bibinfo {author} {\bibfnamefont {J.}~\bibnamefont
  {Steinhauer}},\ }\bibfield  {title} {\bibinfo {title} {{Observation of
  thermal Hawking radiation and its temperature in an analogue black hole}},\
  }\href {https://doi.org/10.1038/s41586-019-1241-0} {\bibfield  {journal}
  {\bibinfo  {journal} {Nature}\ }\textbf {\bibinfo {volume} {569}},\ \bibinfo
  {pages} {688} (\bibinfo {year} {2019})},\ \Eprint
  {https://arxiv.org/abs/1809.00913} {arXiv:1809.00913 [gr-qc]} \BibitemShut
  {NoStop}%
\bibitem [{\citenamefont {Del~Porro}(2024)}]{DelPorro_2024_thesis}%
  \BibitemOpen
  \bibfield  {author} {\bibinfo {author} {\bibfnamefont {F.}~\bibnamefont
  {Del~Porro}},\ }\emph {\bibinfo {title} {{Beyond Lorentz invariance: a
  journey from Analogue to Ho\v{r}ava Gravity}}},\ \href@noop {} {Ph.D.
  thesis},\ \bibinfo  {school} {SISSA} (\bibinfo {year} {2024})\BibitemShut
  {NoStop}%
\bibitem [{\citenamefont {Unruh}(1976)}]{unruh1976notes}%
  \BibitemOpen
  \bibfield  {author} {\bibinfo {author} {\bibfnamefont {W.~G.}\ \bibnamefont
  {Unruh}},\ }\bibfield  {title} {\bibinfo {title} {Notes on black-hole
  evaporation},\ }\href@noop {} {\bibfield  {journal} {\bibinfo  {journal}
  {Physical Review D}\ }\textbf {\bibinfo {volume} {14}},\ \bibinfo {pages}
  {870} (\bibinfo {year} {1976})}\BibitemShut {NoStop}%
\bibitem [{\citenamefont {Jacobson}(2003)}]{Jacobson:2003vx}%
  \BibitemOpen
  \bibfield  {author} {\bibinfo {author} {\bibfnamefont {T.}~\bibnamefont
  {Jacobson}},\ }\bibfield  {title} {\bibinfo {title} {{Introduction to quantum
  fields in curved space-time and the Hawking effect}},\ }in\ \href
  {https://doi.org/10.1007/0-387-24992-3_2} {\emph {\bibinfo {booktitle}
  {{School on Quantum Gravity}}}}\ (\bibinfo {year} {2003})\ pp.\ \bibinfo
  {pages} {39--89},\ \Eprint {https://arxiv.org/abs/gr-qc/0308048}
  {arXiv:gr-qc/0308048} \BibitemShut {NoStop}%
\bibitem [{\citenamefont {Senovilla}\ and\ \citenamefont
  {Torres}(2015)}]{Senovilla_2014}%
  \BibitemOpen
  \bibfield  {author} {\bibinfo {author} {\bibfnamefont {J.~M.~M.}\
  \bibnamefont {Senovilla}}\ and\ \bibinfo {author} {\bibfnamefont
  {R.}~\bibnamefont {Torres}},\ }\bibfield  {title} {\bibinfo {title}
  {{Particle production from marginally trapped surfaces of general
  spacetimes}},\ }\href {https://doi.org/10.1088/0264-9381/32/8/085004}
  {\bibfield  {journal} {\bibinfo  {journal} {Class. Quant. Grav.}\ }\textbf
  {\bibinfo {volume} {32}},\ \bibinfo {pages} {085004} (\bibinfo {year}
  {2015})},\ \bibinfo {note} {[Erratum: Class.Quant.Grav. 32, 189501 (2015)]},\
  \Eprint {https://arxiv.org/abs/1409.6044} {arXiv:1409.6044 [gr-qc]}
  \BibitemShut {NoStop}%
\end{thebibliography}%
\end{document}